%% file: main.tex
\documentclass[10pt,fleqn,a4paper]{article}
\pdfoutput=1
\usepackage{mathtools}
\usepackage{lipsum}
\usepackage{amsmath}
\usepackage{amssymb}
\usepackage{graphicx}
\usepackage[numbers,sort&compress]{natbib}
\usepackage{titlesec}
\usepackage{amsmath}
\usepackage{dsfont}
\usepackage{amsthm}
\usepackage{multicol}
\usepackage{bold-extra}
\usepackage[title]{appendix}
\usepackage[usenames,dvipsnames,svgnames,table]{xcolor}
\usepackage[margin=30pt,font=small,labelfont=bf,labelsep=endash]{caption}
\usepackage{amsthm}
\usepackage{rotating}
\usepackage{hyperref}
\usepackage{slashed} 
\usepackage{array,multirow}
\usepackage{shuffle}
\usepackage{empheq}
\usepackage{marvosym}
\usepackage{yfonts}
\usepackage[title]{appendix}
\usepackage[bottom]{footmisc}
\usepackage{tikz}
\usetikzlibrary{calc,matrix}
\usepackage{stackrel}
\usepackage{textcomp}
\usepackage{fancyvrb}
\usepackage{sectsty}
\usepackage{scalefnt}
\usepackage{setspace}
\usepackage{enumitem}

\newcommand{\cC}{{\cal C}}
\newcommand{\cD}{{\cal D}}
\newcommand{\cE}{{\cal E}}

\newcommand{\cG}{{\cal G}}

\newcommand{\cV}{{\cal V}}

\newcommand{\rd}{\text{d}}

\newcommand{\ie}{{\it i.e.\ }}

\newcommand{\eg}{{\it e.g.\ }}

\newcommand{\Erdos}{{Erd\H{o}s}}

\newcommand{\nAB}[1]{{n^{(#1)}_{\rm \scriptscriptstyle AB}}}

\newcommand{\pin}{{p_{\rm in}}}
\newcommand{\pout}{{p_{\rm out}}}

\newtheorem{mydef}{Definition}

\numberwithin{equation}{section}
\definecolor{myblue}{rgb}{0.97,0.97,0.97}

\begin{document}
\title{\bf\textcolor{black}{Topological aspects of the multi-language phases \\ of the Naming Game on community-based networks}}
\author{
\\[-0.1cm]
{{{Filippo Palombi$^{a}$\footnote{Corresponding
        author. E--mail: {\tt filippo.palombi@enea.it}\newline}\ \  and Simona Toti$^{b}$}}}\\[1.0ex]
 {{\small{$^a$ENEA---Italian National Agency for New Technologies, Energy and}}}\\
{{\small{Sustainable Economic Development,}}}
 {\small {{\it Via Enrico Fermi 45, 00044 Frascati -- Italy}}}\\[.1cm]
 {{\small{$^b$ISTAT---Italian National Institute of Statistics,}}}
 {\small {\it Via Cesare Balbo 16, 00184 Rome -- Italy}}\\[.1cm]
}

\date{\today}

\maketitle
\vskip -0.8cm
\input abstract

\input sect1

\input sect2
\input sect3

\input sect4

\input sect5
\input sect6

\input sect7
\input sect8
\input sect9
\input ackno
\input apndx
\bibliographystyle{hunsrt}
\bibliography{main}

\end{document}

%% file: abstract.tex
\begin{abstract}
  The Naming Game is an agent-based model where individuals communicate to name an initially unnamed object. On a large class of networks continual pairwise interactions lead the system to an ultimate consensus state, in which agents converge on a globally shared name. Soon after the introduction of the model, it was observed in literature that on community-based networks the path to consensus passes through metastable multi-language states. Subsequently, it was proposed to use this feature as a mean to discover communities in a given network. In this paper we show that metastable states correspond to genuine multi-language phases, emerging in the thermodynamic limit when the fraction of links connecting communities drops below critical thresholds. In particular, we study the transition to multi-language states in the stochastic block model and on networks with community overlap. We also examine the scaling of critical thresholds under variations of topological properties of the network, such as the number and relative size of communities and the structure of intra-/inter-community links. Our results provide a theoretical justification for the proposed use of the model as a community-detection algorithm.   
\end{abstract}

%% file: sect1.tex
\section{Introduction}

The emergence of spoken languages and their continuous evolution in human societies are complex phenomena in which interaction and self-organization play an essential r\^ole. Lying at the heart of opinion dynamics~\cite{fortcastlor}, these features have attracted great interest from researchers in statistical physics over the past twenty years. After some attempts to ascribe the origin of language conventions to evolutionary mechanisms~\cite{Niyogi,Nowak:1,Nowak:2,Nowak:3,Smith,Komarova}, in ref.~\cite{Baronchelli:1} a multi-agent model was proposed where the rise of a globally shared language occurs with no underlying guiding principle and no external influence. The model, known as the Naming Game ({\bf NG}), was inspired by the seminal work of refs.~\cite{Luc:1,Luc:2}.

\begin{figure}[t!]
  \centering
  \includegraphics[width=0.95\textwidth]{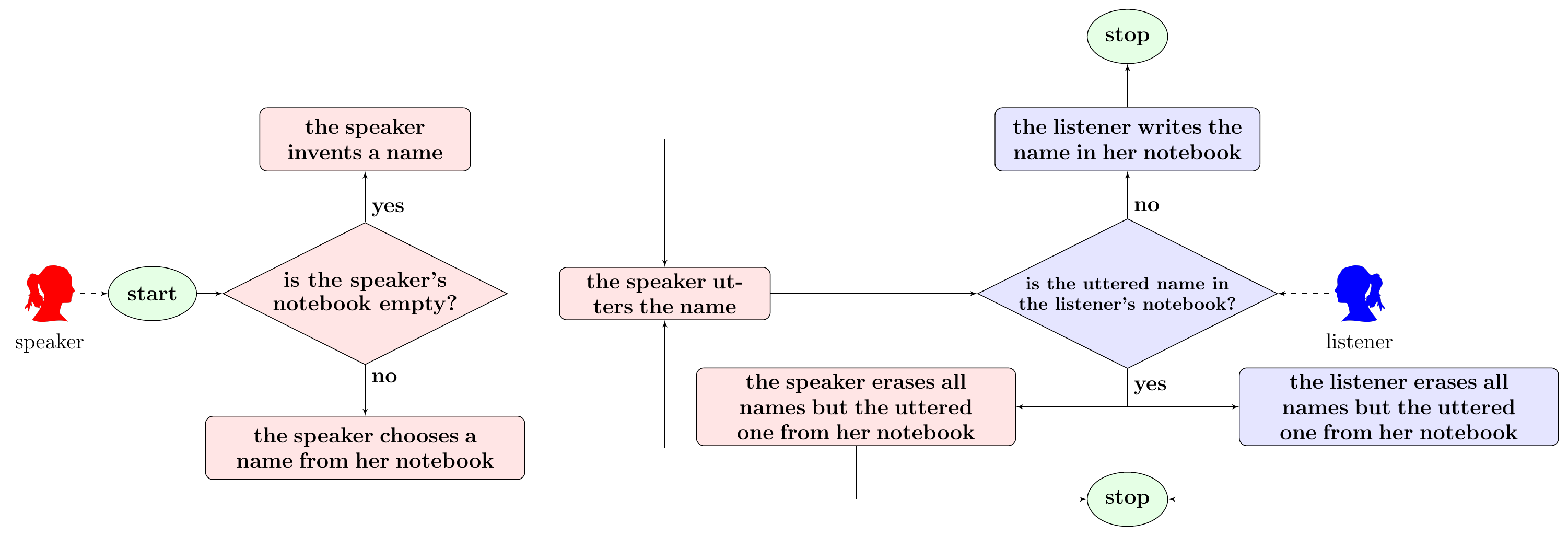}
  \caption{\footnotesize Speaker-listener interaction.\label{fig:gamerule}}
  \vskip -0.2cm
\end{figure}

The NG is a language-game in the sense of ref.~\cite{Wittgenstein}, with agents iteratively communicating to each other conventional names for a target object. Each agent is endowed with a notebook, in which he/she writes names. In the original version of the model all notebooks are initially empty. Elementary interactions involve two agents, playing respectively the r\^ole of \emph{speaker} and \emph{listener}. In each iteration the speaker is chosen randomly among the agents, while the listener is chosen randomly among the speaker's neighbours. The speaker-listener interaction is schematically described by the flowchart reported in Fig.~\ref{fig:gamerule}.

Following ref.~\cite{Baronchelli:1}, the dynamics of the NG was investigated on networks with several topologies, including the fully connected graph~\cite{Baronchelli:1,Baronchelli:5}, low-dimensional regular lattices~\cite{Baronchelli:3},~\Erdos-R\'enyi ({\bf ER}) graphs~\cite{Baronchelli:2}, small-world networks~\cite{Baronchelli:6}, Barab\'asi-Albert ({\bf BA}) networks~\cite{Baronchelli:2,Baronchelli:7}, etc. In all cases, the system was found to evolve dynamically with the number of different competing names initially inflating and then deflating due to self-organization, until the whole population agrees spontaneously on an ultimate name for the target object (consensus). Theoretical predictions derived from the NG have recently been shown to correctly reproduce experimental results in Web-based live games with controlled design~\cite{Centola}.

In ref.~\cite{Baronchelli:2} it was first pointed out that convergence to consensus follows a special pattern on community-based networks. Here, after a  \emph{``creative"} transient during which the number of different competing names inflates, the system relaxes to an equilibrium where different communities reach local consensus on different names. In a finite time dynamical fluctuations break the equilibrium and make the system fall into global consensus. The presence of metastable equilibria was soon realized to be of practical worth. In ref.~\cite{Lu:1} it was shown that local consensus might be used to identify communities in empirical networks. In a sense, this ratified the entrance of the NG into a large family of community detection algorithms~\cite{Schaeffer,Porter,Fortunato:1,Coscia,Newman:book,Xie,Fortunato:2}. More recently, the goodness of the community partition operated by the NG was investigated in terms of quality indicators~\cite{Gubanov} (such as the partition modularity~\cite{Newman}) on the benchmark networks of ref.~\cite{Lancichinetti}. It is important to recall that community detection is a major problem in network science, since modular networks arise in a variety of applicative contexts (see for instance refs.~\cite{Girvan:1,Porter,Flake}). It is also worth noting that the presence of metastable equilibria is not an exclusive feature of the NG. A similar phenomenon is observed in other models of opinion dynamics, such as the majority rule model~\cite{Lambiotte} and  an extension of the Axelrod model~\cite{Candia}. 

While local consensus exists on finite networks only in the form of metastable equilibrium, it becomes fully stable in the thermodynamic limit provided communities are sufficiently isolated. As a consequence, the phase diagram of the model develops a very rich structure. Communities agree or disagree on the ultimate name of their choice depending on how strongly they are connected to each other. Networks on which equal combinations of names survive at equilibrium in the thermodynamic limit correspond to the same multi-language phase. Despite a growing body of literature, a systematic study of the phase structure of the NG on community-based networks is still lacking. Aim of the present paper is to contribute to filling this gap. Studying the phase structure is important in order to identify theoretical limits within which the model can be used effectively as a community detection algorithm. However, phases depend upon all topological properties of communities, including their number, overlaps, relative size, internal topology and the topology of their interconnections. Since an overall parameterization of all such features is not given, we are forced to adopt a case-by-case strategy, where we investigate the transition to multi-language phases on groups of networks with distinct topological features.

We can summarize the results of our study by stating that \emph{i}) steady multi-languages states arise in the thermodynamic limit when links connecting agents in different communities are about 10-20\% or less of those connecting agents within their respective communities and \emph{ii}) multi-language phases look rather robust against changes in the network topology.

Before we start, we mention that multi-language phases are also observed in the NG under variations of its microscopic dynamics. Examples are the introduction of noise in the loss of memory when two agents agree on a given name~\cite{Baronchelli:4} or the introduction of single/opposing committed groups of agents who never change their notebook in time~\cite{Xie:1,Xie:2}. The difference is that communities produce stable multi-language states as a purely topological effect. This is a distinguishing feature of the NG: in other models of opinion dynamics communities are unable to hinder the convergence to global consensus, independently of their degree of isolation, even in the thermodynamic limit. An example is represented by the voter model \cite{Palombi}, where global consensus can be avoided only by introducing zealot agents~\cite{Mobilia:1,Mobilia:2} with competing opinions.

The plan of the paper is as follows. In sect.~2 we set up the notation and introduce the relative inter-community connectedness, a parameter that we use to compare results on different two-community symmetric networks. In sect.~3 we investigate the phase diagram of the  NG in the stochastic block model~\cite{Holland}. In sect.~4, we work out the exact solution to its mean field equations ({\bf MFEs}) in the special case of the planted partition model~\cite{Condon,McSherry} with two communities. In sect.~5 we derive MFEs for the NG on a network made of two overlapping cliques, then we work out an almost fully analytic solution to them. In sect.~6 we study how the phase transition depends on the number of communities in the planted partition model and in sect.~7 we study how it depends on their relative size. In sect.~8 we extend our study to heterogeneous networks by means of Monte Carlo simulations. Finally, we draw our conclusions in\break sect.~9.

%% file: sect2.tex
\section{Relative connectedness in two-community symmetric networks}

We consider a graph $\cG = (\cV,\cE)$ and a partition $\smash{\cV_\cC=\{\cC^{(k)}\}_{k=1}^Q}$ of $\cV$, \ie we assume $\smash{\cV = \cup_{k=1}^Q \cC^{(k)}}$ and $\smash{\cC^{(i)}\cap\cC^{(k)}=\emptyset}$ for $i\ne k$. We let $\smash{N^{(k)} = |\cC^{(k)}|>0}$ and $N=|\cV|$, hence we have $\smash{N=\sum_{k=1}^Q N^{(k)}}$. Then we observe that $\cV_\cC$ induces a partition $\cE_\cC = \smash{\{\cE^{(ik)}\}_{i,k=1}^Q}$ of $\cE$, \ie $\cE = \cup_{ik=1}^Q \cE^{(ik)}$ with
\begin{equation}
  \cE^{(ik)} = \{\,(x,y):\ x\in \cC^{(i)} \text{ and } y\in\cC^{(k)}\}\,.
\end{equation}
We take $(x,y)$ as an ordered pair. This implies by no means that the graph is either directed or undirected, but only that if $(x,y)\in\cE^{(i,k)}$, then $(x,y)\notin\cE^{(k,i)},$ for $i\ne k$. In particular, an undirected graph is obtained by requesting that $(x,y)\in\cE$ iff $(y,x)\in\cE$ and by considering $(x,y) = (y,x)$. In the sequel we always assume undirected graphs with $(x,x)\notin\cE$ for all $x$. We say that $\cV_\cC$ displays an explicit community structure ({\bf ECS}) provided
\begin{equation}
  |\cE^{(i,k)}|\ll\min(|\cE^{(ii)}|,|\cE^{(kk)}|)\,,\qquad \text{ for all } i\ne k\,.
  \label{eq:ecs}
\end{equation}
If eqs.~(\ref{eq:ecs}) are fulfilled, we interpret the sets $\cC^{(k)}$ as communities of agents. 
Although restrictive, the above ECS conditions leave several topological features of $\cG$ totally unspecified. For instance, in static network models (which are, however, unfit to describe realistic networks~\cite{BarabasiNS}) the topology is defined by assigning the $Q+1$ deterministic parameters $Q$, $\{N^{(k)}\}_{k=1}^Q$ and the edge probability laws
\begin{equation}
\mathfrak{p}^{(ik)}(x,y) = \text{prob}\left\{(x,y)\in \cE^{(ik)}\,\biggl|\, x\in\cC^{(i)} \text{ and } y\in\cC^{(k)}\right\}\,,\qquad i,k=1,\ldots,Q\,.
\end{equation}
In principle the functions $\smash{\mathfrak{p}^{(ik)}(x,y)}$ may be arbitrarily complex. They may depend explicitly on the community indexes $(ik)$, \ie for each choice of these they may depend upon different discrete or continuous parameters. Indeed, different static network models correspond to different settings of the above degrees of freedom. As such, they allow to explore (limited) subsets of the wider \emph{ensemble} defined by ECS conditions.

A relevant question is how to compare results for an agent-based model running on different network models, when these are defined in terms of different parameters. Unfortunately, there exists no universal answer to such question. Yet, for two-community networks which are symmetric under exchange of community indexes, we can use a simple indicator that allows to make comparisons. The indicator measures the relative extent to which communities are connected to each other rather than to themselves. To define it, we start from the notion of node degree, which counts the number of neighbours of a given node, and extend it to entire communities. We first introduce the inner average degree of the $i$th community
\begin{equation}
\langle\kappa^{(i)}_\text{in}\rangle = \frac{1}{N^{(i)}}\left\langle\sum_{x,y\in\cC^{(i)}}\mathbf{1}_{\cE^{(i,i)}}(x,y)\right\rangle = \frac{2\langle|\cE^{(ii)}|\rangle}{N^{(i)}}\,, \qquad i=1,2\,,
  \label{eq:gammain}
\end{equation}
where $\mathbf{1}_A(x)$ denotes the indicator function of $A$ (\ie $\mathbf{1}_A(x) = 1$ if $x\in A$ and $\mathbf{1}_A(x) = 0$ otherwise) and the symbol $\langle\, \cdot\, \rangle$ represents an average over the corresponding network model. By definition, we have $\langle \kappa^{(1)}_\text{in}\rangle = \langle \kappa^{(2)}_\text{in}\rangle$ on two-community symmetric networks. Then, we introduce the outer average degree of the $i$th to $k$th community
\begin{equation}
  \langle \kappa^{(i,k)}_\text{out}\rangle = \frac{1}{N^{(i)}}\left\langle\sum_{x\in\cC^{(i)}}\sum_{y\in\cC^{(k)}}\mathbf{1}_{\cE^{(i,k)}}(x,y)\right\rangle = \frac{\langle|\cE^{(ik)}|\rangle}{N^{(i)}}\,,\qquad i\ne k \,,
  \label{eq:gammaout}
\end{equation}
and again we observe that  $\langle \kappa^{(12)}_\text{out}\rangle = \langle \kappa^{(21)}_\text{out}\rangle$ on two-community symmetric networks. Finally, we define the relative inter-community connectedness as the ratio
\begin{equation}
  \gamma_\text{out/in} = \frac{\langle\kappa^{(12)}_{\text{out}}\rangle}{\langle\kappa^{(1)}_{\text{in}}\rangle} = \frac{1}{2}\frac{\langle|\cE^{(12)}|\rangle}{\langle|\cE^{(11)}|\rangle}\,.
  \label{eq:intercomconnect}
\end{equation}
ECS conditions are fulfilled provided $\gamma_{\text{out/in}}\ll 1$. We notice that $\gamma_\text{out/in}$ has a rather general valence in that either of eqs.~(\ref{eq:gammain})--(\ref{eq:gammaout}) depends by no means on the specific topology of $\cE^{(11)}$ and $\cE^{(12)}$. Unfortunately, there is no unambiguous way to generalize $\gamma_{\text{out/in}}$ to networks with two asymmetric and/or more than two communities. Such a generalization goes beyond our aims here.  

%% file: sect3.tex
\section{$Q$-ary Naming Game in the Stochastic Block Model}

As a first step we investigate the dynamics of the NG in the Stochastic Block Model~({\bf SBM})~\cite{Holland}. In the SBM we consider $Q$ communities with $N^{(i)}=N/Q$ for $i=1,\ldots, Q$. We introduce a set of $Q(Q+1)/2$ parameters $\{p^{(ik)}\}_{i\le k}^{1\ldots Q}$ and we assume $\mathfrak{p}^{(ik)}(x,y) = p^{(ik)}$ for all $i,k$. For $Q=2$ the SBM yields asymmetric networks whenever $p^{(11)}\ne p^{(22)}$. Hence, $\gamma_{\text{out/in}}$ is in general not well defined. 

As mentioned in sect.~1, in the original version of the NG \cite{Baronchelli:1} agents have empty notebooks at the beginning of the game, hence they invent names. After a while the number of different competing names observed across the network peaks at a value which is $\text{O}(N/2)$. Then, it decreases. If we identify the state of an agent with his/her notebook, we see that the number of allowed agent states (notebooks containing all possible combinations of the competing names) inflates exponentially just in the initial stage of the dynamics. This makes studying the system rather impractical beyond numerical simulations. In order to avoid such a complication, we resort to a trick which was first introduced in ref.~\cite{Baronchelli:9}: instead of starting the game with empty notebooks, we assign precisely one name to each agent. As a result the \emph{``creative''} transient disappears, while the left side of Fig.~\ref{fig:gamerule} reduces to a single square, with the speaker choosing randomly a name from his/her notebook and uttering it. Depending on how many different names we distribute across the network, the trick allows to set the overall dimension of the state space of the system. In particular, in ref.~\cite{Baronchelli:4} each agent was randomly assigned one of two names, respectively represented by letters $A$ and $B$.

Since we are interested in community-based networks, we find it preferable to prepare the initial state of the system with agents in a given community being assigned a common name and with different communities being assigned different names. We let $A_k$ represent the name initially assigned to $\cC^{(k)}$. Then we introduce a \emph{Rosetta} notebook\footnote{Evidently, this name evokes the famous stone rediscovered near the town of Rashid (Rosetta, Egypt) by Napoleon's army in 1799. The stone contained versions of the same text in Greek, Demotic and Hieroglyphic. As such, it served as a language translation tool.} $\cD = \{A_1,\ldots,A_Q\}$ and we let $S(\cD) = \{D:\, D\subset\cD\}$. At time $t\ge 0$ an agent $x$ has a certain notebook $D$, hence $D$ represents the state of $x$ at time $t$. We count the number of agent states $|S(\cD)|$ in full generality by counting all notebooks $D$ with $1\le |D|\le Q$ names. There are precisely
\vskip 0.1cm
\begin{itemize}[itemsep=0.4em]
\item{$Q$ notebooks with one name,}
\item{$\frac{1}{2!}Q(Q-1)$ notebooks with two names,}
\item{$\frac{1}{3!}Q(Q-1)(Q-2)$ notebooks with three names,}
\item[$\vdots$]{}
\item{$\frac{1}{Q!}Q(Q-1)(Q-2)\ldots 1 = 1$ notebooks with $Q$ names,}
\end{itemize}
\vskip 0.1cm
with the factorials at denominator ensuring that the inclusion of states differing by a permutation of names is avoided in the counting. By adding all the above numbers, we get
\begin{equation}
  |S(\cD)| = \sum_{k=1}^Q \frac{1}{k!}Q(Q-1)\cdots (Q-k+1) = \sum_{k=1}^Q \frac{Q!}{k!(Q-k)!} = \sum_{k=1}^Q {Q \choose k} = 2^Q - 1\,.
  \label{eq:numstates}
\end{equation}
We conclude that the number of agent states still increases exponentially with the number of communities; nevertheless, eq.~(\ref{eq:numstates}) represents the minimum one must cope with to study multi-language phases with no substantial restriction.

\subsection{Mean field equations}

MFEs describe correctly the dynamics of the system in the thermodynamic limit (where stochastic fluctuations become increasingly negligible). In the SBM we define this by letting $N\to\infty$ with $Q=\text{const.}$, $N^{(i)}/N = \text{const.}$ and $p^{(ik)}=\text{const.}$ for all $i,k$. To derive MFEs we need to take into account and correctly weigh all possible agent-agent interactions yielding an increase/decrease of the fraction of agents in a given state. For each notebook $D$ we introduce local densities
\begin{equation}
  n^{(i)}_D = \frac{\text{no. of agents with notebook $D$ belonging to $\cC^{(i)}$}}{N^{(i)}}\,.
  \label{eq:comdens}
\end{equation}
At each time the vectors $\{n^{(i)}_D\}_{D\in S(\cD)}$ fulfill simplex conditions separately for each community, that is
to say state densities are constrained by equations
\begin{equation}
  \sum_{D\in S(\cD)} n^{(i)}_D = 1\,,\qquad i = 1,\ldots,Q\,.
\end{equation}
Hence, there is one redundant state per community, whose density we represent in terms of the remaining ones via the corresponding simplex equation. We are free to choose the \emph{Rosetta} notebook $\cD$ as redundant state for all communities. If we let $\bar S(\cD) = S\setminus \cD$, then we have
\begin{equation}
  n^{(i)}_\cD = 1 - \sum_{D \in \bar S(\cD)} n^{(i)}_D\,,\qquad i = 1,\ldots,Q\,.
\end{equation}
Following this choice, we introduce the essential state vector
\begin{equation}
  \bar n = \{n^{(i)}_D:\ D\in \bar S(\cD) \ \ \text{and} \ \ i=1,\ldots, Q\}\,.
\end{equation}
The domain of $\bar n$ is the Cartesian product of $Q$ simplices. Taken as a whole, $\bar n(t)$ provides a full kinematic description of the state of the system at time $t$. Its trajectory in state space is mathematically described by a set of stochastic differential equations, governing the dynamics of the system under the joint action of deterministic drift and random diffusion terms. MFEs follow as the result of switching off all diffusion terms. They read 
\begin{equation}
  \frac{\rd n^{(i)}_D}{\rd t} = f^{(i)}_D(\bar n)\,,\qquad D\in \bar S(\cD)\,.
  \label{eq:mfebsm}
\end{equation}
The function $f^{(i)}_D$ yields the overall transition rate for the agent state $D$ in the $i$th community. It includes positive and negative contributions, each corresponding to an interaction involving two neighbouring agents belonging to $\cC^{(i)}$ or rather an agent belonging to $\cC^{(i)}$ and a neighbour lying somewhere else. We group terms contributing to $f^{(i)}_D$ in two different ways, namely
\vskip -0.7cm
\begin{equation}
  f^{(i)}_D \, = \, f^{(i,+)}_D - f^{(i,-)}_D \, = \, f^{(ii)}_D + \sum_{k\ne i}^{1\ldots Q}f^{(ik)}_D\,,  
\end{equation}
\vskip -0.5cm
\noindent where $f^{(i,+)}_D$ collects all positive contributions, $f^{(i,-)}_D$ all negative ones and $f^{(ik)}_D$ all contributions involving agents who belong respectively to the $i$th and $k$th communities. The first representation shows that $D$ is a steady state in the $i$th community provided the balance $f^{(i,+)}_D = f^{(i,-)}_D$ is exactly fulfilled. The second one allows to count easily the number of dimensions of the phase space of the system. Indeed, $f^{(ik)}_D$ is proportional to the probability of picking up an agent $x$ in the $i$th community and a neighbour $x'$ of $x$ in the $k$th one. This probability amounts to
\begin{equation}
  \pi^{(ik)} \,=\, \text{prob}\left\{x\in\cC^{(i)},\,x'\in\cC^{(k)}\right\} \ = \ \frac{1}{Q}\,\frac{p^{(ik)}}{\sum_{\ell=1}^Q p^{(i\ell)}} \ = \ \frac{1}{Q}\,\frac{\nu^{(ik)}}{1 + \sum_{\ell\ne i}^{1\ldots Q}\nu^{(i\ell)}}\,,
\end{equation}
with $\nu^{(ik)} = p^{(ik)}/p^{(ii)}$. When we look for a steady solution to eqs.~(\ref{eq:mfebsm}) we annihilate all derivatives on the left hand side. Since the denominator of $\pi^{(ik)}$ is the same for all $k$ with fixed $i$, we just factorize all such denominators and leave them out. We thus see that the only parameters a steady solution depends on are precisely the constants $\{\nu^{(ik)}\}_{i\ne k}$. Since these are all independent, we conclude that the phase space of the model has $Q(Q-1)$ dimensions. 

\begin{table}[!t]
  \begin{center}
    \small
    \begin{tabular}{|c|c|c|c|c|c|}
      \hline
      before interaction & after interaction & \multicolumn{4}{c|}{conditional transition rates} \\
      \hline
      \\[-2.8ex]
      $S^{(i)} \to L^{(k)}$ & $S^{(i)} - L^{(k)}$ & $\Delta n^{(i)}_{A_1}$ & $\Delta n^{(i)}_{A_2}$ & $\Delta n^{(k)}_{A_1}$ & $\Delta n^{(k)}_{A_2}$\\
      \hline\\[-2.8ex]
      $\phantom{A_2}A_1\stackbin{A_1}{\to}A_1\phantom{A_1}$ & $\phantom{A_2}A_1 - A_1\phantom{A_2}$ & 0 & 0 & 0 & 0 \\
      $\phantom{A_2}A_1\stackbin{A_1}{\to}A_2\phantom{A_1}$ & $\phantom{A_2}A_1 - A_1A_2$ & 0 & 0 & 0 & $-n^{(i)}_{A_1}n^{(k)}_{A_2}$ \\
      $\phantom{A_2}A_1\stackbin{A_1}{\to}A_1A_2$ & $\phantom{A_2}A_1 - A_1\phantom{A_2}$ & 0 & 0 & $n^{(i)}_{A_1}n^{(k)}_{A_1A_2}$ & 0  \\
      $\phantom{A_1}A_2\stackbin{A_2}{\to}A_1\phantom{A_2}$ & $\phantom{A_1}A_2 - A_1A_2$ & 0 & 0 & $-n^{(i)}_{A_2}n^{(k)}_{A_1}$ & 0 \\
      $\phantom{A_1}A_2\stackbin{A_2}{\to}A_2\phantom{A_1}$ & $\phantom{A_1}A_2 - A_2\phantom{A_1}$ & 0 & 0 & 0 & 0 \\
      $\phantom{A_1}A_2\stackbin{A_2}{\to}A_1A_2$ & $\phantom{A_1}A_2 - A_2\phantom{A_1}$ & 0 & 0 & 0 & $n^{(i)}_{A_2}n^{(k)}_{A_1A_2}$ \\
      $A_1A_2\stackbin{A_1}{\to}A_1\phantom{A_2}$ & $\phantom{A_2}A_1 - A_1\phantom{A_2}$ & $\frac{1}{2}n^{(i)}_{A_1A_2}n^{(k)}_{A_1}$ & 0 & 0 & 0 \\
      $A_1A_2\stackbin{A_1}{\to}A_2\phantom{A_1}$ & $A_1A_2 - A_1A_2$ & 0 & 0 & 0 & $-\frac{1}{2}n^{(i)}_{A_1A_2}n^{(k)}_{A_2}$ \\
      $A_1A_2\stackbin{A_1}{\to}A_1A_2$ & $\phantom{A_2}A_1 - A_1\phantom{A_2}$ & $\frac{1}{2}n^{(i)}_{A_1A_2}n^{(k)}_{A_1A_2}$ & 0 & $\frac{1}{2}n^{(i)}_{A_1A_2}n^{(k)}_{A_1A_2}$ & 0 \\
      $A_1A_2\stackbin{A_2}{\to}A_1\phantom{A_2}$ & $A_1A_2 - A_1A_2$ & 0 & 0 & $-\frac{1}{2}n^{(i)}_{A_1A_2}n^{(k)}_{A_1}$ & 0 \\
      $A_1A_2\stackbin{A_2}{\to}A_2\phantom{A_1}$ & $\phantom{A_1}A_2 - A_2\phantom{A_1}$ & 0 &  $\frac{1}{2}n^{(i)}_{A_1A_2}n^{(k)}_{A_2}$ & 0 & 0 \\
      $A_1A_2\stackbin{A_2}{\to}A_1A_2$ & $\phantom{A_1}A_2 - A_2\phantom{A_1}$ & 0 & $\frac{1}{2}n^{(i)}_{A_1A_2}n^{(k)}_{A_1A_2}$ & 0 & $\frac{1}{2}n^{(i)}_{A_1A_2}n^{(k)}_{A_1A_2}$ \\[1.0ex]
      \hline
    \end{tabular}
    \caption{\footnotesize Conditional transition rates for speaker-listener interactions. Labels $S^{(i)}$ and $L^{(k)}$ denote respectively a speaker in $\cC^{(i)}$ and a listener in $\cC^{(k)}$, $i,k=1,2$.\label{tab:binaryNG}}
  \end{center}
  \vskip -0.4cm
\end{table}

Using the representation $f^{(i)}_D = f^{(i,+)}_D - f^{(i,-)}_D$ is more convenient for calculational purposes. All speaker-listener interactions of a given type generate algebraic expressions differing only in the community indexes carried by $\{\pi^{(ik)}\}$. Such expressions can be easily grouped together. To work out $f^{(i,\pm)}_D$, it is advisable to first enumerate its contributions. In full generality we let 
\begin{equation}
  f^{(i,\pm)}_D = \sum_{\alpha=1}^{N_{\pm,D}}f^{(i,\pm,\alpha)}_D\,,
\end{equation}
where $f_D^{(i,\pm,\alpha)}$ includes all interactions of the $\alpha$th type increasing/decreasing $n^{(i)}_D$ and $N_{\pm,D}$ denotes the overall number of interaction types yielding an increase/decrease of $n^{(i)}_D$. As we just noticed, each contribution to $f_D^{(i,\pm,\alpha)}$ is proportional to $\pi^{(ik)}$ for some $k$. The proportionality factor yields the conditional transition rate $\Delta n^{(i)}_D$ of an interaction between agents $x$ and $x'$ given $x\in\cC^{(i)}$ and $x'\in\cC^{(k)}$. Only in the specific case of the binary NG, where $S(\cD) = \left\{\{A_1\},\{A_2\},\{A_1,A_2\}\right\}$, can such conditional rates be simply enumerated and calculated with paper and pencil. Indeed, these have concise and well known expressions. For the reader's convenience we report them all in Table~\ref{tab:binaryNG} (to keep the notation simple, here as well as in the sequel we allow expressions such as $n^{(i)}_{A}$ in place of $\smash{n^{(i)}_{\{A\}}}$). For $Q>2$ the number of contributions increases. For the sake of readability, we refer the reader to App. A for a complete derivation of MFEs.

\subsection{Phase diagram for $Q=2$}

As explained above, the phase space of the NG in the SBM corresponds to the 1st orthant of the $Q(Q-1)$-dimensional Euclidean space generated by  parameters $\{\nu^{(ik)}\}_{i\ne k}^{1\ldots Q}$. Recall that we start the game with initial configuration
\begin{equation}
  n^{(k)}_D(0) = \left\{\begin{array}{ll} 1 & \text{if } D = \{A_k\}\,, \\[1.0ex]
  0 & \text{otherwise}\,,\end{array}\right.\qquad k=1,\ldots,Q\,.
  \label{eq:initcond}
\end{equation}
After a while the system reaches an equilibrium state where a certain number of names are left out in favour of others. Surviving names are found not necessarily only in their original communities, but also in other ones over which they spread along the dynamics. If $n^{(k)}_{A_\ell}(t\to\infty)\simeq 1$ for $\ell\ne k$, we say that $A_\ell$ has \emph{colonized} $\cC^{(k)}$. Phases correspond to all possible ways names colonize communities. In principle, their total number is given by
\begin{equation}
  \text{no. of phases } \, =  \, \sum_{k=1}^Q{Q\choose k}\sum_{\substack{n_1\ldots n_k=0 \\[1.0ex] n_1 + \ldots + n_k = Q-k}}^{Q-k}\frac{(Q-k)!}{n_1!\ldots n_k!} \, = \, \sum_{k=1}^Q {Q\choose k}k^{Q-k}\,.
  \label{eq:nophases}
\end{equation}
Indeed, assume that  $k$ names survive in the final state. The number of ways to choose them out of a set of $Q$ names is ${Q\choose k}$, which explains the presence of the binomial coefficient in eq.~(\ref{eq:nophases}). We have to sum over $k=1,\ldots,Q$ to take into account all possibilities. The $k$ surviving names certainly dominate their respective communities, so we are left with the task of distributing them across the remaining $Q-k$ ones. The number of ways to distribute $n_1$ copies of the first name, $n_2$ copies of the second one and so forth is  $(Q-k)!/(n_1!\ldots n_k!)$, with the factorials at denominator removing unwanted repetitions. The total number of phases is finally obtained by summing over all possible choices of $n_1,\ldots,n_k$. The rightmost expression in eq.~(\ref{eq:nophases}) simply follows from the multinomial theorem. In Table~\ref{tab:phases}, we report the number of phases for the lowest few values of $Q$. Each phase occupies a sharply bounded region in the phase space. As the reader may notice, the phase structure of the model becomes increasingly complex as $Q$ increases. 

\begin{table}[!t]
  \begin{center}
    \begin{tabular}{c||c|c|c|c|c|c|c}
      \hline\hline
      Q$\phantom{\bigr|^f}$ & 2 & 3 & 4 & 5 & 6 & 7 & 8 \\
      \hline
      \text{no. of phases$\phantom{\bigr|^f}$} & 3 & 10 & 41 & 196 & 1057 & 6322 & 41393 \\
      \hline\hline
    \end{tabular}
    \vskip 0.2cm
    \caption{\footnotesize Number of phases in the SBM with $Q$ communities.\label{tab:phases}}
  \end{center}
    \vskip -0.4cm
\end{table}

\begin{figure}[t!]
\centering
\hskip -0.8cm\includegraphics[width=0.84\textwidth]{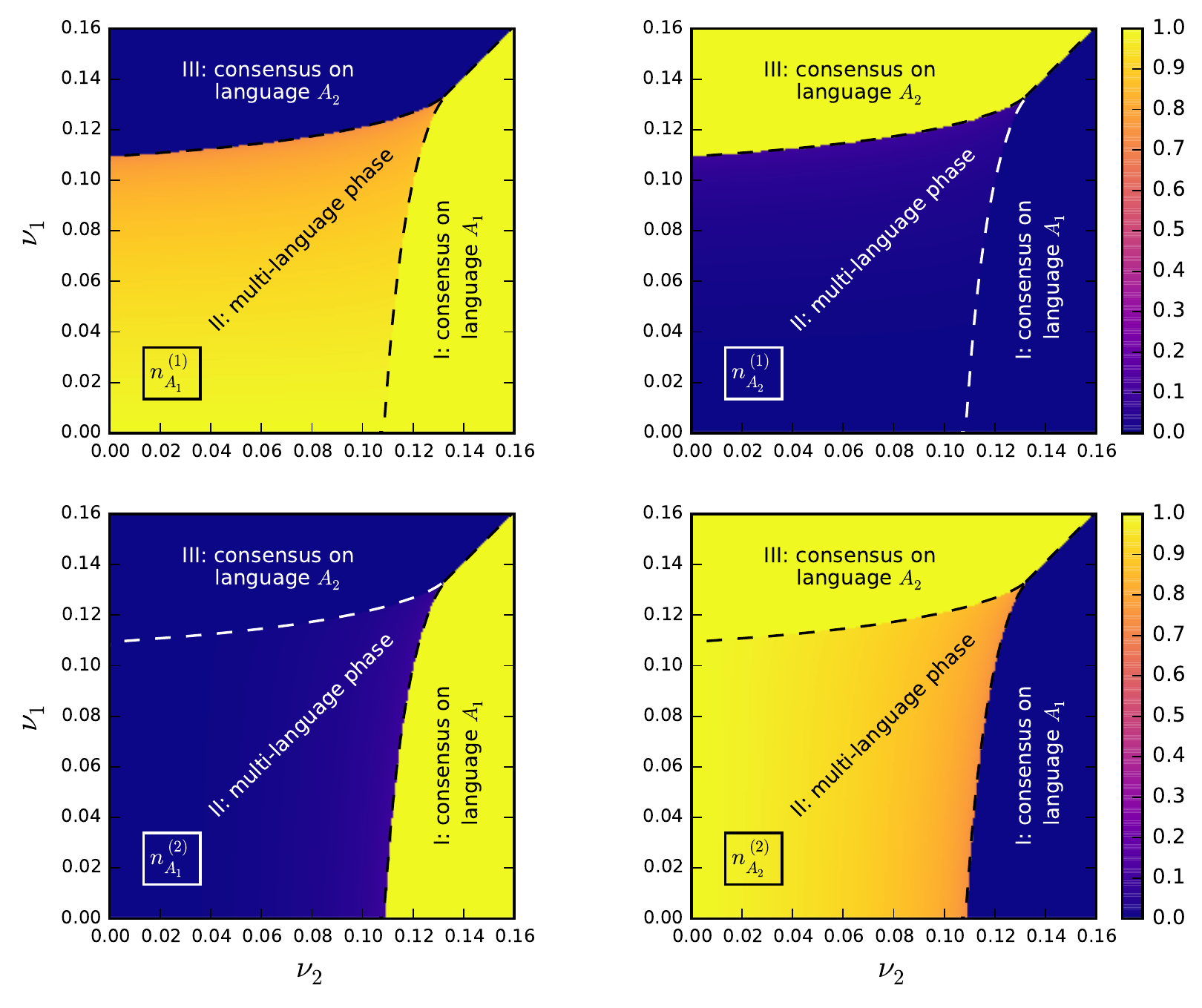}
\vskip -0.2cm
\caption{\footnotesize Phase diagram in the SBM with $Q=2$.\label{fig:phases}}
\vskip -0.2cm
\end{figure}

The only case where the phase diagram can be easily studied is for $Q=2$. For notational simplicity, we introduce symbols $\nu_1 = p^{(12)}/p^{(11)}$ and $\nu_2 = p^{(12)}/p^{(22)}$ in place of $\nu^{(12)}$ and $\nu^{(21)}$ respectively. Notice that $\nu_1$ and $\nu_2$ increase when the  inter-community links become denser and also when the intra-community ones rarefy. MFEs can be easily worked out, either thanks to Table~\ref{tab:binaryNG} or via the code provided in App.~B. They read
\begin{align}
  \frac{\rd n^{(1)}_{A_1}}{\rd t} & = \pi^{(11)}\left\{ n^{(1)}_{A_1}n^{(1)}_{A_1A_2} + (n^{(1)}_{A_1A_2})^2 - n^{(1)}_{A_1}n^{(1)}_{A_2}\right\}\nonumber \\[0.0ex]
  & + \pi^{(12)}\left\{ \frac{3}{2}n^{(1)}_{A_1A_2}n^{(2)}_{A_1} - \frac{1}{2}n^{(1)}_{A_1}n^{(2)}_{A_1A_2} + n^{(1)}_{A_1A_2}n^{(2)}_{A_1A_2} - n^{(1)}_{A_1}n^{(2)}_{A_2} \right\}\,,
  \label{eq:sbmfrst}\\[1.5ex]
  \frac{\rd n^{(1)}_{A_2}}{\rd t} & = \pi^{(11)}\left\{ n^{(1)}_{A_2}n^{(1)}_{A_1A_2} + (n^{(1)}_{A_1A_2})^2 - n^{(1)}_{A_1}n^{(1)}_{A_2} \right\}\nonumber \\[0.0ex]
  & + \pi^{(12)}\left\{\frac{3}{2}n^{(1)}_{A_1A_2}n^{(2)}_{A_2}-\frac{1}{2}n^{(1)}_{A_2}n^{(2)}_{A_1A_2}+n^{(1)}_{A_1A_2}n^{(2)}_{A_1A_2}-n^{(1)}_{A_2}n^{(2)}_{A_1}  \right\}\,,
  \label{eq:sbmscnd}\\[1.5ex]
  \frac{\rd n^{(2)}_{A_1}}{\rd t} & = \pi^{(22)}\left\{ n^{(2)}_{A_1}n^{(2)}_{A_1A_2} + (n^{(2)}_{A_1A_2})^2 - n^{(2)}_{A_1}n^{(2)}_{A_2}\right\}\nonumber \\[0.0ex]
  & + \pi^{(21)}\left\{ \frac{3}{2}n^{(2)}_{A_1A_2}n^{(1)}_{A_1} - \frac{1}{2}n^{(2)}_{A_1}n^{(1)}_{A_1A_2} + n^{(2)}_{A_1A_2}n^{(1)}_{A_1A_2} - n^{(2)}_{A_1}n^{(1)}_{A_2} \right\}\,,
  \label{eq:sbmthrd}\\[1.5ex]
  \frac{\rd n^{(2)}_{A_2}}{\rd t} & = \pi^{(22)}\left\{ n^{(2)}_{A_2}n^{(2)}_{A_1A_2} + (n^{(2)}_{A_1A_2})^2 - n^{(2)}_{A_1}n^{(2)}_{A_2} \right\}\nonumber \\[0.0ex]
  & + \pi^{(21)}\left\{\frac{3}{2}n^{(2)}_{A_1A_2}n^{(1)}_{A_2}-\frac{1}{2}n^{(2)}_{A_2}n^{(1)}_{A_1A_2}+n^{(2)}_{A_1A_2}n^{(1)}_{A_1A_2}-n^{(2)}_{A_2}n^{(1)}_{A_1}  \right\}\,.
  \label{eq:sbmfrth}
\end{align}
The phase diagram of the model, obtained by integrating eqs.~(\ref{eq:sbmfrst})--(\ref{eq:sbmfrth}) numerically, is reported in Fig.~\ref{fig:phases}. We observe three different phases: in region I the system converges to a global consensus state where $A_1$ colonizes $\cC^{(2)}$, in region III it converges to the opposite global consensus state, with $A_2$ colonizing $\cC^{(1)}$, while region II corresponds to a multi-language phase. Here large fractions of both communities keep speaking their original language without ever converging to global consensus. It is interesting to observe that the phase structure of the model is qualitatively similar to that obtained for the binary NG on a fully connected graph when competing committed groups of agents are introduced, see Fig.~1 of ref.~\cite{Xie:2}. Nevertheless, the phase structure here is fully induced by the network topology. Moreover, the cusp of region II, that we shall derive exactly in next section, is located at $\nu_1=\nu_2= 0.1321\ldots$, while in ref.~\cite{Xie:2} it is located at $p_\text{A}=p_\text{B}=0.1623\ldots$.

%% file: sect4.tex
\section{Binary dynamics in the Planted Partition Model with $Q=2$}

The Planted Partition Model~({\bf PPM})~\cite{Condon,McSherry} includes all networks of the SBM generated by letting $p^{(ii)}=\pin$ for $i=1,\ldots,Q$ and $p^{(ik)} = \pout$ for $i\ne k$. Networks in the PPM are fully symmetric under exchange of community indexes. For $Q=2$ we have
\begin{align}
  & \langle\kappa^{(1)}_{\text{in}}\rangle = \langle\kappa^{(2)}_{\text{in}}\rangle = \frac{2}{N}\frac{N}{2}\pin\left(\frac{N}{2}-1\right) = \pin\left(\frac{N}{2}-1\right)\,,\\[4.0ex]
 & \langle\kappa^{(12)}_{\text{out}}\rangle = \langle\kappa^{(21)}_{\text{out}}\rangle = \frac{2}{N}\frac{N}{2}\pout\frac{N}{2} = \pout\frac{N}{2}\,,
 \end{align}
 hence
 \begin{equation}
   \gamma_\text{out/in} = \frac{\pout}{\pin} \frac{1}{\left(1-2/N\right)} \simeq \pout/\pin \equiv \nu\,.
   \label{eq:PPMgamma}
 \end{equation}
ECS conditions are fulfilled by graphs with $\nu\ll 1$. In this limit the PPM is in absolute the simplest \emph{ensemble} of community-based networks.

\begin{figure}[t!]
  \centering
  \hskip -0.8cm\includegraphics[width=0.48\textwidth]{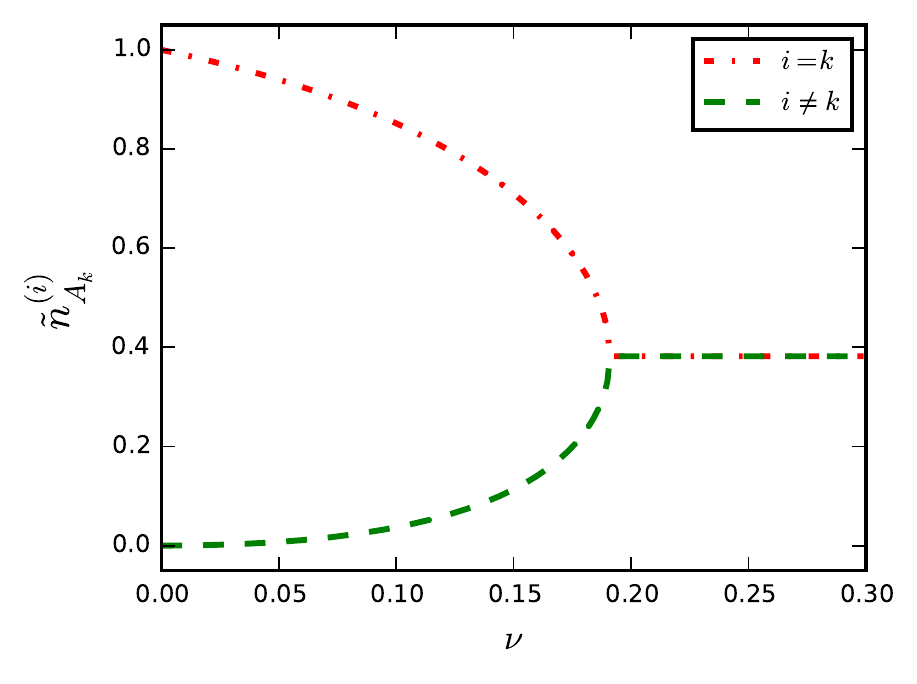}
  \hskip 0.2cm\includegraphics[width=0.48\textwidth]{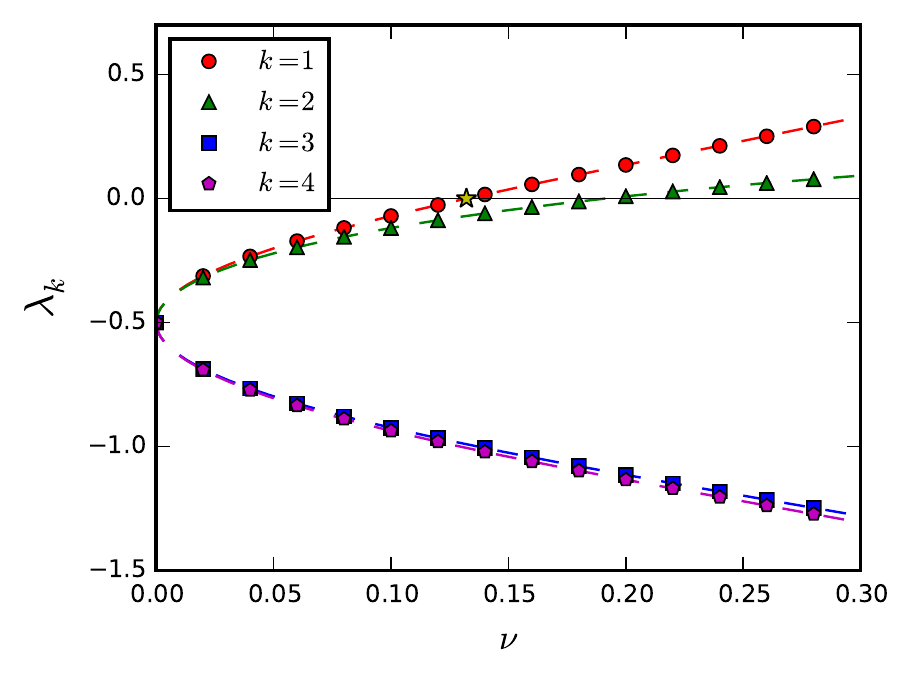}\,.
  \vskip -0.2cm
  \caption{\footnotesize (\emph{left}) Symmetric steady solution to MFEs in the PPM with $Q=2$. (\emph{right}) Eigenvalues of linearized MFEs. The critical point $\nu_c$ is represented by a star.}
  \label{fig:ppmstability}
\end{figure}

For $Q=2$ the phase space of the system corresponds to the bisecting line of Fig.~\ref{fig:phases}, where $\nu_1=\nu_2\equiv\nu$. When the system relaxes to equilibrium, all derivatives on the l.h.s. of eqs.~(\ref{eq:sbmfrst})--(\ref{eq:sbmfrth}) vanish. Therefore, steady densities are determined by a system of algebraic equations. We want to show that the latter admit a symmetric solution $\tilde n = \{\tilde n^{(1)}_{A_1},\tilde n^{(1)}_{A_2},\tilde n^{(2)}_{A_1},\tilde n^{(2)}_{A_2}\}$ with $\tilde n^{(1)}_{A_1}=\tilde n^{(2)}_{A_2}=x$ and $\tilde n^{(1)}_{A_2}=\tilde n^{(2)}_{A_1}=y$.  This turns out to be stable only for $\nu$ lying within region II of Fig.~\ref{fig:phases}. When $\nu$ lies outside it, the symmetric solution becomes unstable under small density perturbations. In this region the symmetry is broken by dynamical fluctuations, leading to global consensus on $A_1$ or $A_2$ with equal probability. Depending on how large $\nu$ is, instabilities may be triggered by density fluctuations occurring along one specific direction or spanning an entire plane in state space, as we shall see in a while. As a result of our \emph{ansatz} two of MFEs become redundant, so we are left with
\begin{align}
  \label{eq:ppmfrst}
  & x(1-x-y) + (1-x-y)^2 -xy \nonumber\\[0.0ex]
  & \hskip 2.0cm + \nu\left\{\frac{3}{2}y(1-x-y) - \frac{1}{2}x(1-x-y) + (1-x-y)^2 - x^2\right\} = 0\,,\\[4.0ex]
  \label{eq:ppmscnd}
  & y(1-x-y) + (1-x-y)^2 -xy \nonumber\\[0.0ex]
  & \hskip 2.0cm + \nu\left\{\frac{3}{2}x(1-x-y) - \frac{1}{2}y(1-x-y) + (1-x-y)^2 - y^2\right\} = 0\,,
\end{align}
We let $u=x-y$ and $v = 1-x-y$. Adding and subtracting the above two equations yields the equivalent system
\begin{align}
  \label{eq:ppmthrd}
  & \left\{v(1-v) + 2v^2 -\frac{1}{2}[(1-v)^2 - u^2]\right\} + \nu \left\{v(1-v) + 2v^2 - \frac{1}{2}[(1-v)^2+u^2]\right\}=0\,,\\[3.0ex]
  \label{eq:ppmfrth}
  & uv + \nu\left\{-\frac{3}{2}uv - \frac{1}{2}uv - u(1-v)\right\} = 0 \qquad \Leftrightarrow \qquad u\left[v - \nu(1+v)\right] = 0\,.
\end{align}
In particular, eq.~(\ref{eq:ppmfrth}) has two solutions: \emph{i}) $u=0$ and \emph{ii}) $u\ne 0$, $v = \nu/(1-\nu)$. These hold separately for $\nu$ belonging to disjoint subintervals of $[0,1]$. If we assume first that $u\ne 0$, from eq.~(\ref{eq:ppmthrd}) it follows that 

\begin{equation}
  u^2 = -2\frac{1+\nu}{1-\nu}\left\{\frac{v^2}{2} + 2v - \frac{1}{2}\right\}\,.
  \label{eq:ueq}
\end{equation}
Inserting $v=\nu/(1-\nu)$ into this yields
\begin{equation}
  u(\nu) = \pm \sqrt{\frac{1+\nu}{(1-\nu)^3}\left(4\nu^2 - 6\nu + 1\right)}\,.
\end{equation}
To ensure that $u(\nu)$ is real, we must have $0<\nu\le \hat\nu =  (3-\sqrt{5})/4 = 0.190983\ldots$
Moreover, from eq.~(\ref{eq:ueq}) we see that $u=0$ entails $v = \sqrt{5}-2 = 0.236068\ldots$ This represents a solution for $\nu>\hat\nu$. Therefore, with initial conditions $n^{(1)}_{A_1} = n^{(2)}_{A_2}=1$ and $n^{(1)}_{A_2} = n^{(2)}_{A_1} = 0$, the symmetric steady solution is given by
\begin{equation}
\left\{\begin{array}{ll}
  \tilde n^{(i)}_{A_i}(\nu) & \hskip -0.2cm = \dfrac{1}{2}\left\{\dfrac{1-2\nu}{1-\nu} + \sqrt{\dfrac{(1+\nu)}{(1-\nu)^3}\left(4\nu^2 - 6\nu + 1\right)}  \right\}\,,\\[4.0ex]
  \tilde n^{(i)}_{A_{3-i}}(\nu) & \hskip -0.2cm = \dfrac{1}{2}\left\{\dfrac{1-2\nu}{1-\nu} - \sqrt{\dfrac{(1+\nu)}{(1-\nu)^3}\left(4\nu^2 - 6\nu + 1\right)}  \right\}\,,
\end{array}\right.\quad \text{ for } \ \nu\le \hat\nu \ \text{ and } \ i=1,2\,,
  \label{eq:symsol}
\end{equation}
and
\begin{equation}
  \tilde n^{(1)}_{A_1}(\nu) = \tilde n^{(1)}_{A_2}(\nu) = n^{(2)}_{A_1}(\nu) = \tilde n^{(2)}_{A_2}(\nu) = \frac{3-\sqrt{5}}{2}\,,\quad \ \text{ for } \ \nu\ge \hat\nu\,.
\end{equation}
In Fig.~\ref{fig:ppmstability} (\emph{left}) we plot the symmetric steady densities in $\cC^{(1)}$ vs $\nu$. We notice that both $\tilde n^{(1)}_{A_1}(\nu)$ and $\tilde n^{(1)}_{A_2}(\nu)$ have discontinuous first order derivatives for $\nu=\hat\nu$.

\subsection{Stability of the symmetric steady solution}

In order to investigate the stability of the symmetric solution, we consider densities deviating by a small amount from $\tilde n$, \ie we let $n^{(i)}_{\scriptscriptstyle X} = \tilde n^{(i)}_{\scriptscriptstyle X} +\, \epsilon^{(i)}_{\scriptscriptstyle X}$ for $i=1,2$ and ${X} = {\rm A_1,A_2}$. Then we examine the conditions under which all deviations $\{\epsilon^{(i)}_{\scriptscriptstyle X}(t)\}_{X=A_1,A_2}^{i=1,2}$ vanish simultaneously as $t\to\infty$. By inserting such perturbations into MFEs and by expanding in Taylor series at leading order, we obtain linearized MFEs 
\vskip -0.5cm
\begin{equation}
  \frac{\rd\epsilon^{(i)}_{\scriptscriptstyle X}}{\rd t} = \sum_{Y=A_1,A_2}\sum_{k=1,2} \epsilon^{(k)}_{\scriptscriptstyle Y} \frac{\partial f_{\scriptscriptstyle X}^{(i)}}{\partial n^{(k)}_{\scriptscriptstyle Y}}(\tilde n)\,,\qquad i=1,2\ \text { and }\ X = A_1,A_2\,.
\end{equation}
The stability matrix $\Lambda^{(i,{\scriptscriptstyle X})}_{(k,{\scriptscriptstyle Y})} = \partial f^{(i)}_{\scriptscriptstyle X}/\partial n^{(k)}_{\scriptscriptstyle Y}(\tilde n)$ has constant elements, depending on the components of $\tilde n$ and the relative  connectedness $\nu$ (but not on $\{\epsilon^{(i)}_{\scriptscriptstyle X}\}$). In particular, we find
\begin{alignat}{2}
  \sigma\Lambda^{(1,{\scriptscriptstyle A_1})}_{(1,{\scriptscriptstyle A_1})} & = - 1 - \frac{3}{2}\nu + \frac{\nu}{2}\tilde n^{(2)}_{A_2}\,, & 
  \sigma\Lambda^{(1,{\scriptscriptstyle A_1})}_{(1,{\scriptscriptstyle A_2})} & =  -2  - \nu - \frac{\nu}{2} \tilde n^{(2)}_{A_1} + 2\tilde n^{(1)}_{A_2}  + \nu \tilde n^{(2)}_{A_2}\,,\\[1.0ex]
  \sigma\Lambda^{(1,{\scriptscriptstyle A_1})}_{(2,{\scriptscriptstyle A_1})} & =  \frac{1}{2}\nu -\frac{\nu}{2}\tilde n^{(1)}_{A_2}\,, & 
  \sigma\Lambda^{(1,{\scriptscriptstyle A_1})}_{(2,{\scriptscriptstyle A_2})} & =  -\nu + \frac{\nu}{2}\tilde n^{(1)}_{A_1}+\nu\tilde n^{(1)}_{A_2}\, 
\end{alignat}
\begin{alignat}{2}
  \sigma\Lambda^{(1,{\scriptscriptstyle A_2})}_{(1,{\scriptscriptstyle A_1})} & =  -2 - \nu  - \frac{\nu}{2} \tilde n^{(2)}_{A_2}  + 2 \tilde n^{(1)}_{A_1} + \nu \tilde n^{(2)}_{A_1}\,,\qquad& 
  \sigma\Lambda^{(1,{\scriptscriptstyle A_2})}_{(1,{\scriptscriptstyle A_2})} & = -1 -\frac{3}{2}\nu + \frac{\nu}{2} \tilde n^{(2)}_{A_1}\, \\[1.0ex]
  \sigma\Lambda^{(1,{\scriptscriptstyle A_2})}_{(2,{\scriptscriptstyle A_1})} & = -\nu  + \frac{\nu}{2} \tilde n^{(1)}_{A_2} + \nu\tilde n^{(1)}_{A_1}\,, & 
  \sigma\Lambda^{(1,{\scriptscriptstyle A_2})}_{(2,{\scriptscriptstyle A_2})} & = \frac{\nu}{2} - \frac{\nu}{2}\tilde n^{(1)}_{A_1}\,, \\[1.0ex]
  \sigma\Lambda^{(2,{\scriptscriptstyle A_1})}_{(1,{\scriptscriptstyle A_1})} & = \frac{\nu}{2} - \frac{\nu}{2}\tilde n^{(2)}_{A_2}\,,& 
  \sigma\Lambda^{(2,{\scriptscriptstyle A_1})}_{(1,{\scriptscriptstyle A_2})} & =  -\nu + \frac{\nu}{2}\tilde n^{(2)}_{A_1} + \nu\tilde n^{(2)}_{A_2} \,, \\[1.0ex]
  \sigma\Lambda^{(2,{\scriptscriptstyle A_1})}_{(2,{\scriptscriptstyle A_1})} & = -1-\frac{3}{2}\nu + \frac{\nu}{2}\tilde n^{(1)}_{A_2}  \,,& 
  \sigma\Lambda^{(2,{\scriptscriptstyle A_1})}_{(2,{\scriptscriptstyle A_2})} & = -2 -\nu - \frac{\nu}{2}\tilde n^{(1)}_{A_1} + 2 \tilde n^{(2)}_{A_2}  + \nu\tilde n^{(1)}_{A_2} \,, \\[1.0ex]
  \sigma\Lambda^{(2,{\scriptscriptstyle A_2})}_{(1,{\scriptscriptstyle A_1})} & =  -\nu +\frac{\nu}{2}\tilde n^{(2)}_{A_2}+ \nu \tilde n^{(2)}_{A_1} \,, & 
  \sigma\Lambda^{(2,{\scriptscriptstyle A_2})}_{(1,{\scriptscriptstyle A_2})} & =  \frac{\nu}{2} - \frac{\nu}{2}\tilde n^{(2)}_{A_1} \,, \\[1.0ex]
  \sigma\Lambda^{(2,{\scriptscriptstyle A_2})}_{(2,{\scriptscriptstyle A_1})} & =  -2 -\nu - \frac{\nu}{2}\tilde n^{(1)}_{A_2} + 2\tilde n^{(2)}_{A_1} + \nu\tilde n^{(1)}_{A_1} \,, & 
  \sigma\Lambda^{(2,{\scriptscriptstyle A_2})}_{(2,{\scriptscriptstyle A_2})} & =  -1 -\frac{3}{2}\nu + \frac{\nu}{2}\tilde n^{(1)}_{A_1} \,, 
\end{alignat}
\vskip 0.2cm
\noindent with $\sigma=2(1+\nu)$. It is possible to work out the four eigenvalues of $\Lambda$ exactly, either by paper-and-pencil calculations or via a simple Maple$^\text{TM}$ script. Rather exceptionally, their algebraic expressions are sufficiently concise to allow us to report them in full. Indeed, we have
\begin{align}
  \lambda_1 & =  \frac{1}{4}\frac{3\nu^2 - 2 + \sqrt{\nu^4 -20\nu^3 + 8\nu^2 + 28\nu}}{1-\nu^2} \,,\\[3.7ex]
  \lambda_2 & =  \frac{1}{4}\frac{\nu^2 -\nu - 2 + \sqrt{17\nu^4 - 26\nu^3 - 15\nu^2 + 28\nu}}{1-\nu^2}\,,\\[3.7ex]
  \lambda_3 & =  \frac{1}{4}\frac{ 3\nu^2 - 2 - \sqrt{\nu^4 -20\nu^3 + 8\nu^2 + 28\nu}}{1-\nu^2} \,,\\[3.7ex]
  \lambda_4 & =  \frac{1}{4}\frac{\nu^2 -\nu - 2 - \sqrt{17\nu^4 - 26\nu^3 - 15\nu^2 + 28\nu}}{1-\nu^2}\,.
\end{align}
The behaviour of these eigenvalues as functions of $\nu$ is reported in Fig.~\ref{fig:ppmstability} (\emph{right}). For sufficiently small $\nu$ all of them are negative, thus granting that the symmetric steady solution is stable. In fact, the transition to the multi-language phase occurs when the eigenvalue $\lambda_1$ shifts from negative to positive values~\cite{Arnold}. The critical point $\gamma_\text{out/in,\,c} = \nu_\text{c}$, in correspondence of which we have $\lambda_1=0$, can be calculated exactly. The equation $\lambda_1(\nu)=0$ is indeed equivalent to a quartic equation for $\nu$ with four real simple roots. Among these two are negative and one is larger than one. The fourth root, that we just identify with $\nu_\text{c}$, is given by 
\begin{align}
  \nu_\text{c} & =   \frac{\sqrt{19}}{2}\sin\left[-\frac{1}{3}\arctan\left(2\frac{\sqrt{2694}}{99}\right)+\frac{\pi}{3} \right]  -\frac{\sqrt{57}}{6}\sin\left[\frac{1}{3}\arctan\left(2\frac{\sqrt{2694}}{99}\right)+ \frac{\pi}{6} \right] - \frac{1}{2}\nonumber\\[3.0ex]
  & = 0.132122756\ldots  
\label{eq:ppmnuc}
\end{align}
Actually, among all network models that we consider in the present paper, the PPM is the only one where a calculation of the critical connectedness can be performed analytically to the very end.

The eigenvector $v_1$ of $\Lambda$ corresponding to $\lambda_1$ becomes a direction of instability for the symmetric steady solution for $\nu>\nu_\text{c}$. In other words,  the projection of the perturbation vector along $v_1$ diverges asymptotically.  From Fig.~\ref{fig:ppmstability} (\emph{right}) we observe that also $\lambda_2$ shifts to positive values at some point. In particular, it can be shown that $\lambda_2=0$ for $\nu=\hat \nu$. Therefore, as anticipated, the eigenvector $v_2$ of $\Lambda$ corresponding to $\lambda_2$ represents a second direction of instability  for $\nu>\hat \nu$. 

\subsection{Numerical integration of mean field equations}

It is worthwhile discussing at this point the integration of MFEs. We can take advantage of the analytic solution presented above to fix the details of our numerical recipe, so as to be confident that numerical integration yields correct results when applied to network models for which no analytic solution is available (for instance the SBM with $Q=2$ and $\nu_1 \ne \nu_2$, discussed in sect.~3). First of all, we notice that  for $\nu_1=\nu_2=\nu$ eqs.~(\ref{eq:sbmfrst})--(\ref{eq:sbmfrth}) are symmetric under the exchange $n^{(1)}_{A_1}\leftrightarrow n^{(2)}_{A_2}$, $n^{(1)}_{A_2}\leftrightarrow n^{(2)}_{A_1}$. Since the initial state densities, eq.~(\ref{eq:initcond}), are symmetric too and no dynamical fluctuations are encoded in MFEs, nothing can break the exchange symmetry, hence numerical solutions always converge to symmetric steady densities.

To let the system fall into global consensus, we have two possibilities. One is to break the exchange symmetry explicitly in the initial conditions. For instance, we can introduce a contamination of $A_2$ within~$\cC^{(1)}$ by letting
\begin{alignat}{3}
  \begin{array}{ll}n^{(1)}_{A_1}(0) = 1-\epsilon\,,& \qquad n^{(2)}_{A_1}(0) = 0\,,\\[3.3ex]
    n^{(1)}_{A_2}(0) = \epsilon\,,& \qquad n^{(2)}_{A_2}(0) = 1\,,
  \end{array}
  \label{eq:PPMasyminitcond}
\end{alignat}
with $0<\epsilon\ll 1$. For $\nu>\nu_\text{c}$ such a perturbation makes the system converge with certainty to global consensus on $A_2$. However, the contamination affects the results of numerical integration. More specifically, it modifies the duration of metastable states, thus changing the value of the critical connectedness by terms~$\text{O}(\epsilon)$. To get rid of this effect, we must integrate numerically MFEs for a sequence of decreasing values of $\epsilon$ and then extrapolate to $\epsilon\to 0^+$.

Albeit legitimate, the above approach has the drawback that symmetry breaking is implicit in the initial conditions, while MFEs are kept fully symmetric. An opposite possibility is to leave initial conditions unchanged and assume that $\cC^{(1)}$ and $\cC^{(2)}$ have different size. For instance, we can let $N^{(2)} = (1+\epsilon)N^{(1)}$ for $0<\epsilon\ll 1$, so that $\cC^{(2)}$ is slightly larger than $\cC^{(1)}$ (for $\nu>\nu_\text{c}$ the system is then expected to converge to global consensus on $A_2$). This assumption modifies the coefficients $\{\pi^{(ik)}\}$. Indeed, the probability of picking up an agent belonging to $\cC^{(1)}$ is now $N^{(1)}/N = (1/2)(1-\epsilon/2) + \text{O}(\epsilon^2)$, while the probability of picking up one belonging to $\cC^{(2)}$ is $N^{(2)}/N = (1/2)(1+\epsilon/2) + \text{O}(\epsilon^2)$.  Therefore, the exchange symmetry is explicitly broken in MFEs. As previously, numerical estimates of the critical connectedness are biased by terms $\text{O}(\epsilon)$, hence we must extrapolate results to $\epsilon\to 0^+$. All in all, the above two approaches for breaking the exchange symmetry are equivalent. 

\begin{figure}[t!]
  \centering
  \hskip -0.8cm\includegraphics[width=0.45\textwidth]{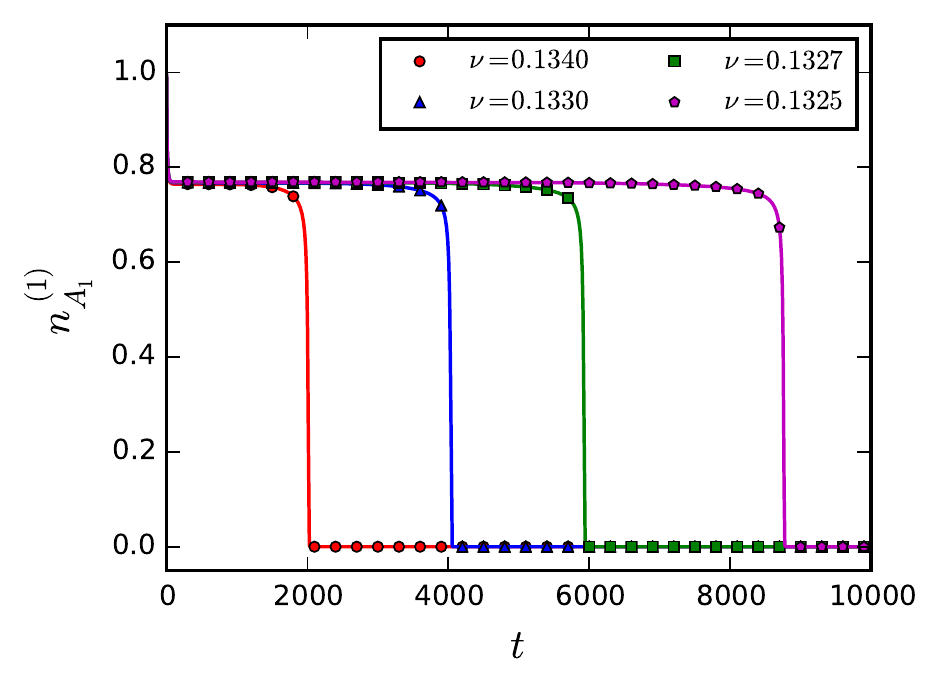}
  \hskip 0.2cm\includegraphics[width=0.45\textwidth]{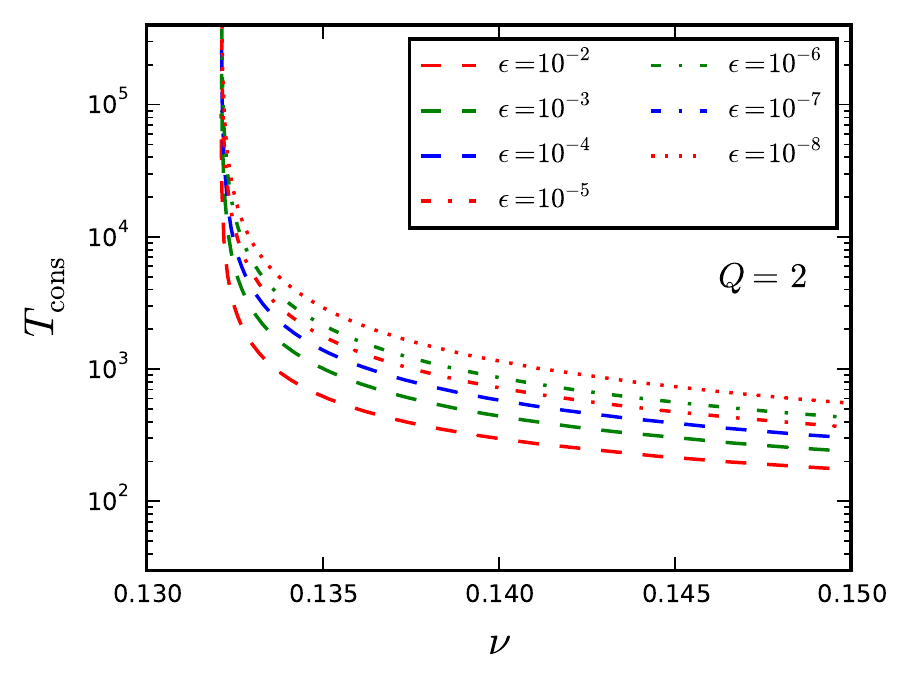}\,.
  \caption{\footnotesize (\emph{left}) Numerical integration of MFEs in the PPM; (\emph{right}) Time to consensus vs.~$\nu$ for several values of the symmetry breaking parameter $\epsilon$.}
  \vskip 0.2cm
  \label{fig:numint}  
\end{figure}

Apart from this issue, we discretize MFEs according to the Euler method~\cite{Atkinson} with step size $\rd t = 0.1$. In Fig.~\ref{fig:numint} (\emph{left}) we show examples of numerical integrations for a handful of values of $\nu$. The plot has been obtained with asymmetric initial conditions corresponding to $\epsilon=1.0\times 10^{-4}$. As anticipated, we observe the presence of metastable states followed by collapse to global consensus. These states have a finite duration depending on $\nu$. In particular, we see from the plot that the closer $\nu$ to $\nu_\text{c}$, the longer metastable states persist, until for $\nu<\nu_c$ they become truly stable. In Fig.~\ref{fig:numint} (\emph{right}) we plot the time to consensus $T_\text{cons}$ as a function of~$\nu$. The collapse of metastable states to global consensus takes a finite time $\Delta t$. Therefore, we need to define $T_\text{cons}$ operatively by setting a threshold. Throughout the paper we define $T_\text{cons}$ as the lowest value of $t$ for which $n^{(1)}_{A_1}(t)<1.0\times 10^{-4}$. This introduces a systematic error, which is however negligible to all purposes, since $\Delta t / T_\text{cons}\to 0$ as $\nu\to\nu_\text{c}$. The dependence of $T_{\rm cons}$ upon $\nu$ is well described by the function
\begin{equation}
  T_\text{cons}(\nu,\epsilon) = \left\{\begin{array}{ll} \dfrac{A(\epsilon)}{[\nu-\nu_\text{c}(\epsilon)]^{\gamma(\epsilon)}} & \text{if } \ \ \nu>\nu_c\, \\[3.0ex]
  +\infty & \text{otherwise}\,.\end{array}\right.
  \label{eq:model}
\end{equation}
with $\nu_\text{c}(\epsilon)$ and $\gamma(\epsilon)$ converging as $\epsilon \to 0^+$. In Table~\ref{tab:fitpars} we report estimates of the parameters $A,\nu_\text{c},\gamma$, obtained upon fitting data produced by numerical integrations to eq.~(\ref{eq:model}). In particular, the critical exponent $\gamma(\epsilon)$ converges to $\gamma(0^+) \simeq 0.96$ (for a definition of critical exponents see ref.~\cite{Stanley}), while $\nu_\text{c}(\epsilon)$ converges to the exact value, eq.~(\ref{eq:ppmnuc}).

\begin{table}[!t]
  \begin{center}
    \small
    \begin{tabular}{c|r|r|r}
      \hline
      \hline
      $\epsilon$ & $A(\epsilon)\ \ \ $ & $\nu_\text{c}(\epsilon)\ \ \ $ & $\gamma(\epsilon)\ \ \ $ \\
      \hline\\[-2.8ex]
      $1.0\times 10^{-2}$ & 8.214(1)  & 0.1321161(2) & 0.74205(3) \\
      $1.0\times 10^{-3}$ & 6.537(1)  & 0.1321222(2) & 0.86468(3) \\
      $1.0\times 10^{-4}$ & 6.920(1)  & 0.1321227(2) & 0.90872(3) \\
      $1.0\times 10^{-5}$ & 7.729(1)  & 0.1321228(2) & 0.93087(3) \\
      $1.0\times 10^{-6}$ & 8.523(1)  & 0.1321228(2) & 0.94602(3) \\
      $1.0\times 10^{-7}$ & 9.730(1)  & 0.1321229(2) & 0.95210(3) \\
      $1.0\times 10^{-8}$ & 10.790(1) & 0.1321229(2) & 0.95840(3) \\[0.3ex]
      \hline
      \hline
    \end{tabular}
    \vskip 0.1cm
    \caption{\footnotesize Estimates of fit parameters for $T_\text{cons}(\nu,\epsilon)$.\label{tab:fitpars}}
  \end{center}
    \vskip -0.4cm
\end{table}
In Fig.~\ref{fig:PPMsimul} (\emph{left}) we show the equilibrium densities $n^{(1)}_{A_k}(\infty)$ as obtained from numerical integration of MFEs with asymmetric initial conditions corresponding to $\epsilon=1.0\times 10^{-4}$. They are in perfect agreement with the symmetric steady solution derived in sect.~4.1 for $\nu<\nu_\text{c}$. We conclude that our numerical recipe introduces no relevant systematic error in the calculation. 

\subsection{Finite size effects}

\begin{figure}[t!]
  \centering
  \hskip -0.8cm\includegraphics[width=0.45\textwidth]{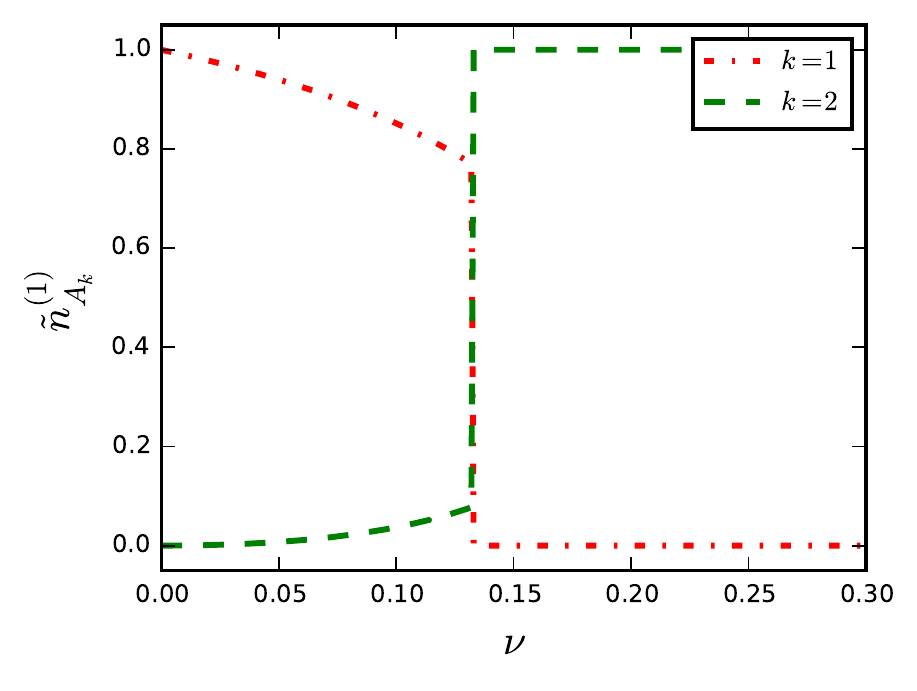}
  \hskip 0.2cm\includegraphics[width=0.45\textwidth]{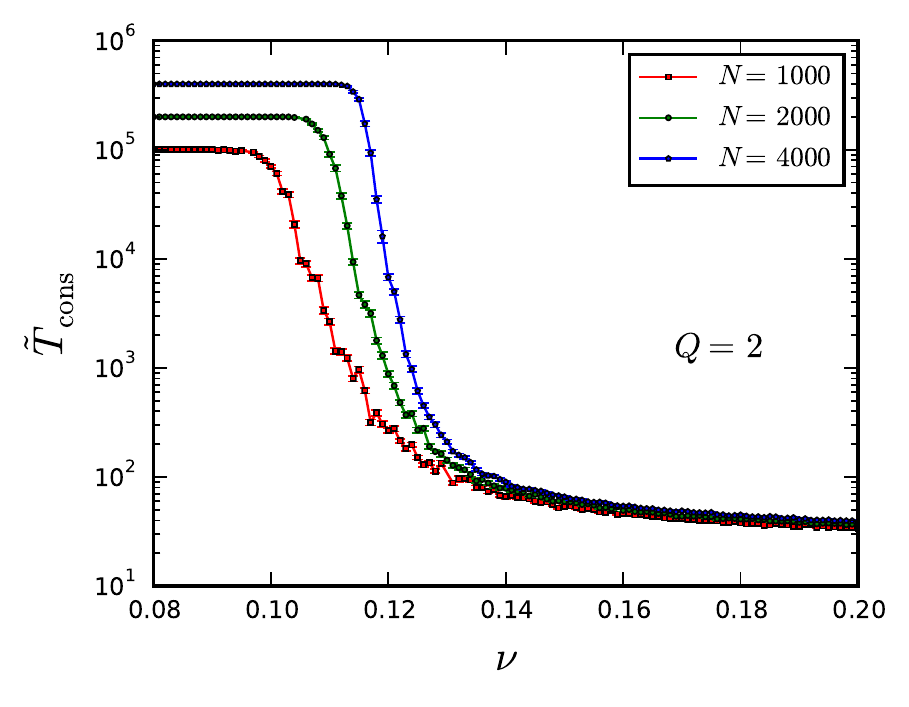}\,.
  \vskip -0.2cm
  \caption{\footnotesize (\emph{left}) Equilibrium densities in $\cC^{(1)}$ with initial conditions as in eq.~(\ref{eq:PPMasyminitcond}); (\emph{right}) Bounded time to consensus from simulations in the PPM with $Q=2$.}
  \label{fig:PPMsimul}
  \vskip -0.2cm
\end{figure}

So far we studied the model in the mean field approximation. This is known to work well only in the thermodynamic limit. In Monte Carlo simulations networks are necessarily made of a finite number of agents. Moreover, due to computational limitations this number is never exceedingly large. The main effect induced by the finiteness of the network is a blurring of the phase transition. The critical connectedness $\nu_\text{c}$ disappears on small networks together with the multi-language phase. The coexistence of different local languages within communities becomes a purely metastable phenomenon, independently of $\nu$. Dynamical fluctuations of state densities are always able to trigger a collapse to global consensus after a finite time since the game start. Of course, the lower $\nu$ the longer the system persists in the metastable phase. To quantify this, we average $\text{O}(100)$ independent Monte Carlo measures of the time to consensus for several choices of $N$ and $\nu$. In particular, we let $p^{(11)} = p^{(22)} = 1$ in all our numerical tests, hence the communities that we simulate are actually cliques. Since measuring time to consensus becomes increasingly costly as $\nu$ decreases, we need to set up a threshold beyond which the stochastic dynamics is forcedly arrested. We introduce the bounded time to consensus
\begin{equation}
  \tilde T_\text{cons}(N,\nu) = \min\left\{T_\text{cons}(N,\nu),100\,N\right\}\,.
  \label{eq:TconsPPM}
\end{equation}
In Fig.~\ref{fig:PPMsimul} (\emph{right}) we show the behaviour of $\tilde T_\text{cons}$ in a range of $\nu$ around the critical point $\nu_\text{c}$ for $N=1000,\,2000,\,4000$. We observe that $\tilde T_\text{cons}$  is essentially the same for all values of $N$ if $\nu\gg\nu_\text{c}$. As $\nu$ approaches $\nu_\text{c}$ from above $\tilde T_\text{cons}$ begins to rise and the increase is steeper for larger values of $N$. Finally, we see that $\tilde T_\text{cons}$ keeps finite for $\nu<\nu_\text{c}$ even though it takes soon large values as $\nu$ decreases. In principle it is possible to reproduce the observed curves thanks to a numerical technique that allows to build quasi-stationary solutions of the Master Equation for stochastic processes with absorbing states~\cite{Dickman:1,Dickman:2}. Although this technique has been applied to the NG in other contexts~\cite{Baronchelli:8,Xie:1} with very good results, its use here goes beyond our aims. We conclude by noting that in the crossover region, \ie in the range of $\nu$ across which $\tilde T_\text{cons}(N,\nu)$ rises from $\text{O}(100)$ to the upper bound $100\cdot N$, the behaviour of $\tilde T_\text{cons}(N,\nu)$ is well described by the function
\begin{equation}
  \tilde T_\text{cons}(N,\nu) \propto \exp\left\{N(\nu_\text{c}-\nu)^\beta\right\}\,,
  \label{eq:thmodelT}
\end{equation}
with $\beta\simeq 1.5$, in analogy with the findings of ref.~\cite{Xie:1}.

%% file: sect5.tex
\section{Binary Naming Game on two overlapping cliques}

\begin{figure}[t!]
  \centering
  \includegraphics[width=0.31\textwidth]{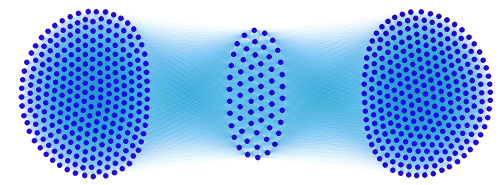}
  \vskip 0.4cm
  \caption{\footnotesize A network with two overlapping cliques, $N=600$ and $N_\text{ov} = 60$.\label{fig:overlap}}
  \vskip 0.2cm
\end{figure}

In order to investigate how the coexistence of multi-language states in the NG is affected by the presence of agents belonging simultaneously to different communities, we consider a graph $\cG = \cC^{(1)}\cup\cC^{(2)}$ made of two partially overlapping cliques, having size $N^{(1)}=N^{(2)}=N/2$. We recall that $\cC^{(k)}$ is a clique provided $\mathfrak{p}^{(kk)}(x,y)=1$ for all $x,y\in\cC^{(k)}$ and $x\ne y$. We split $\cC^{(1)}$ and $\cC^{(2)}$ into two disjoint groups of nodes respectively, \ie we let
\begin{equation}
  \cC^{(1)} = \cC^{(1)}_\text{in}\cup\cC^{(1)}_\text{ov}\,,\qquad \cC^{(2)} = \cC^{(2)}_\text{in}\cup\cC^{(2)}_\text{ov}\,,
  \label{eq:ovgroups}
\end{equation}
with $\cC^{(i)}_\text{in}$ and $\cC^{(i)}_\text{ov}$ fulfilling
\begin{align}
  i\text{)} & \qquad (x,y)\notin\cE\text{ for all } x\in\cC^{(1)}_\text{in}\text{ and for all } y\in\cC^{(2)}\,,\nonumber
\end{align}
\begin{align}
  ii\text{)} & \qquad (x,y)\notin\cE\text{ for all } x\in\cC^{(2)}_\text{in}\text{ and for all } y\in\cC^{(1)}\,,\nonumber\\[2.0ex]
 iii\text{)} & \qquad (x,y)\in\cE\text{ for all } x\in\cC^{(1)}_\text{ov}\text{ and for all } y\in\cC^{(2)}\,,\nonumber\\[2.0ex]
 iv\text{)} & \qquad (x,y)\in\cE\text{ for all } x\in\cC^{(2)}_\text{ov}\text{ and for all } y\in\cC^{(1)}\,.\nonumber
\end{align}
We also assume $|\cC^{(1)}_\text{in}|=|\cC^{(2)}_\text{in}|=N_\text{in}$ and $|\cC^{(1)}_\text{ov}| = |\cC^{(2)}_\text{ov}| = N_\text{ov}/2$. Therefore, we have $N = 2N_\text{in} + N_\text{ov}$. It will be noticed that these networks have no stochastic elements\footnote{With little effort we could consider a generalization where edges exist with probabilities $\mathfrak{p}^{(kk)}(x,y)<1$ for $x,y\in\cC^{(k)}$. We prefer to restrict our study to cliques, as we wish to investigate how the overlap affects the multi-language phase of the NG with no additional degree of freedom.}. An example corresponding to $N=600$ and $N_\text{ov} = 60$ is reported in Fig.~\ref{fig:overlap}. The connectedness parameters are given by
\begin{align}
  & \langle\kappa^{(1)}_{\text{in}}\rangle = \langle\kappa^{(2)}_{\text{in}}\rangle = \frac{2}{N}\frac{N}{2}\left(\frac{N}{2}-1\right) = \frac{N}{2}-1\,,\\[2.0ex]
  & \langle\kappa^{(12)}_{\text{out}}\rangle = \langle\kappa^{(21)}_{\text{out}}\rangle = \frac{2}{N}\frac{N_\text{ov}}{2}\frac{N}{2} = \frac{N_\text{ov}}{2}\,,
\end{align}
hence
\begin{equation}
  \gamma_\text{out/in}= \frac{N_\text{ov}}{N}\frac{1}{1-2/N} \simeq \frac{N_\text{ov}}{N} = \frac{N_\text{ov}}{2N_\text{in}+N_\text{ov}} = \frac{\omega}{2+\omega}\,,
  \label{eq:OVgamma}
\end{equation}
with $\omega = N_\text{ov}/N_\text{in}$. ECS conditions are fulfilled provided $\omega\ll 1$.

It is possible to study the binary NG on such networks with the same approach used for the PPM. However, we first need to clarify how the overlap contributes to shaping MFEs. So far we considered communities as groups of dynamically homogeneous agents. Accordingly, we identified the state densities~$n^{(i)}_D$, introduced in eq.~(\ref{eq:comdens}), as fundamental degrees of freedom of the system. When agents belonging to more than one community are present and they are likewise connected to all of these, assigning  such agents to one community or another becomes ambiguous, hence it is advisable to treat them separately. Indeed, in our case we can distinguish precisely three groups of dynamically homogeneous agents, namely $\cC^{(1)}_\text{in}$, $\cC^{(2)}_\text{in}$ and $\cC_{\text{ov}} = \cC^{(1)}_\text{ov}\cup\cC^{(2)}_\text{ov}$. Correspondingly, it makes sense to define state densities
\begin{align}
  n^{(i)}_D & = \frac{\text{no. of agents with notebook $D$ belonging to $\cC^{(i)}_{\text{in}}$}}{N_\text{in}}\,,\qquad \text{ for } \ i=1,2\,, \ \text{ and } \ D\in\bar S(\cD)\,, \\[3.0ex]
  n^{(\text{o})}_D & = \frac{\text{no. of agents with notebook $D$ belonging to $\cC_\text{ov}$}}{N_{\text{ov}}}\,,\qquad \text{ for } \ D\in\bar S(\cD)\,.
\end{align}
Possible agent-agent interactions and corresponding conditional rates are still those listed in Table~\ref{tab:binaryNG}, but the probabilities of picking up an agent $x$ and a neighbour $x'$ of $x$ belonging to combinations of the above groups must be specified over again. In particular, here we let
\begin{align}
  \pi^{(ii)} & = \text{prob}\left\{x\in\cC^{(i)}_\text{in},\ x'\in \cC^{(i)}_\text{in}\right\} = \frac{1}{(1+\omega)(2+\omega)}\,, \\[0.5ex]
  \pi^{(i\text{o})} & = \text{prob}\left\{x\in\cC^{(i)}_\text{in},\ x'\in\cC_\text{ov}\right\} = \frac{\omega}{(1+\omega)(2+\omega)}\,, \\[0.5ex]
  \pi^{(\text{o}i)} & = \text{prob}\left\{x\in\cC_\text{ov},\ x'\in\cC^{(i)}_\text{in}\right\} = \frac{\omega}{(2+\omega)^2}\,, \\[0.5ex]
  \pi^{(\text{oo})} & = \text{prob}\left\{x\in\cC_\text{ov},\ x'\in\cC_\text{ov}\right\} = \frac{\omega^2}{(2+\omega)^2}\,,
\end{align}
and we notice that $\omega = {\pi^{(1\text{o})}}/{\pi^{(11)}} = {\pi^{(2\text{o})}}/{\pi^{(22)}} = {\pi^{(\text{oo})}}/{\pi^{(\text{o}1)}} ={\pi^{(\text{oo})}}/{\pi^{(\text{o}2)}}$. The above probabilities include all possible pairings, indeed they fulfill
\begin{equation}
  \pi^{(11)} + \pi^{(1\text{o})} + \pi^{(22)} + \pi^{(2\text{o})} + \pi^{(\text{oo})} + \pi^{(\text{o}1)} + \pi^{(\text{o}2)} = 1\,.
\end{equation}
From the above definitions we easily recognize that the system is governed by MFEs
\begin{alignat}{2}
  \frac{\rd n^{(1)}_{A_1}}{\rd t} & = \pi^{(11)}\left\{ n^{(1)}_{A_1}n^{(1)}_{A_1A_2} + (n^{(1)}_{A_1A_2})^2 - n^{(1)}_{A_1}n^{(1)}_{A_2}\right\} \nonumber \\[0.0ex]
  & + \pi^{(1\text{o})}\left\{ \frac{3}{2}n^{(1)}_{A_1A_2}n^{(\text{o})}_{A_1} - \frac{1}{2}n^{(1)}_{A_1}n^{(\text{o})}_{A_1A_2} + n^{(1)}_{A_1A_2}n^{(\text{o})}_{A_1A_2} - n^{(1)}_{A_1}n^{(\text{o})}_{A_2} \right\}\,,
  \label{eq:ovfrsteq}\\[1.0ex]
  \frac{\rd n^{(1)}_{A_2}}{\rd t} & = \pi^{(11)}\left\{ n^{(1)}_{A_2}n^{(1)}_{A_1A_2} + (n^{(1)}_{A_1A_2})^2 - n^{(1)}_{A_1}n^{(1)}_{A_2} \right\} \nonumber \\[0.0ex]
  & + \pi^{(1\text{o})}\left\{\frac{3}{2}n^{(1)}_{A_1A_2}n^{(\text{o})}_{A_2}-\frac{1}{2}n^{(1)}_{A_2}n^{(\text{o})}_{A_1A_2}+n^{(1)}_{A_1A_2}n^{(\text{o})}_{A_1A_2}-n^{(1)}_{A_2}n^{(\text{o})}_{A_1}  \right\}\,,
  \label{eq:ovscndeq}\\[1.0ex]
  \frac{\rd n^{(2)}_{A_1}}{\rd t} & = \pi^{(22)}\left\{ n^{(2)}_{A_1}n^{(2)}_{A_1A_2} + (n^{(2)}_{A_1A_2})^2 - n^{(2)}_{A_1}n^{(2)}_{A_2}\right\} \nonumber\\[0.0ex]
  & + \pi^{(2\text{o})}\left\{ \frac{3}{2}n^{(2)}_{A_1A_2}n^{(\text{o})}_{A_1} - \frac{1}{2}n^{(2)}_{A_1}n^{(\text{o})}_{A_1A_2} + n^{(2)}_{A_1A_2}n^{(\text{o})}_{A_1A_2} - n^{(2)}_{A_1}n^{(\text{o})}_{A_2} \right\}\,,
  \label{eq:ovthrdeq} \\[1.0ex]
  \frac{\rd n^{(2)}_{A_2}}{\rd t} & = \pi^{(22)}\left\{ n^{(2)}_{A_2}n^{(2)}_{A_1A_2} + (n^{(2)}_{A_1A_2})^2 - n^{(2)}_{A_1}n^{(2)}_{A_2} \right\} \nonumber\\[0.0ex]
  & + \pi^{(2\text{o})}\left\{\frac{3}{2}n^{(2)}_{A_1A_2}n^{(\text{o})}_{A_2}-\frac{1}{2}n^{(2)}_{A_2}n^{(\text{o})}_{A_1A_2}+n^{(2)}_{A_1A_2}n^{(\text{o})}_{A_1A_2}-n^{(2)}_{A_2}n^{(\text{o})}_{A_1}  \right\}\,,
  \label{eq:ovfrtheq}\\[1.0ex]
  \frac{\rd n^{(\text{o})}_{A_1}}{\rd t} & = \pi^{(\text{oo})}\left\{ n^{(\text{o})}_{A_1}n^{(\text{o})}_{A_1A_2} + (n^{(\text{o})}_{A_1A_2})^2 - n^{(\text{o})}_{A_1}n^{(\text{o})}_{A_2}\right\} \nonumber\\[0.0ex]
  & + \pi^{(\text{o}1)}\left\{ \frac{3}{2}n^{(\text{o})}_{A_1A_2}n^{(1)}_{A_1} - \frac{1}{2}n^{(\text{o})}_{A_1}\nAB{\text{1}} + n^{(\text{o})}_{A_1A_2}\nAB{\text{1}} - n^{(\text{o})}_{A_1}n^{(1)}_{A_2} \right\}\phantom{\,,} \nonumber\\[0.0ex]
  & + \pi^{(\text{o}2)}\left\{ \frac{3}{2}n^{(\text{o})}_{A_1A_2}n^{(2)}_{A_1} - \frac{1}{2}n^{(\text{o})}_{A_1}\nAB{\text{2}} + n^{(\text{o})}_{A_1A_2}\nAB{\text{2}} - n^{(\text{o})}_{A_1}n^{(2)}_{A_2} \right\}\,,
  \label{eq:ovfitheq}\\[1.0ex]
  \frac{\rd n^{(\text{o})}_{A_2}}{\rd t} & = \pi^{(\text{oo})}\left\{ n^{(\text{o})}_{A_2}n^{(\text{o})}_{A_1A_2} + (n^{(\text{o})}_{A_1A_2})^2 - n^{(\text{o})}_{A_1}n^{(\text{o})}_{A_2} \right\} \nonumber\\[0.0ex]
  & + \pi^{(\text{o}1)}\left\{\frac{3}{2}n^{(\text{o})}_{A_1A_2}n^{(1)}_{A_2}-\frac{1}{2}n^{(\text{o})}_{A_2}n^{(1)}_{A_1A_2}+n^{(\text{o})}_{A_1A_2}n^{(1)}_{A_1A_2}-n^{(\text{o})}_{A_2}n^{(1)}_{A_1} \right\}\phantom{\,,}\nonumber\\[0.0ex]
  & + \pi^{(\text{o}2)}\left\{\frac{3}{2}n^{(\text{o})}_{A_1A_2}n^{(2)}_{A_2}-\frac{1}{2}n^{(\text{o})}_{A_2}n^{(2)}_{A_1A_2}+n^{(\text{o})}_{A_1A_2}n^{(2)}_{A_1A_2}-n^{(\text{o})}_{A_2}n^{(2)}_{A_1}  \right\}\,,
  \label{eq:ovsitheq}
\end{alignat}
In analogy with sect.~4, we can show that these admit a symmetric steady solution $\tilde n = \{\tilde n^{(1)}_{A_1},\tilde n^{(1)}_{A_2},\tilde n^{(2)}_{A_1},$ $n^{(2)}_{A_2},\tilde n^{(\text{o})}_{A_1},\tilde n^{(\text{o})}_{A_2}\}$ with $\tilde n^{(1)}_{A_1} = \tilde n^{(2)}_{A_2} = x$, $\tilde n^{(1)}_{A_2}=\tilde n^{(2)}_{A_1} = y$ and $\tilde n^{(\text{o})}_{A_1} = \tilde n^{(\text{o})}_{A_2}=z$. Moreover, here too there exists a finite critical threshold $\omega_c$, such that the symmetric steady solution is stable for $\omega<\omega_c$ and unstable for $\omega>\omega_c$. In particular, if the game starts with initial conditions
\begin{alignat}{3}
  \begin{array}{lll} n^{(1)}_{A_1}(0) = 1-\epsilon\,,& \qquad n^{(2)}_{A_1}(0) = 0\,, & \qquad n^{(\text{o})}_{A_1} = 1/2\,,\\[2.0ex]
    n^{(1)}_{A_2}(0) = \epsilon\,,& \qquad n^{(2)}_{A_2}(0) = 1\,, & \qquad n^{(\text{o})}_{A_2} = 1/2\,,
  \end{array}
  \label{eq:ovinitcond}
\end{alignat}
we find that for $\omega<\omega_\text{c}$ the system relaxes to a stable symmetric equilibrium with $A_1$ and $A_2$ prevailing respectively in $\cC^{(1)}$ and $\cC^{(2)}$, while for $\omega>\omega_\text{c}$ the system converges to global consensus on $A_2$, due to the symmetry breaking induced by dynamical fluctuations.

\begin{figure}[t!]
  \centering
  \hskip -0.8cm\includegraphics[width=0.48\textwidth]{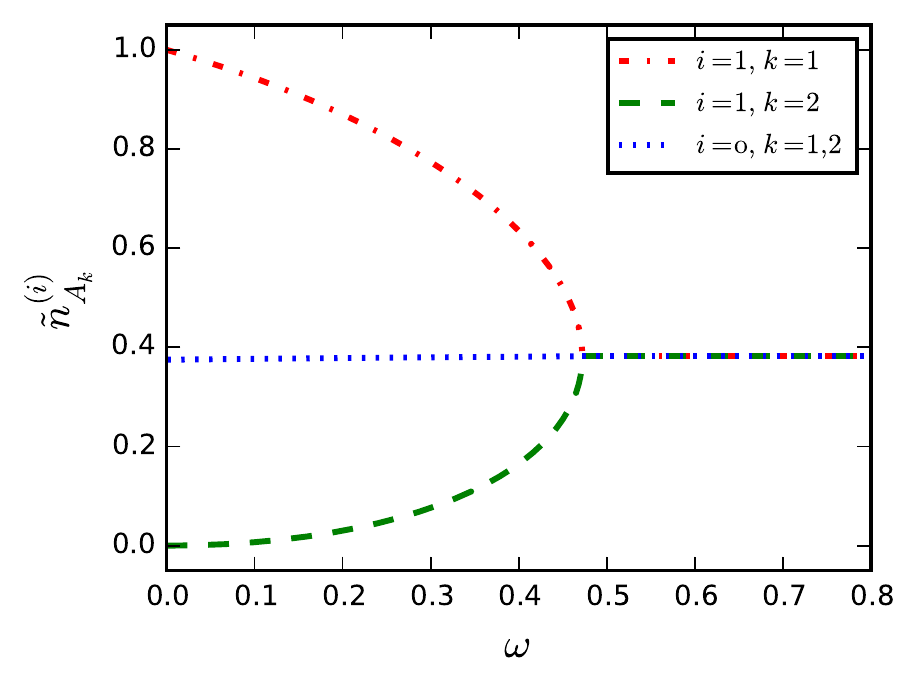}
  \hskip 0.2cm\includegraphics[width=0.48\textwidth]{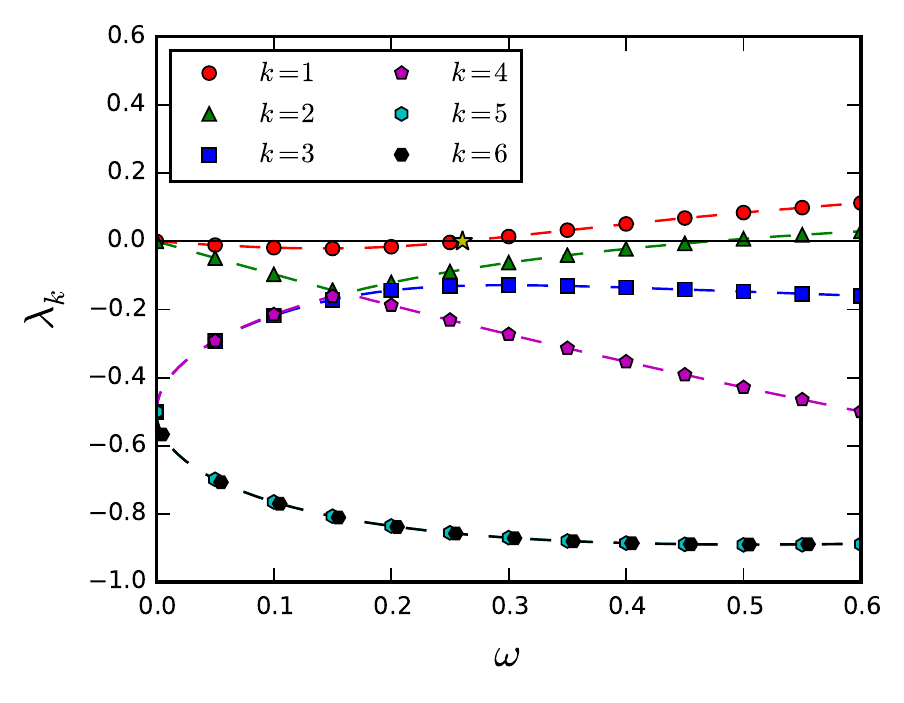}\,.
  \vskip -0.2cm
  \caption{\footnotesize (\emph{left}) Symmetric steady solution to MFEs on a network with two overlapping cliques. (\emph{right}) Eigenvalues of the linearized MFEs. The critical point $\omega_c$ is denoted by a star.}
  \label{fig:ovstability}
  \vskip -0.1cm
\end{figure}

Now, as a consequence of the exchange symmetry of our \emph{ansatz} the unknown density values $x,y,z$ are fully determined by algebraic equations
\begin{align}
  \label{eq:ov1eq}
  & x(1-x-y) + (1-x-y)^2 - xy \nonumber\\[0.0ex]
  & \hskip 2.0cm + \omega\left\{\frac{3}{2}(1-x-y)z - \frac{1}{2}x(1-2z) + (1-x-y)(1-2z) -xz \right\} = 0\,, \\[1.0ex]
  \label{eq:ov2eq}
  & y(1-x-y) + (1-x-y)^2 - xy \nonumber\\[1.0ex]
  & \hskip 2.0cm + \omega\left\{\frac{3}{2}(1-x-y)z - \frac{1}{2}y(1-2z) + (1-x-y)(1-2z) -yz \right\} = 0\,, \\[1.0ex]
  & \omega(z^2 - 3z + 1) +  \left\{\frac{3}{2}(x+y)(1-2z) - z(1-x-y) + 2(1-x-y)(1-2z) - (x+y)z \right\} = 0\,.
  \label{eq:ov3eq}
\end{align}
Again we let $u = x-y$ and $v = 1-x-y$. Then we observe that adding and subtracting eqs.~(\ref{eq:ov1eq})--(\ref{eq:ov2eq}) yields the equivalent system
\begin{align}
  \label{eq:ov4eq}
& v(1-v) + 2v^2 - \frac{1}{2}[(1-v)^2-u^2] + \omega\left\{ 3vz - \frac{1}{2}(1-v)(1-2z) + 2v(1-2z)-(1-v)z\right\} = 0\,,
\\[2.0ex]
  \label{eq:ov5eq}
& uv + \omega\left\{ -\frac{1}{2}u(1-2z) - uz \right\} = uv  -\frac{1}{2}\omega u = 0\,,\\[2.0ex]
  \label{eq:ov6eq}
& \omega(z^2-3z+1)+\left\{\frac{3}{2}(1-v)(1-2z)-zv+2v(1-2z)-(1-v)z\right\}\,.
\end{align}
Eq.~(\ref{eq:ov5eq}) has solutions: \emph{i}) $u=0$ and \emph{ii}) $u\ne 0$, $v = \omega/2$. Similar to what we found in sect.~4, these hold within disjoint intervals of $\omega$. We focus first on the second solution. Specifically, since eq.~(\ref{eq:ov6eq}) depends on $v$ but not on $u$, inserting $v=\omega/2$ into it yields immediately an equation for $z$ alone, namely
\begin{equation}
z^2 - \left(\frac{7}{2} + \frac{4}{\omega}\right)z + \frac{5}{4} + \frac{3}{2\omega} = 0\,.
\end{equation}
This has positive solution
\begin{equation}
  z(\omega) = \frac{7}{4} + \frac{2}{\omega} - \frac{1}{4}\frac{\sqrt{29\omega^2 + 88\omega + 64}}{\omega}\,.
  \label{eq:ovzsol}
\end{equation}
Despite the presence of inverse powers of $\omega$, $z(\omega)$ keeps always finite, as can be seen by expanding the square root on the right hand side in Taylor series. Indeed we have $\lim_{\omega\to 0^+}z(\omega) = 3/8$. By inserting the values just determined for $v(\omega)$ and $z(\omega)$ into eq.~(\ref{eq:ov4eq}), we get
\begin{equation}
u(\omega) = \pm\sqrt{1 + \omega - \omega^2 - \frac{1}{4}\omega\Omega}\,.
\end{equation}
with $\Omega = \sqrt{29\omega^2 + 88\omega + 64}$. In order for $u(\omega)$ to be real, it must be $0\le\omega\le \hat\omega = 2\sqrt{5}-4=0.472136\ldots$. More precisely, it can be shown that the equation $u(\omega)=0$ is equivalent to a quartic equation in $\omega$ with four real simple roots. Among these, only $\hat \omega$ is positive. For $\omega>\hat\omega$, \ie for $u=0$, eq.~(\ref{eq:ovzsol}) holds no more. In this region the unknowns $v$ and $z$ are jointly determined by eqs.~(\ref{eq:ov4eq}) and (\ref{eq:ov6eq}). These admit constant solutions, \ie not depending on $\omega$. Indeed, $z$ and $v$ are separately determined by $z^2-3z+1=0$ and $v(1-v) + 2v^2 - (1-v)^2/2=0$, yielding respectively $z=(3-\sqrt{5})/2$ and $v=1/(2+\sqrt{5})$. With some algebra we find that the symmetric steady solution, corresponding to initial conditions as specified in eq.~(\ref{eq:ovinitcond}) with $\epsilon=0$, is given by
\begin{equation}
  \left\{\begin{array}{ll}
  \tilde n^{(i)}_{A_i}(\omega) & \hskip -0.25cm = \dfrac{1}{2}\left\{ 1 +  \sqrt{1+\omega - \omega^2 - \dfrac{\omega}{4}\Omega}-\dfrac{\omega}{2}\right\}\,,\\[2.0ex]
  \tilde n^{(i)}_{A_{3-i}}(\omega) & \hskip -0.25cm = \dfrac{1}{2}\left\{ 1 -  \sqrt{1+\omega - \omega^2 - \dfrac{\omega}{4}\Omega}-\dfrac{\omega}{2}\right\}\,,\\[2.0ex]
  \tilde n^{(\text{o})}_{A_i}(\omega) & \hskip -0.25cm = \dfrac{7}{4} + \dfrac{2}{\omega}-\dfrac{1}{4}\dfrac{\Omega}{\omega}\,,
  \end{array}\right.\qquad \text{ for }\ \omega\le\hat\omega \ \text{ and } \ i=1,2\,,
  \label{eq:ovstabone}
\end{equation}
and
\begin{equation}
  \tilde n^{(i)}_{A_k}(\omega) = \frac{3-\sqrt{5}}{2}\,,\qquad \text{ for } \ \omega>\hat\omega\,, \ \ i=1,2,\text{o} \ \text{ and } \ k=1,2\,.
  \label{eq:ovstabtwo}
\end{equation}
In Fig.~\ref{fig:ovstability} (\emph{left}) we plot the symmetric steady densities in $\cC^{(1)}_\text{in}$ and $\cC_\text{ov}$ vs. $\omega$. Results for $n^{(i)}_{A_k}$, $i,k=1,2$ are qualitatively similar to those reported in Fig.~\ref{fig:ppmstability}. Also in this case we see that both $\tilde n^{(1)}_{A_1}(\omega)$, $\tilde n^{(1)}_{A_2}(\omega)$ and  $\tilde n^{(\text{o})}_{A_k}$  have discontinuous first order derivatives for $\omega=\hat\omega$.

\subsection{Stability of the symmetric steady solution}

The existence of a critical threshold $\omega_\text{c}$ above which the symmetric steady solution $\tilde n$ becomes unstable is again revealed by a stability analysis similar to that performed in sect.~4.1. The algebra is just a bit harder here. In particular, although the game rules are unchanged and the network is still made of two interacting communities, the state space of the system is now larger: we have six coupled equations in six unknown variables. As a result, the stability matrix has $6\times 6$ entries $\Lambda^{(i,{\scriptscriptstyle X})}_{(k,{\scriptscriptstyle Y})} = \partial f^{(i)}_{\scriptscriptstyle X}/\partial n^{(k)}_{\scriptscriptstyle Y}(\tilde n)$ corresponding to $i,k = 1,2,\text{o}$ and $X,Y = A_1,A_2$. The lack of a direct interaction between agents belonging to $\cC^{(1)}_\text{in}$ and $\cC^{(2)}_\text{in}$ makes some of these matrix elements vanish. Concretely, we find
\begingroup\makeatletter\def\f@size{9}\check@mathfonts
\begin{alignat}{3}
  \rho_1\Lambda^{(1,{\scriptscriptstyle A_1})}_{(1,{\rm\scriptscriptstyle A_1})} & = - 1 - \frac{3}{2}\omega + \frac{\omega}{2}\tilde n^{(\text{o})}_{A_2}\,, &
   \rho_1\Lambda^{(1,{\rm\scriptscriptstyle A_1})}_{(1,{\rm\scriptscriptstyle A_2})} & =  -2  - \omega - \frac{\omega}{2} \tilde n^{(\text{o})}_{A_1} + 2\tilde n^{(1)}_{A_2} + \omega \tilde n^{(\text{o})}_{A_2}\,,\\[1.1ex]
  \rho_1\Lambda^{(1,{\rm\scriptscriptstyle A_1})}_{(2,{\rm\scriptscriptstyle A_1})} & = 0\,, & 
  \rho_1\Lambda^{(1,{\rm\scriptscriptstyle A_1})}_{(2,{\rm\scriptscriptstyle A_2})} & = 0\,, \\[1.1ex]
  \rho_1\Lambda^{(1,{\rm\scriptscriptstyle A_1})}_{(\text{o},{\rm\scriptscriptstyle A_1})} & =  \frac{1}{2}\omega -\frac{\omega}{2}\tilde n^{(1)}_{A_2}\,, & 
  \rho_1\Lambda^{(1,{\rm\scriptscriptstyle A_1})}_{(\text{o},{\rm\scriptscriptstyle A_2})} & =  -\omega + \frac{\omega}{2}\tilde n^{(1)}_{A_1}+\omega\tilde n^{(1)}_{A_2}\,, \\[1.1ex]
  \rho_1\Lambda^{(1,{\rm\scriptscriptstyle A_2})}_{(1,{\rm\scriptscriptstyle A_1})} & = -2 - \omega  - \frac{\omega}{2} \tilde n^{(\text{o})}_{A_2}  + 2 \tilde n^{(1)}_{A_1} + \omega \tilde n^{(\text{o})}_{A_1}\,,& 
  \hskip 0.8cm \rho_1\Lambda^{(1,{\rm\scriptscriptstyle A_2})}_{(1,{\rm\scriptscriptstyle A_2})} & = -1 -\frac{3}{2}\omega + \frac{\omega}{2} \tilde n^{(\text{o})}_{A_1}\, \\[1.1ex]
  \rho_1\Lambda^{(1,{\rm\scriptscriptstyle A_2})}_{(2,{\rm\scriptscriptstyle A_1})} & = 0\,, & 
  \rho_1\Lambda^{(1,{\rm\scriptscriptstyle A_2})}_{(2,{\rm\scriptscriptstyle A_2})} & = 0\,, \\[1.1ex]
  \rho_1\Lambda^{(1,{\rm\scriptscriptstyle A_2})}_{(\text{o},{\rm\scriptscriptstyle A_1})} & = -\omega  + \frac{\omega}{2} \tilde n^{(1)}_{A_2} + \omega\tilde n^{(1)}_{A_1}\,, & 
  \rho_1\Lambda^{(1,{\rm\scriptscriptstyle A_2})}_{(\text{o},{\rm\scriptscriptstyle A_2})} & = \frac{\omega}{2} - \frac{\omega}{2}\tilde n^{(1)}_{A_1}\,,\\[1.1ex]
  \rho_1\Lambda^{(2,{\rm\scriptscriptstyle A_1})}_{(1,{\rm\scriptscriptstyle A_1})} & = 0\,, & 
  \hskip 0.8cm  \rho_1\Lambda^{(2,{\rm\scriptscriptstyle A_1})}_{(1,{\rm\scriptscriptstyle A_2})} & = 0\,,\\[1.1ex]
  \rho_1\Lambda^{(2,{\rm\scriptscriptstyle A_1})}_{(2,{\rm\scriptscriptstyle A_1})} & = - 1 - \frac{3}{2}\omega + \frac{\omega}{2}\tilde n^{(\text{o})}_{A_2}\,, & 
  \rho_1\Lambda^{(2,{\rm\scriptscriptstyle A_1})}_{(2,{\rm\scriptscriptstyle A_2})} & = - 2 - \omega - \frac{\omega}{2} \tilde n^{(\text{o})}_{A_1} + 2\tilde n^{(2)}_{A_2} + \omega \tilde n^{(\text{o})}_{A_2}\,,\\[1.1ex]
  \rho_1\Lambda^{(2,{\rm\scriptscriptstyle A_1})}_{(\text{o},{\rm\scriptscriptstyle A_1})} & = \frac{1}{2}\omega -\frac{\omega}{2}\tilde n^{(2)}_{A_2}\,, & 
  \hskip 1.0cm \rho_1\Lambda^{(2,{\rm\scriptscriptstyle A_1})}_{(\text{o},{\rm\scriptscriptstyle A_2})} & = -\omega + \frac{\omega}{2}\tilde n^{(2)}_{A_1}+\omega\tilde n^{(2)}_{A_2}\,, \\[1.1ex]
  \rho_1\Lambda^{(2,{\rm\scriptscriptstyle A_2})}_{(1,{\rm\scriptscriptstyle A_1})} & = 0\,, & 
  \rho_1\Lambda^{(2,{\rm\scriptscriptstyle A_2})}_{(1,{\rm\scriptscriptstyle A_2})} & = 0\,, \\[1.1ex]
  \rho_1\Lambda^{(2,{\rm\scriptscriptstyle A_2})}_{(2,{\rm\scriptscriptstyle A_1})} & = -2 - \omega  - \frac{\omega}{2} \tilde n^{(\text{o})}_{A_2}  + 2 \tilde n^{(2)}_{A_1} + \omega \tilde n^{(\text{o})}_{A_1}\,,& 
  \rho_1\Lambda^{(2,{\rm\scriptscriptstyle A_2})}_{(2,{\rm\scriptscriptstyle A_2})} & = -1 -\frac{3}{2}\omega + \frac{\omega}{2} \tilde n^{(\text{o})}_{A_1}\, \\[1.1ex]
  \rho_1\Lambda^{(2,{\rm\scriptscriptstyle A_2})}_{(\text{o},{\rm\scriptscriptstyle A_1})} & = -\omega  + \frac{\omega}{2} \tilde n^{(2)}_{A_2} + \omega\tilde n^{(2)}_{A_1}\,, & 
  \rho_1\Lambda^{(2,{\rm\scriptscriptstyle A_2})}_{(\text{o},{\rm\scriptscriptstyle A_2})} & = \frac{\omega}{2} - \frac{\omega}{2}\tilde n^{(2)}_{A_1}\,,\\[1.1ex]
  \rho_2\Lambda^{(\text{o},{\rm\scriptscriptstyle A_1})}_{(1,{\rm\scriptscriptstyle A_1})} & = \frac{1}{2}  - \frac{1}{2}\tilde n^{(\text{o})}_{A_2}\,, & 
   \rho_2\Lambda^{(\text{o},{\rm\scriptscriptstyle A_1})}_{(1,{\rm\scriptscriptstyle A_2})} & = -1 +\frac{1}{2}\tilde n^{(\text{o})}_{A_1} + \tilde n^{(\text{o})}_{A_2}\,, \\[1.0ex]
  \rho_2\Lambda^{(\text{o},{\rm\scriptscriptstyle A_1})}_{(2,{\rm\scriptscriptstyle A_1})} & = \frac{1}{2} - \frac{1}{2}\tilde n^{(\text{o})}_{A_2}\,, & \hskip 2.1cm
  \rho_2\Lambda^{(\text{o},{\rm\scriptscriptstyle A_1})}_{(2,{\rm\scriptscriptstyle A_2})} & = -1 +\frac{1}{2}\tilde n^{(\text{o})}_{A_1} + \tilde n^{(\text{o})}_{A_2}\,,\\[1.1ex]
  \rho_2\Lambda^{(\text{o},{\rm\scriptscriptstyle A_1})}_{(\text{o},{\rm\scriptscriptstyle A_1})} & = -3-\omega + \frac{1}{2}[\tilde n^{(1)}_{A_2} +\tilde n^{(2)}_{A_2}]\,, &
  \rho_2\Lambda^{(\text{o},{\rm\scriptscriptstyle A_1})}_{(\text{o},{\rm\scriptscriptstyle A_2})} & = -2 -2\omega -\frac{1}{2}[\tilde n^{(1)}_{A_1}+\tilde n^{(2)}_{A_1}] \nonumber\\[1.1ex]
  & & & +[\tilde n^{(1)}_{A_2}+\tilde n^{(2)}_{A_2}] + 2\omega\tilde n^{(\text{o})}_{A_2}\,, \\[1.3ex]
  \rho_2\Lambda^{(\text{o},{\rm\scriptscriptstyle A_2})}_{(1,{\rm\scriptscriptstyle A_1})} & = -1 + \tilde n^{(\text{o})}_{A_1} + \frac{1}{2}\tilde n^{(\text{o})}_{A_2}\,, & 
  \rho_2\Lambda^{(\text{o},{\rm\scriptscriptstyle A_2})}_{(1,{\rm\scriptscriptstyle A_2})} & = \frac{1}{2} - \frac{1}{2}\tilde n^{(\text{o})}_{A_1}\,, \\[1.1ex]
  \rho_2\Lambda^{(\text{o},{\rm\scriptscriptstyle A_2})}_{(2,{\rm\scriptscriptstyle A_1})} & = -1 + \tilde n^{(\text{o})}_{A_1} + \frac{1}{2}\tilde n^{(\text{o})}_{A_2}\,, & 
  \rho_2\Lambda^{(\text{o},{\rm\scriptscriptstyle A_2})}_{(2,{\rm\scriptscriptstyle A_2})} & = \frac{1}{2}- \frac{1}{2}\tilde n^{(\text{o})}_{A_1}\,,\\[1.1ex]
  \rho_2\Lambda^{(\text{o},{\rm\scriptscriptstyle A_2})}_{(\text{o},{\rm\scriptscriptstyle A_1})} & = -2-2\omega-\frac{1}{2}[\tilde n^{(1)}_{A_2}+\tilde n^{(2)}_{A_2}] &  \hskip 2.1cm \rho_2\Lambda^{(\text{o},{\rm\scriptscriptstyle A_2})}_{(\text{o},{\rm\scriptscriptstyle A_2})} & = -3-\omega +\frac{1}{2}[\tilde n^{(1)}_{A_1}+\tilde n^{(2)}_{A_1}]\,,  \nonumber\\[1.0ex]
   & + [\tilde n^{(1)}_{A_1}+\tilde n^{(2)}_{A_1}] + 2\omega\tilde n^{(\text{o})}_{A_1}\,, &
\end{alignat}
\endgroup
with $\rho_1=(1+\omega)(2+\omega)=1/\pi^{(11)}=1/\pi^{(22)}$ and $\rho_2=\omega^{-1}(2+\omega)^2=1/\pi^{(\text{o}1)}=1/\pi^{(\text{o}2)}$. In order to calculate the eigenvalues $\{\lambda_k\}_{k=1}^6$ of $\Lambda$ we need to solve the secular equation $\det(\Lambda - \lambda\mathds{1})=0$.  With the components of $\tilde n$ depending on $\omega$ either explicitly (in the form of direct and inverse powers of the latter) and implicitly via $\Omega$, the secular determinant turns out to be a polynomial of sixth degree in $\lambda$ with rational coefficient functions in $\omega$ and $\Omega$. Luckily, the determinant factorizes into cubic polynomials, \ie we have
\begin{equation}
  \det(\Lambda-\lambda\mathds{1}) = \frac{p_1(\lambda)p_2(\lambda)}{64(1+\omega)^4(2+\omega)^8}\,,
\end{equation}
with
\begingroup\makeatletter\def\f@size{9}\check@mathfonts
\begin{align}
  p_1(\lambda) & = [
    128
    +512\,\omega
    +832\,{\omega}^{2}
    +704\,{\omega}^{3}
    +328\,{\omega}^{4}    
    +80\,{\omega}^{5}
    +8\,{\omega}^{6} ]\lambda^3
  \nonumber\\
  & \hskip -0.5cm + [80\,\omega
    +200\,{\omega}^{2}
    +180\,{\omega}^{3}
    +70\,{\omega}^{4}
    +10\,{\omega}^{5}
    +(16
    +56\,\omega
    +84\,{\omega}^{2}
    +66\,{\omega}^{3}
    +26\,{\omega}^{4}
    +4\,{\omega}^{5})\Omega
  ]\lambda^2 \nonumber\\
  & \hskip -0.5cm + [32
    +64\,\omega
    +24\,{\omega}^{2}
    +10\,{\omega}^{3}
    +15\,{\omega}^{4}
    +5\,{\omega}^{5}
    +(-8\,\omega +10\,{\omega}^{3}+4\,{\omega}^{4})\Omega
    +(2\,\omega+3\,{\omega}^{2} +{\omega}^{3})\,{\Omega}^{2}
  ]\lambda\nonumber\\
  & \hskip -0.5cm + 4\,\omega\,\Omega+4\,{\omega}^{2}\,\Omega-4\,{\omega}^{3}\Omega-{\omega}^{2}{\Omega}^{2}\,,
\end{align}
and
\begin{align}
  p_2(\lambda) & = 
  [128
    +512\,\omega
    +832\,{\omega}^{2}
    +704\,{\omega}^{3}
    +328\,{\omega}^{4}
    +80\,{\omega}^{5}
    +8\,{\omega}^{6} ]\lambda^3 \nonumber\\
  & \hskip -0.5cm +[240\,\omega
    +760\,\omega^2
    +940\,\omega^3
    +570\,\omega^4
    +170\,\omega^5
    +20\,\omega^6
    -(16\,
    +24\,\omega\,
    -12\,\omega^2\,
    -38\,\omega^3
    -22\,\omega^4\,
    -4\,\omega^5\,)\Omega ]\lambda^2 \nonumber\\
  & \hskip -0.5cm +[32
    +128\,\omega
    +300\,\omega^2
    +384\,\omega^3
    +213\,\omega^4
    +41\,\omega^5
    +(4\,\omega\,
    +2\,\omega^2\,
    -8\,\omega^3\,
    -4\,\omega^4\,)\Omega \nonumber\\
    & \hskip -0.5cm -(2\,\omega\,
    +3\,\omega^2\,
    +\omega^3)\Omega^2 ]\lambda
  -(8\,\omega
  +52\,\omega^2
  +56\,\omega^3
  +16\,\omega^4)
  -(8\,\omega\,
  +5\,\omega^2\,
  -2\,\omega^3\, )\,\Omega
  +(\omega\,
  +\omega^2)\,\Omega^2\,.
\end{align}
\endgroup
Therefore, the eigenvalues of $\Lambda$ are simply given by the roots of $p_1$ and $p_2$ separately. The behaviour of the eigenvalues as functions of $\omega$ is shown in Fig.~\ref{fig:ovstability} (\emph{right}). All eigenvalues are negative for sufficiently small $\omega$, hence $\tilde n$ represents here a stable multi-language steady state. At some point as $\omega$ increases (precisely for $\omega=\omega_\text{c}$), $\lambda_1$ shifts from negative to positive values, thus determining a sudden change of phase, with the system  converging to global consensus in a finite time. For an even larger value of $\omega$ (more precisely for $\omega=\hat\omega$), also $\lambda_2$ shifts to positive values. It is possible to find analytic expressions for all the eigenvalues of $\Lambda$. Below we report only $\lambda_1$, since this is related to the critical threshold $\omega_\text{c}$. We have
\begin{equation}
  \lambda_1 = \frac{1}{12(1+\omega)(2+\omega)^2}\left[a + \frac{ (1+\omega)^2(2+\omega)^2\left( 3(2+\omega)\sqrt{b}+c\right)^{2/3}+d}{(1+\omega)^2(2+\omega)^2\left(3(2+\omega)\sqrt{b}+c\right)^{1/3}}\right]\,,
\end{equation}
with the coefficient functions $a,b,c,d$ being given respectively by
\begin{align}
  a & = -30\,\omega-35\,{\omega}^{2}-10\,{\omega}^{3}-2\,\Omega+\omega\,
  \Omega+2\,{\omega}^{2}\,\Omega\,, \\[2.0ex]
  b & = 196608
+2162688\,\omega
+10736640\,{\omega}^{2}
+28446720\,{\omega}^{3}
+42713856\,{\omega}^{4}
+32143872\,{\omega}^{5}\nonumber\\
& -5234928\,{\omega}^{6}
-36468864\,{\omega}^{7}
-28560744\,{\omega}^{8}
+127224\,{\omega}^{9}
+10823661\,{\omega}^{10}
+4106838\,{\omega}^{11}\nonumber\\
& -808959\,{\omega}^{12}
-755556\,{\omega}^{13}
-120300\,{\omega}^{14}
-129024\,\omega\,\Omega
-2365440\,{\omega}^{2}\,\Omega
-9703680\,{\omega}^{3}\Omega\nonumber\\
& -16942464\,{\omega}^{4}\Omega
-13454016\,{\omega}^{5}\Omega
-531888\,{\omega}^{6}\Omega
+9250152\,{\omega}^{7}\Omega
+8316420\,{\omega}^{8}\Omega
+2124942\,{\omega}^{9}\Omega\nonumber\\
& -796212\,{\omega}^{10}\Omega
-426270\,{\omega}^{11}\Omega
+16944\,{\omega}^{12}\Omega
+21720\,{\omega}^{13}\Omega
-3072\,{\Omega}^{2}
-165888\,\omega{\Omega}^{2}\nonumber\\
& -440064\,{\omega}^{2}{\Omega}^{2}
+521856\,{\omega}^{3}{\Omega}^{2}
+2961120\,{\omega}^{4}{\Omega}^{2}
+3698208\,{\omega}^{5}{\Omega}^{2}
+1102452\,{\omega}^{6}{\Omega}^{2}
-1830852\,{\omega}^{7}{\Omega}^{2}\nonumber\\
& -2180301\,{\omega}^{8}{\Omega}^{2}
-849546\,{\omega}^{9}{\Omega}^{2}
-15249\,{\omega}^{10}{\Omega}^{2}
+66012\,{\omega}^{11}{\Omega}^{2}
+11148\,{\omega}^{12}{\Omega}^{2}
+14592\,\omega\,{\Omega}^{3}\nonumber\\
& +64896\,{\omega}^{2}{\Omega}^{3}
-12672\,{\omega}^{3}{\Omega}^{3}
-326976\,{\omega}^{4}{\Omega}^{3}
-447696\,{\omega}^{5}{\Omega}^{3}
-83904\,{\omega}^{6}{\Omega}^{3}
+307404\,{\omega}^{7}{\Omega}^{3}\nonumber\\
& +309864\,{\omega}^{8}{\Omega}^{3}
+116052\,{\omega}^{9}{\Omega}^{3}
+13392\,{\omega}^{10}{\Omega}^{3}
-816\,{\omega}^{11}{\Omega}^{3}
+1920\,\omega\,{\Omega}^{4}
+8976\,{\omega}^{2}{\Omega}^{4}\nonumber\\
& +25824\,{\omega}^{3}{\Omega}^{4}
+42912\,{\omega}^{4}{\Omega}^{4}
+28200\,{\omega}^{5}{\Omega}^{4}
-16689\,{\omega}^{6}{\Omega}^{4}
-42798\,{\omega}^{7}{\Omega}^{4}
-30489\,{\omega}^{8}{\Omega}^{4}\nonumber
\end{align}
\begin{align}
\phantom{b} & -9468\,{\omega}^{9}{\Omega}^{4}
-1044\,{\omega}^{10}{\Omega}^{4}
-192\,\omega\,{\Omega}^{5}
-912\,{\omega}^{2}{\Omega}^{5}
-1848\,{\omega}^{3}{\Omega}^{5}
-1380\,{\omega}^{4}{\Omega}^{5}
+1278\,{\omega}^{5}{\Omega}^{5}\nonumber\\
& +3612\,{\omega}^{6}{\Omega}^{5}
+3210\,{\omega}^{7}{\Omega}^{5}
+1344\,{\omega}^{8}{\Omega}^{5}
+216\,{\omega}^{9}{\Omega}^{5}
-12\,{\omega}^{2}{\Omega}^{6}
-60\,{\omega}^{3}{\Omega}^{6}
-135\,{\omega}^{4}{\Omega}^{6}\nonumber\\
& -174\,{\omega}^{5}{\Omega}^{6}
-135\,{\omega}^{6}{\Omega}^{6}
-60\,{\omega}^{7}{\Omega}^{6}
-12\,{\omega}^{8}{\Omega}^{6}\,,
\end{align}
\begin{align}
c & =
5184\,\omega
+15264\,{\omega}^{2}
+23760\,{\omega}^{3}
+30960\,{\omega}^{4}
+18540\,{\omega}^{5}
-7838\,{\omega}^{6}
-15393\,{\omega}^{7}
-6810\,{\omega}^{8}\nonumber\\
& -1000\,{\omega}^{9}
+576\,\Omega
-1440\,\omega\,\Omega
-9720\,{\omega}^{2}\,\Omega
-13968\,{\omega}^{3}\Omega
-7278\,{\omega}^{4}\Omega
+2904\,{\omega}^{5}\Omega
+6483\,{\omega}^{6}\Omega\nonumber\\
& +3402\,{\omega}^{7}\Omega
+600\,{\omega}^{8}\Omega
+144\,\omega\,{\Omega}^{2}
+696\,{\omega}^{2}{\Omega}^{2}
+660\,{\omega}^{3}{\Omega}^{2}
-288\,{\omega}^{4}{\Omega}^{2}
-867\,{\omega}^{5}{\Omega}^{2}
-558\,{\omega}^{6}{\Omega}^{2}\nonumber\\
& -120\,{\omega}^{7}{\Omega}^{2}
-24\,\omega\,{\Omega}^{3}
-18\,{\omega}^{2}{\Omega}^{3}
+22\,{\omega}^{3}{\Omega}^{3}
+45\,{\omega}^{4}{\Omega}^{3}
+30\,{\omega}^{5}{\Omega}^{3}
+8\,{\omega}^{6}{\Omega}^{3}
-8\,{\Omega}^{3}\,,
\end{align}
\vskip -0.0cm
\noindent and
\vskip -0.7cm
\begin{align}
d & =
-768
-5376\,\omega
-15312\,{\omega}^{2}
-22752\,{\omega}^{3}
-16760\,{\omega}^{4}
-440\,{\omega}^{5}
+10835\,{\omega}^{6}
+10180\,{\omega}^{7}
+4571\,{\omega}^{8}\nonumber\\
& +1054\,{\omega}^{9}
+100\,{\omega}^{10}
+384\,\omega\,\Omega
+1424\,{\omega}^{2}\,\Omega
+1656\,{\omega}^{3}\Omega
-308\,{\omega}^{4}\Omega
-2614\,{\omega}^{5}\Omega
-2792\,{\omega}^{6}\Omega\nonumber\\
& -1438\,{\omega}^{7}\Omega
-376\,{\omega}^{8}\Omega
-40\,{\omega}^{9}\Omega
+16\,{\Omega}^{2}
80\,\omega\,{\Omega}^{2}
+192\,{\omega}^{2}{\Omega}^{2}
+300\,{\omega}^{3}{\Omega}^{2}
+331\,{\omega}^{4}{\Omega}^{2}
+252\,{\omega}^{5}{\Omega}^{2}\nonumber\\
& +123\,{\omega}^{6}{\Omega}^{2}
+34\,{\omega}^{7}{\Omega}^{2}
+4\,{\omega}^{8}{\Omega}^{2}\,.
\end{align}
Due to the complex algebraic structure of $\lambda_1$ it is not possible to solve the equation $\lambda_1(\omega)=0$ exactly. We find numerically $\omega_\text{c} =  0.26065807\ldots$ and accordingly $\gamma_\text{out/in,\,c} = \omega_\text{c}/(2+\omega_\text{c}) = 0.1153019\ldots$

A comparison with the results of sect.~4 shows that the critical connectedness on the network with two overlapping cliques is rather close to that observed in the PPM. This leads us to conclude that, with reference to the stochastic dynamics of the NG, few agents with many links to more than one community make a tie comparable with many agents who tightly belong to a single community and have sparse links to the other. 

\subsection{Finite size effects}

\begin{figure}[t!]
  \centering
  \hskip 0.2cm\includegraphics[width=0.48\textwidth]{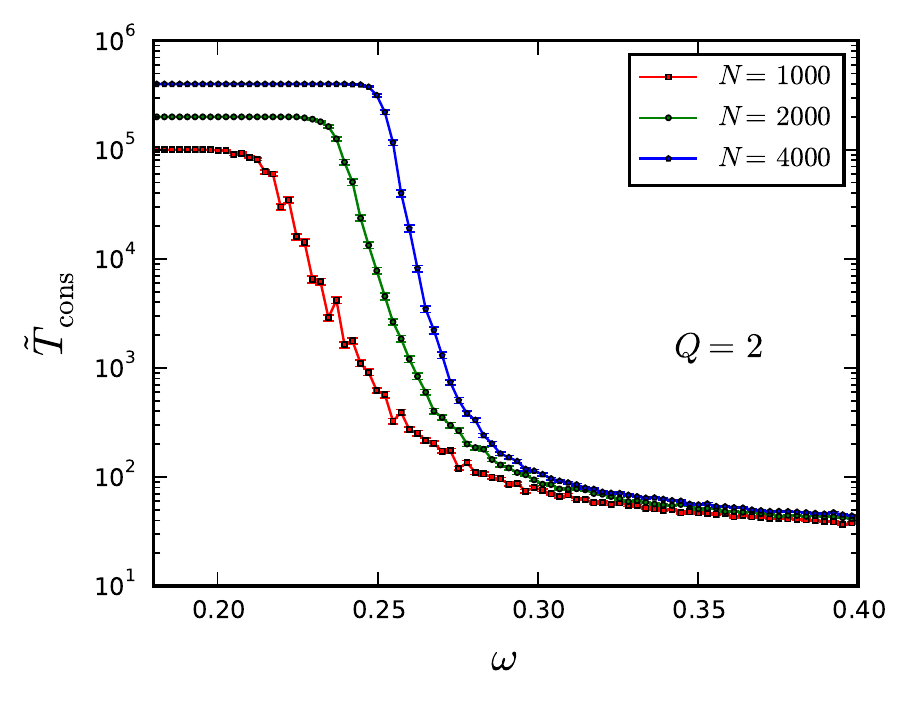}\,.
  \vskip -0.3cm
  \caption{\footnotesize Bounded time to consensus from simulations on a network with two overlapping cliques.}
  \label{fig:OVsimul}
  \vskip 0.1cm
\end{figure}

In Fig.~\ref{fig:OVsimul} we show the behaviour of the bounded time to consensus $\tilde T_\text{cons}(N,\omega)$ as a function of $\omega$, as obtained from  Monte Carlo simulations on finite networks with $N=1000,\,2000,\,4000$ (the exit threshold is set to~$100\cdot N$, like in eq.~(\ref{eq:TconsPPM})). Similar to Fig.~\ref{fig:PPMsimul} (\emph{right}), also in this case $\tilde T_\text{cons}$ is seen to increase as $\omega$ decreases and, in perfect analogy, the rise is steeper in correspondence of larger values of $N$. The main (and only) difference is that the appearance of long-lasting metastable states occurs here for values of $\omega$ which are about twice the values of $\nu$ observed therein, in accordance with $\omega_\text{c}\simeq 2\nu_\text{c}$. This confirms the validity of the analysis we made in the previous pages.

We conclude this section by commenting that since long-lasting multi-language metastable states are observed in a finite volume only for $\gamma_\text{out/in}$ below its critical threshold, having $\gamma_\text{out/in}(\nu_\text{c})\simeq \gamma_\text{out/in}(\omega_\text{c})$ yields a qualitative indication that the NG, when used as a community detection algorithm on empirical networks, is equivalently robust (in the language of ref.~\cite{Lambiotte} we would say that it has an equivalent resolution) in finding overlapping or non-overlapping communities, $\gamma_\text{out/in}$ being the same. This result represents the main lesson we learn from the algebraic exercises of sect.~4 and the present one.  

%% file: sect6.tex
\section{Dependence of $\nu_\text{c}$ upon $Q$ in the Planted Partition Model}

As seen in sect.~3, the $Q$-ary NG in the SBM develops an exponentially large number of phases as $Q$\break increases. As a result, studying the the phase diagram becomes soon unfeasible. Permutational symmetry certainly helps reduce the complexity of the problem, however the boundary surface of single phases is anyway expected to depend on $Q$ to some degree. Moreover, phase diagrams corresponding to different values of $Q$ live in Euclidean spaces with different dimensions, hence direct comparisons are ---strictly speaking--- ill-defined. In spite of this, something about the phase structure of the model can be said.  For all $Q>2$ the phase diagram is bounded by coordinate planes delimiting the 1st orthant of the $[Q(Q-1)]$-dimensional Euclidean space. We recall that for $i\ne k$ the $(\nu^{(ik)},\nu^{(ki)})$-plane is just the set of points with coordinates $\{\nu^{(\ell,m)}=0\}_{(\ell,m)\ne (i,k),(k,i)}$. Such points correspond physically to networks where all communities but the $i$th and $k$th ones are disconnected. Therefore, on the coordinate planes we fall back into the case $Q=2$. We conclude that the geometric structure of the phase diagram, at the boundaries of its domain, is precisely that of Fig.~\ref{fig:phases}.

More difficult is to establish the structure of phases in the bulk of the phase diagram. For instance, we know from Fig.~\ref{fig:phases} that for $Q=2$ the cusp of region II represents the point with maximum Euclidean distance from the origin, for which the system does not converge to global consensus. The formalism developed in sect.~3 and App. A allows us to show that for $Q>2$ the NG in the PPM is equally characterized by a critical threshold $\nu_\text{c}(Q)$. The reader may wonder whether it is true as well that in the SBM for $Q>2$ the point $\nu^{(12)} = \nu^{(21)} = \nu^{(13)} = \ldots =  \nu_\text{c}(Q)$ is that with maximum Euclidean distance from the origin, for which language coexistence is observed. We shall see in a while that the answer is negative. In fact, $\nu_\text{c}(Q)$ turns out to be a monotonically decreasing function of $Q$. 

\begin{figure}[t!]
  \centering
  \includegraphics[width=0.87\textwidth]{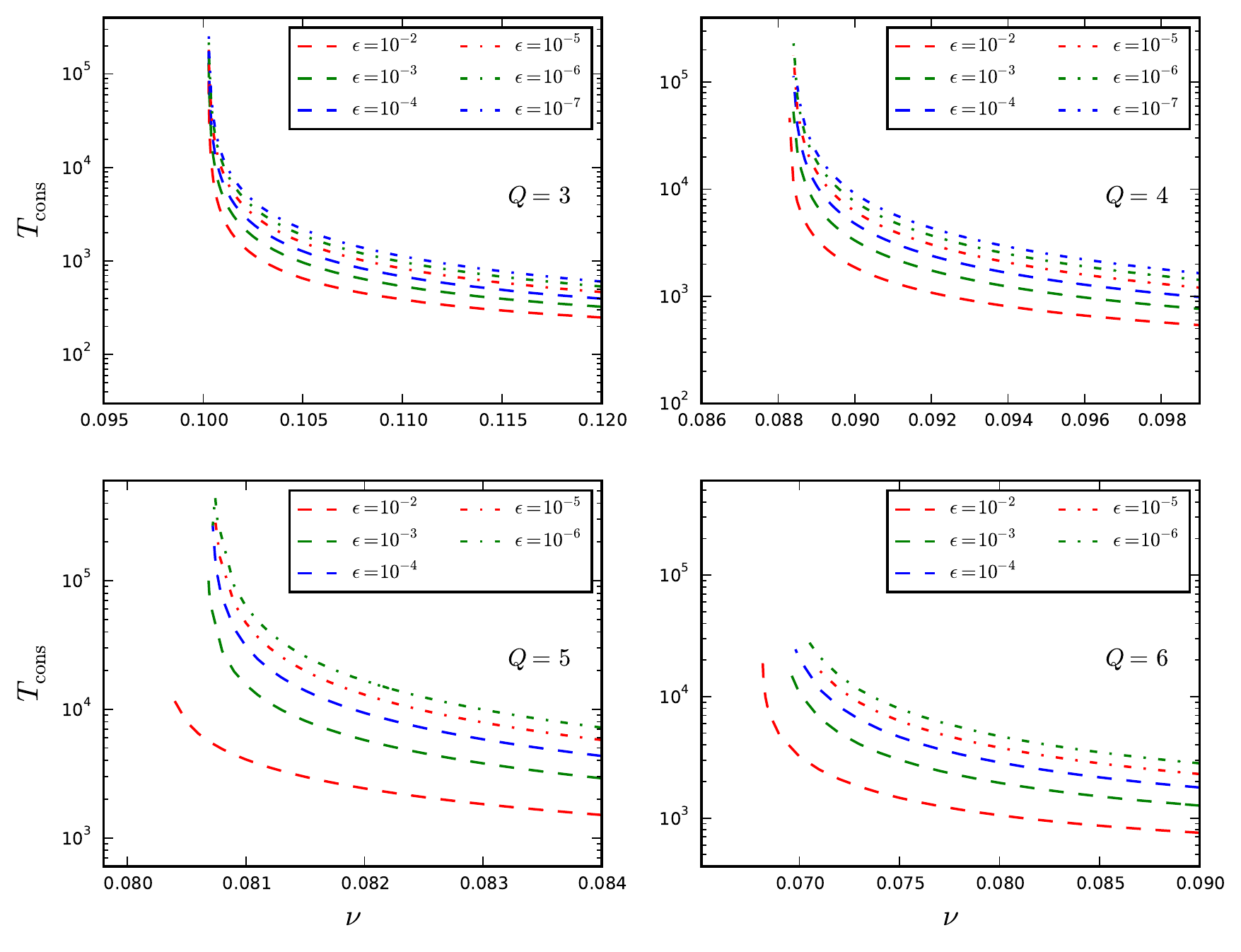}\,.
  \vskip -0.3cm
  \caption{\footnotesize Time to consensus in the PPM with $Q>2$.}
  \label{fig:TCPPMQQ}
  \vskip -0.2cm
\end{figure}

For $Q>2$ the dynamics of the NG in the PPM is still described by eqs.~(\ref{eq:mfebsm}) with all $\nu^{(ik)}=\nu=p_\text{out}/p_\text{in}$. Owing to the large number of degrees of freedom, we are unable to work out the symmetric steady state of the system exactly, as we did in sect.~4. Since $|\bar S(\cD)|=2^Q-2$, the overall number of unknowns (or equivalently coupled equations) amounts to $Q(2^Q-2)$, \eg for $Q=6$ we have 372 coupled equations.  The algebraic complexity of the transition rates $\{f_D^{(i)}\}$ increases with $Q$, too. For these reasons, we can only study the system numerically. Since in the PPM communities have all the same connectedness, either each of them keeps speaking its original language forever or the system goes to global consensus, with one single name colonizing the whole network. In order to let the system reach global consensus, we introduce an asymmetric perturbation $\epsilon$ in eq.~(\ref{eq:initcond}) that favours $A_1$, namely we integrate MFEs with initial conditions
\begin{equation}
  \left\{\begin{array}{ll} n^{(1)}_{A_1}\hskip -0.01em = 1\,,& \\[3.0ex] n^{(1)}_D = 0 & \text{ for } \ D\ne A_1\,,\end{array}\right.\quad \text{ and } \quad
  \left\{\begin{array}{ll} n^{(k)}_{A_k}\hskip -0.01em = 1-\epsilon\,,& \\[3.0ex] n^{(k)}_{A_1} = \epsilon &\,,\\[3.0ex]
    n^{(k)}_{D} = 0 & \text{ for } \ D\ne A_1,A_k\,,  \end{array}\right.\,\quad \text{ for }\ k\ne 1\,.
\end{equation}

In Fig.~\ref{fig:TCPPMQQ}, we show the behaviour of $T_\text{cons}$ as a function of $\nu$ for $Q=3,\ldots,6$ and for several values of~$\epsilon$. Numerical integration becomes demanding for $Q\gtrsim 5$, which is why the rise of $T_\text{cons}$ in  the plot at bottom right (corresponding to $Q=6$) is cut off. A glance to the four plots reveals that $\nu_\text{c}(Q)$ decreases as $Q$ increases. The dependence of $T_\text{cons}$ upon $\nu$ is still well described by eq.~(\ref{eq:model}). In Table~\ref{tab:fitparsQQ} we report estimates of the parameters $A,\nu_\text{c},\gamma$, obtained from fits to the theoretical model. Interestingly, the critical exponent $\gamma(0^+)$ appears to be independent of $Q$ (actually, for $Q=5,6$ we observe numerical instabilities in the fits due to a variety of factors, including round-off errors arising in numerical integration of MFEs and a gradual enhancement of the systematic error associated to the finiteness of $\Delta t/T_\text{cons}$ as $Q$ increases, see sect.~4).

\begin{table}[!t]
\begingroup\makeatletter\def\f@size{8}\check@mathfonts
  \begin{center}
    \begin{tabular}{c|r|r|r||r|r|r}
      \cline{2-7}\\[-2.7ex]\cline{2-7}
       & \multicolumn{3} {c||} {$Q=3$} & \multicolumn{3} {c} {$Q=4$} \\
      \hline
      $\epsilon$ & $A(\epsilon)\ \ \ $ & $\nu_\text{c}(\epsilon)\ \ \ $ & $\gamma(\epsilon)\ \ \ $ & $A(\epsilon)\ \ \ $ & $\nu_\text{c}(\epsilon)\ \ \ $ & $\gamma(\epsilon)\ \ \ $ \\
      \hline\\[-2.2ex]
      $1.0\times 10^{-2}$ & $8.524(2)$  & $0.100257(4)$ & $0.8162(2)$ & $22.72(4)$ & $0.088284(6)$ & $0.6902(1)$ \\
      $1.0\times 10^{-3}$ & $8.348(3)$  & $0.100252(4)$ & $0.8871(2)$ & $18.35(5)$ & $0.088345(6)$ & $0.8128(2)$ \\
      $1.0\times 10^{-4}$ & $9.894(4)$  & $0.100249(4)$ & $0.9100(3)$ & $17.72(5)$ & $0.088357(6)$ & $0.8733(2)$ \\
      $1.0\times 10^{-5}$ & $11.56(4)$  & $0.100247(4)$ & $0.9233(3)$ & $19.06(5)$ & $0.088360(6)$ & $0.9032(3)$ \\
      $1.0\times 10^{-6}$ & $12.34(5)$  & $0.100245(4)$ & $0.9409(4)$ & $20.59(6)$ & $0.088361(6)$ & $0.9235(4)$ \\
      $1.0\times 10^{-7}$ & $14.08(5)$  & $0.100244(4)$ & $0.9462(4)$ & $22.55(6)$ & $0.088361(6)$ & $0.9366(4)$ \\
      \hline
       & \multicolumn{3} {c||} {$Q=5$} & \multicolumn{3} {c} {$Q=6$} \\
      \hline
      $\epsilon$ & $A(\epsilon)\ \ \ $ & $\nu_\text{c}(\epsilon)\ \ \ $ & $\gamma(\epsilon)\ \ \ $ & $A(\epsilon)\ \ $ & $\nu_\text{c}(\epsilon)\, \ \ $ & $\gamma(\epsilon)\ \ \ $ \\
      \hline\\[-2.2ex]
      $1.0\times 10^{-2}$ & $52.1(3)$  & $0.08027(2)$ & $0.6028(3)$ & $93.6(4)$& $0.0681(4)$ & $0.5519(1)$ \\
      $1.0\times 10^{-3}$ & $52.3(4)$  & $0.08066(2)$ & $0.7045(8)$ & $75.4(4)$& $0.0689(4)$ & $0.7258(1)$ \\
      $1.0\times 10^{-4}$ & $44.5(4)$  & $0.08070(2)$ & $0.804(8)$ & $68.0(4)$& $0.0690(4)$ & $0.8254(1)$ \\
      $1.0\times 10^{-5}$ & $34.3(5)$  & $0.08067(2)$ & $0.92(2)$ & $72.9(4)$& $0.0690(4)$ & $0.8710(1)$ \\
      $1.0\times 10^{-6}$ & $22.5(7)$  & $0.08064(2)$ & $1.03(6)$ & $77.3(4)$& $0.0690(4)$ & $0.9045(1)$ \\
      \hline
      \hline
    \end{tabular}
    \vskip 0.2cm
    \caption{\footnotesize Estimates of fit parameters for $T_\text{cons}(\nu,\epsilon)$ for $Q>2$.\label{tab:fitparsQQ}}
  \end{center}
\endgroup
\vskip -0.4cm
\end{table}

In Fig.~\ref{fig:ppmcritpoints} (\emph{left}) we plot $\nu_\text{c}(Q)$ vs. $1/Q$. We observe an approximately linear behaviour, distorted however by a mild modulation in correspondence of the largest values of $Q$. This makes it difficult to extrapolate $\nu_\text{c}(Q)$ for $Q\to\infty$ (we leave this as an open problem). The scaling law $\nu_\text{c}(Q)\cdot Q \simeq \text{const}.$ looks pretty natural in consideration that communities  are equally connected to each other in the PPM: since the overall number of inter-community links connecting one community to the rest of the network increases proportionally to $Q-1$ for fixed $\nu$, the critical connectedness is expected to decrease correspondingly. The absence of anomalous scaling, such as $\nu_\text{c}(Q)\cdot Q^\alpha \simeq \text{const}.$ with $\alpha>1$, is a signal of robustness of multi-language phases against variations of $Q$. The observed scaling and our previous considerations about the boundaries of the phase diagram suggest as well that the joint union of all multi-language phases in the SBM is reverse convex on a large scale, \ie it is progressively squeezed towards the origin of the phase space as we approach the ``bisecting'' line $\{\nu^{(ik)} = \nu\}_{i\ne k}$. However, this picture needs further investigation to be confirmed or disproved.

To conclude, in Fig.~\ref{fig:ppmcritpoints} (\emph{right}) we show the behaviour of $\tilde T_\text{cons}$ as a function of $\nu$ for $Q=3$, as obtained from Monte Carlo simulations. We chose $N$ such that communities have the same size as in Fig.~\ref{fig:PPMsimul} (which corresponds to $Q=2$). The plot shows that the crossover region lies in a range of $\nu$ that is shifted to the left with respect to Fig.~\ref{fig:PPMsimul}, in agreement with the predictions of mean field theory.

%% file: sect7.tex
\section{Effects induced by a change of the relative size of communities}

Another important aspect of the problem is the way and the extent to which a change in the relative size of communities affects locally and/or globally the geometric structure of the phase diagram. For instance, consider the SBM and let $\sigma$ be (a piece of) some critical surface separating two phases. For fixed $N$, $\sigma$ moves across the phase space as $N^{(1)}$, \ldots, $N^{(Q)}$ are modified  continuously under the constraint $N^{(1)}+\ldots+N^{(Q)}=N$.  The question is whether and how $\sigma$ shifts, rotates, contracts and/or expands as a function of $\{N^{(k)}\}$. The problem depends on $Q-1$ continuous variables, thus answering in full generality is not easy.

\begin{figure}[t!]
  \centering
  \frame{\includegraphics[width=0.63\textwidth]{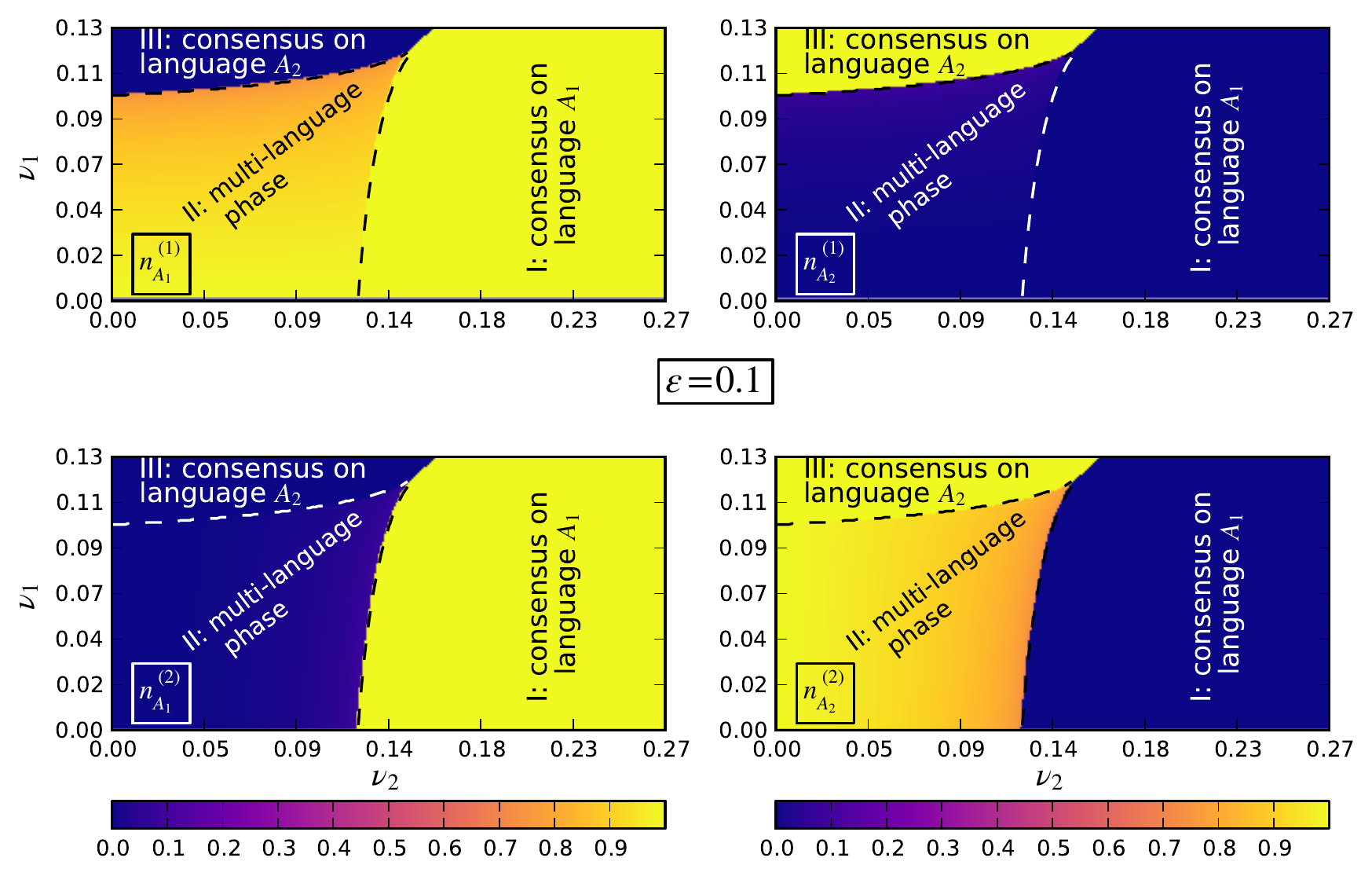}}\\[2.0ex]
  \frame{\includegraphics[width=0.63\textwidth]{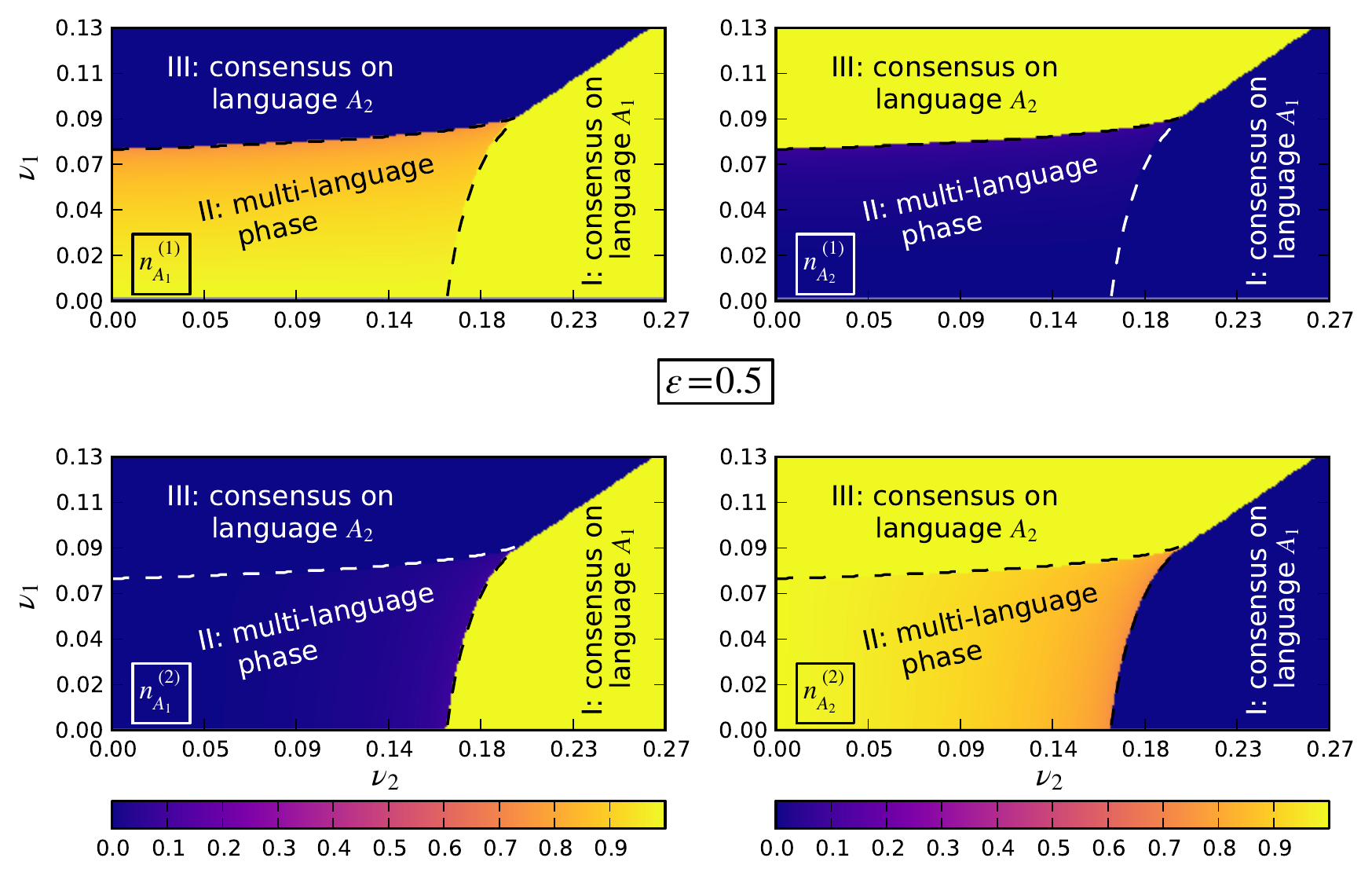}}\\[2.0ex]
  \frame{\includegraphics[width=0.63\textwidth]{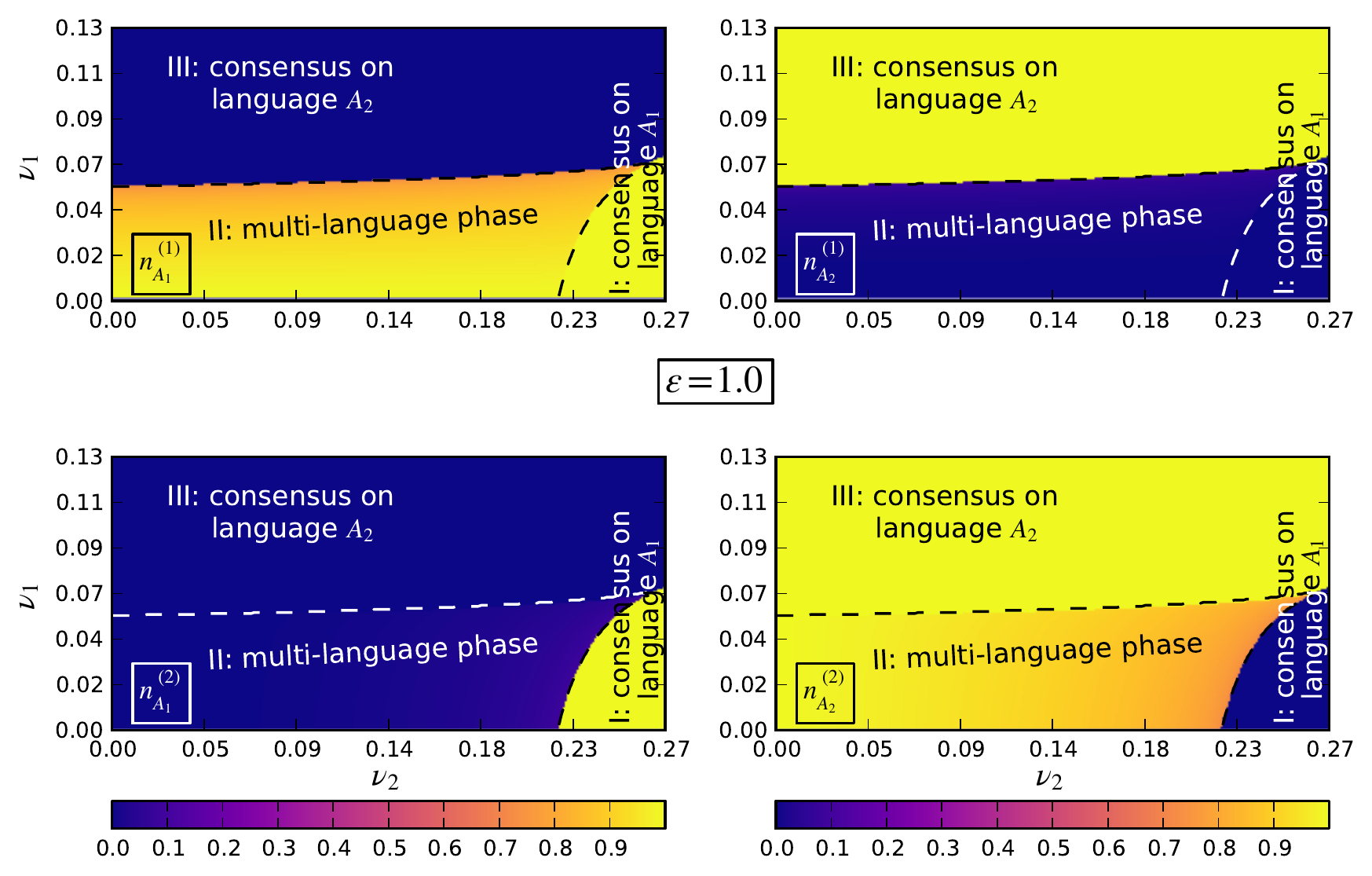}}\\[5.0ex]
  \caption{\footnotesize Phase diagram of the NG in the SBM with $Q=2$ and $N^{(2)}=(1+\epsilon)N^{(1)}$ for $\epsilon = 0.1,0.5,1.0$.}
  \label{fig:phdiffsize}
\end{figure}

To keep the theoretical framework as simple as possible, we assume $Q=2$. In this case, we have only one additional parameter. More precisely, we let $N^{(2)} = (1+\epsilon)N^{(1)}$. We assume $\epsilon>0$, hence $\cC^{(1)}$ is smaller than $\cC^{(2)}$. We also let $\nu_1 = p^{(12)}/p^{(11)}$ and $\nu_2 = p^{(12)}/p^{(22)}$, as we also did in sect. 3. Since the network is no more symmetric under exchange of community indexes, $\gamma_{\text{out/in}}$ is not well defined. The dynamics of the binary NG is still ruled by eqs.~(\ref{eq:sbmfrst})--(\ref{eq:sbmfrth}), but the probabilities of picking up an agent $x$ and a neighbour $x'$ of $x$ in one community or the other are now given by
\begin{align}
  \pi^{(11)} & = \text{prob}\left\{x\in\cC^{(1)},\ x'\in\cC^{(1)}\right\} = \frac{1}{2+\epsilon}\,\frac{1}{1+\nu_1(1+\epsilon)}\,, \\[0.0ex]
  \pi^{(12)} & = \text{prob}\left\{x\in\cC^{(1)},\ x'\in\cC^{(2)}\right\} = \frac{1}{2+\epsilon}\,\frac{\nu_1(1+\epsilon)}{1+\nu_1(1+\epsilon)}\,, \\[0.0ex]
  \pi^{(21)} & = \text{prob}\left\{x\in\cC^{(2)},\ x'\in\cC^{(1)}\right\} = \frac{1+\epsilon}{2+\epsilon}\,\frac{\nu_2}{1+\epsilon+\nu_2}\,, \\[0.0ex]
  \pi^{(22)} & = \text{prob}\left\{x\in\cC^{(2)},\ x'\in\cC^{(2)}\right\} = \frac{1+\epsilon}{2+\epsilon}\,\frac{1+\epsilon}{1+\epsilon+\nu_2}\,.
\end{align}
The above probabilities correctly fulfill $\pi^{(11)}+\pi^{(12)}+\pi^{(21)}+\pi^{(22)}=1$. Since the exchange symmetry is explicitly broken, steady solutions to MFEs are inevitably asymmetric. As such, they are also harder to work out than for $\epsilon=0$. Accordingly, we solve MFEs by numerical integration. In Fig.~\ref{fig:phdiffsize} we show phase diagrams corresponding to $\epsilon = 0.1,0.5,1.0$. We see that region II is progressively squeezed downwards, while it simultaneously expands rightwards, as $\epsilon$ increases. If we approximate region II by a rectangle with sides at $\nu_1=\nu_{1,\text{c}}$ and $\nu_2=\nu_{2,\text{c}}$, then Fig.~\ref{fig:phdiffsize} suggests that $\nu_{1,\text{c}}\cdot N^{(2)} \simeq \text{const.}$ and $\nu_{2,\text{c}}\cdot N^{(1)} \simeq \text{const.}$ For instance, for $\epsilon=1$ we have $N^{(2)} = 2N^{(1)}$ and we find $\nu_{1,\text{c}} \simeq 0.055$, which is about half the value found for $\epsilon = 0$. The above scaling laws look pretty natural in consideration that agents in $\cC^{(1)}$ have a number of neighbours in $\cC^{(2)}$ increasing proportionally to $N^{(2)}$ for fixed $\nu_1$ and the other way round. Hence, $\nu_{1,\text{c}}$ ($\nu_{2,\text{c}}$) is expected to decrease as $N^{(2)}$ ($N^{(1)}$) increases. The absence of anomalous scaling, such as $\nu_{1,\text{c}}\cdot (N^{(2)})^\alpha\simeq \text{const.}$ or $\nu_{2,\text{c}}\cdot (N^{(1)})^\alpha \simeq \text{const.}$ with $\alpha>1$, is a signal of robustness of multi-language phases against variations of the relative size of communities. 

\begin{figure}[t!]
  \centering
  \includegraphics[width=0.45\textwidth]{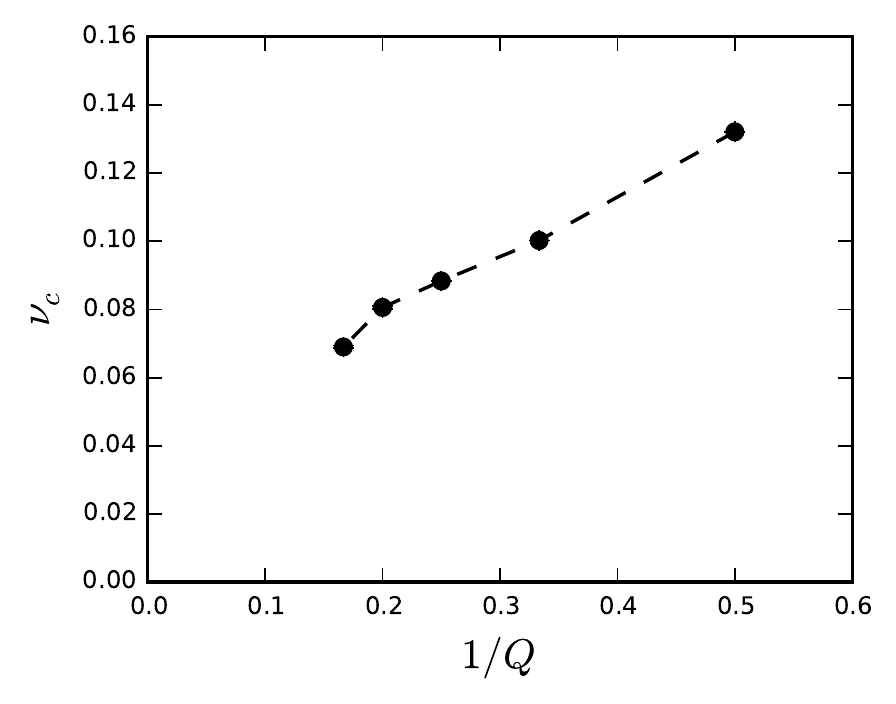}\,.
  \includegraphics[width=0.45\textwidth]{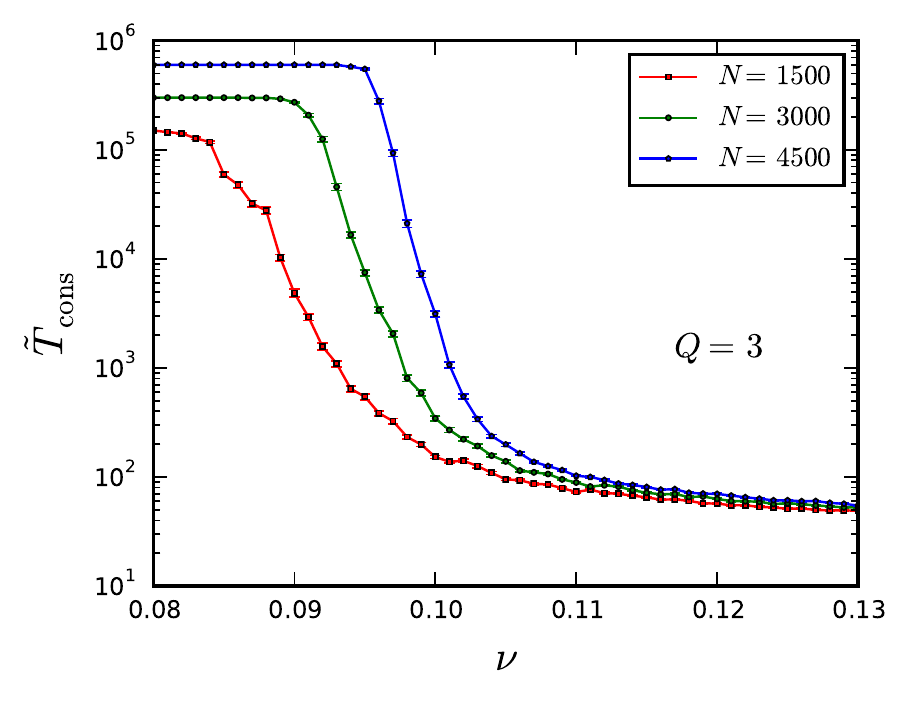}\,.
  \caption{\footnotesize (\emph{left}) Dependence of $\nu_c$ upon the number $Q$ of communities; (\emph{right}) Bounded time to consensus from simulations in the PPM with $Q=3$.}
  \label{fig:ppmcritpoints}
  \vskip -0.2cm
\end{figure}

%% file: sect8.tex
\section{Dependence of $\gamma_\text{out/in,\,c}$ upon the topology of $\{\cE^{(ik)}\}$}

 \begin{figure}[t!]
  \centering
  \hskip 0.2cm\includegraphics[width=0.25\textwidth]{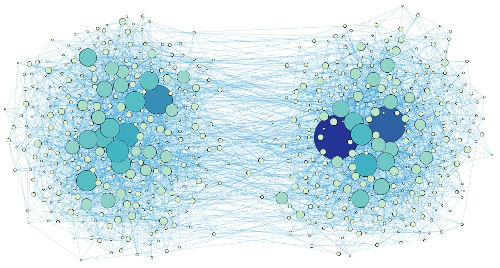}
  \hskip 2.0cm
  \hskip 0.5cm \includegraphics[width=0.35\textwidth]{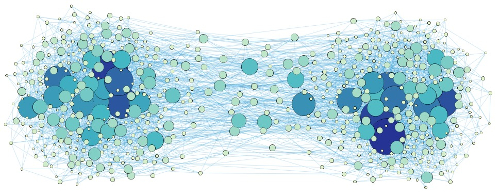}
  \vskip 0.1cm
  \caption{\footnotesize (left) BA communities interconnected by scale-free links (no inter-community assortativity); (right) BA communities interconnected by scale-free links (with inter-community assortativity). In both cases, $N=600$, $\rho=0.1$, the size of each node is proportional to its overall degree and the spatial position of nodes represents the equilibrium configuration of the Force Atlas visualization algorithm~\cite{forceatlas}.\label{fig:banetworks}}
  \vskip -0.2cm
\end{figure}

It is well known that realistic networks are heterogeneous (node degrees display high variability). Networks typically result from growth processes where new nodes join progressively those already in place. As a result, their topology cannot be described by static functions such as $\mathfrak{p}^{(ik)}(x,y)$. In order to examine how the critical point of the multi-language phase depends on the internal topology of communities and their interconnections, we study a network model with two interacting BA communities. Specifically, we fix $N$ and consider first two disjoint BA subgraphs $\cG_\text{BA}(N^{(k)},m_0,m)$~\cite{BarabasiSF}, each made of $N^{(k)}=N/2$ nodes, with $m_0$ and $m$ denoting respectively the number of initial nodes of each subgraph and the number of links each new node establishes, based on preferential attachment, when it joins the subgraph. In particular, we let $m_0=m=5$ in our numerical simulations. Then, we consider three possible definitions of $\cE^{(12)}$, all relying on one parameter~$\rho$: 
 \vskip 0.2cm
 \begin{itemize}[leftmargin=0.4in,rightmargin=0.0in]
 \item[$\cE^{(12)}_\text{ER\phantom{1}}$:]{we statically connect nodes belonging to different communities with probability $\pout(N) = 4m\rho/N$;}
 \item[$\cE^{(12)}_\text{SF1}$:]{starting with no inter-community links, we alternately choose at random a node belonging to one community and connect it to a target node belonging to the other one. The target node is chosen using a variant of preferential attachment where only inter-community links are taken into account when defining the target-node degree distribution. We stop the growth process as soon as $|\cE^{(12)}_\text{SF1}|=m\rho N$. We end up with $\cE^{(12)}_\text{SF1}$ having a scale-free topology. Moreover, there is no inter-community assortativity, \ie nodes with high inner degree in one community do not tend to attach preferably to nodes with high inner degree in the other one;}
 \item[$\cE^{(12)}_\text{SF2}$:]{we generate inter-community links similar to $\cE^{(12)}_\text{SF1}$, the only difference being that, concerning preferential attachment, both intra- and inter-community links are now taken into account when defining the target-node degree distribution. Again, $\cE^{(12)}_\text{SF2}$ develops a scale-free topology. Yet, there is inter-community assortativity in this case.}
 \end{itemize}
 \vskip -0.0cm
 \noindent The connectedness parameters are given by
 \begin{align}
   & \langle\kappa^{(1)}_\text{in}\rangle = \langle\kappa^{(2)}_\text{in}\rangle = \frac{2}{N}\frac{N}{2}2m = 2m\,,\\[0.0ex]
   & \langle\kappa^{(12)}_\text{out}\rangle_{\rm \scriptscriptstyle ER} = \langle\kappa^{(21)}_\text{out}\rangle_{\rm\scriptscriptstyle ER} = \frac{2}{N}\frac{N}{2}\pout(N)\frac{N}{2} = 2m\rho\,,\\[0.0ex]
   & \langle\kappa^{(12)}_\text{out}\rangle_{\rm \scriptscriptstyle SF1} = \langle\kappa^{(21)}_\text{out}\rangle_{\rm\scriptscriptstyle SF1} = \langle\kappa^{(12)}_\text{out}\rangle_{\rm \scriptscriptstyle SF2} = \langle\kappa^{(21)}_\text{out}\rangle_{\rm\scriptscriptstyle SF2} = \frac{2}{N}\frac{N}{2}\frac{2}{N}|\cE^{(12)}| = 2m\rho\,,
 \end{align}
 hence
 \begin{equation}
   \gamma_\text{out/in,\,ER} = \gamma_\text{out/in,\,SF1} = \gamma_\text{out/in,\,SF2} = \rho
 \end{equation}
 Examples of networks with $N=600$, inter-community links generated according to $\cE^{(12)}_\text{SF1}$ or  $\cE^{(12)}_\text{SF2}$ and $\rho=0.1$ are reported in Fig.~\ref{fig:banetworks}.

The main assumption underlying mean field theory is that agents are all equivalent. When links are heterogeneously distributed, this assumption is violated. In such a case, agents with many neighbours may (or may not) turn out to be more influential than agents with few ones, depending on the microscopic dynamics of the system. When this happens, MFEs lose predictive accuracy. Historically, the problem arose first in the context of epidemic spreading and was solved by Pastor-Satorras and Vespignani with the introduction of heterogeneous mean field theory~\cite{Vespignani,VespignaniTwo}. Here, agents with different degrees are treated separately. By analogy, the dynamics of the NG on heterogeneous community-based networks is expected to be accurately described by equations
\begin{equation}
  \frac{\rd n^{(i)}_{D,\kappa}}{\rd t} = f^{(i)}_{D,\kappa}(\bar n)\,,\quad\text{with }\   n^{(i)}_{D,\kappa} = \frac{\text{no. of agents with degree $\kappa$ and notebook $D$ belonging to $\cC^{(i)}$}}{N^{(i)}}\,.
  \label{eq:hmfes}
\end{equation}
State densities $\{\smash{n^{(i)}_{D,\kappa}}\}$ represent a refinement of $\{n^{(i)}_D\}$ in that they fulfill $n^{(i)}_D = \sum_\kappa n^{(i)}_{D,\kappa}$. As such, they also fulfill finer simplex conditions $\smash{\sum_{\kappa}\sum_D n^{(i)}_{D,\kappa}=1}$ for $i=1,\ldots,Q$. The mathematical structure of $\smash{f^{(i)}_{D,\kappa}}$ is similar to that of $f^{(i)}_D$ in standard MFEs, the only difference being that each term contributing to $f^{(i)}_{D,\kappa}$ is proportional to the probability $\pi^{(ik)}_{\kappa\kappa_2}$ of picking up an agent $x$ with degree $\kappa$ in $\cC^{(i)}$ and a neighbour $x'$ of $x$ with degree $\kappa_2$ in $\cC^{(k)}$ for some $k,\kappa_2$. Since agents can have arbitrarily large degrees in the thermodynamic limit, the number of heterogeneous MFEs is virtually infinite. In view of this, it seems reasonable to impose an upper cut-off $\kappa_\text{max}$ to the agent degree. Numerical solutions are then expected to converge as $\kappa_\text{max}\to\infty$.

\begin{table}[!t]
  \begin{center}
    \small
    \begin{tabular}{c|r|r|r}
      \hline
      \hline\\[-2.0ex]
            & $\cE^{(12)}_\text{ER}$ & $\cE^{(12)}_\text{SF1}$ & $\cE^{(12)}_\text{SF2}$ \\[1.0ex]
      \hline\\[-2.0ex]
      $\rho_\text{c}$ & 0.087(1)  &  0.187(5) & 0.127(4) \\[1.0ex]
      $\beta$        & 1.45(5)   &  1.50(9)  & 1.62(8)  \\[0.3ex]
      \hline
      \hline
    \end{tabular}
    \caption{\footnotesize Estimates of fit parameters for $\tilde T_\text{cons}$.\label{tab:fitboundpars}}
  \end{center}
  \vskip -0.5cm
\end{table}

\begin{figure}[t!]
  \centering
  \includegraphics[width=0.48\textwidth]{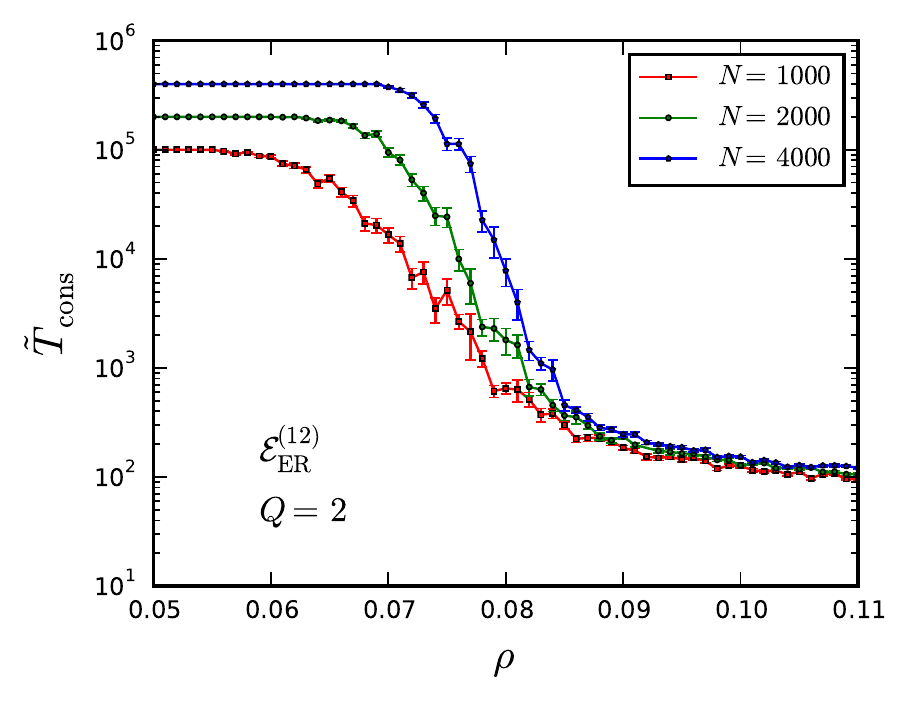}\,.
  \includegraphics[width=0.48\textwidth]{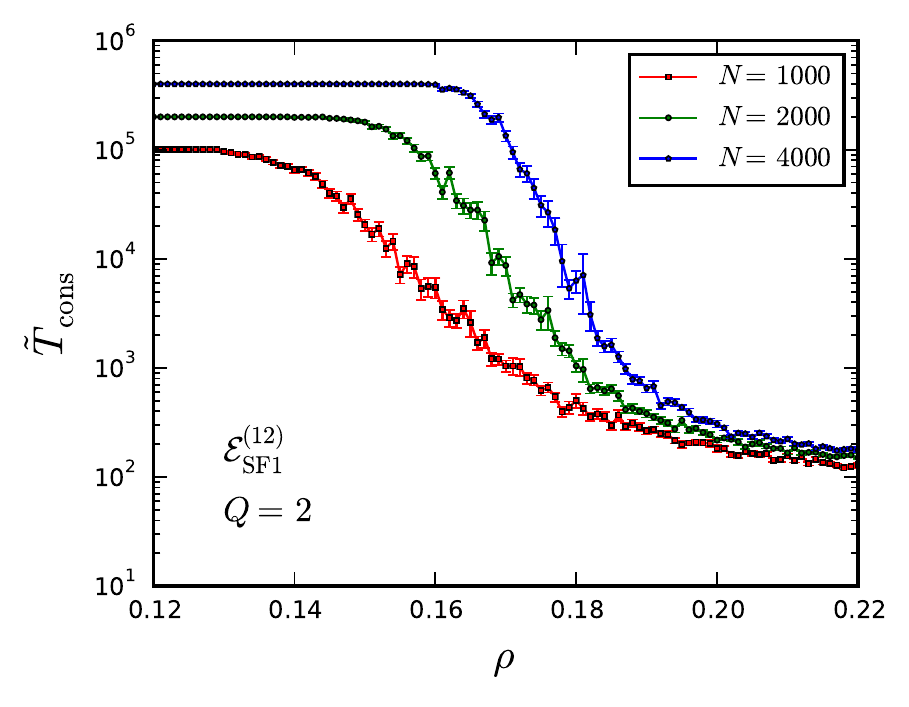}\,.
  \includegraphics[width=0.48\textwidth]{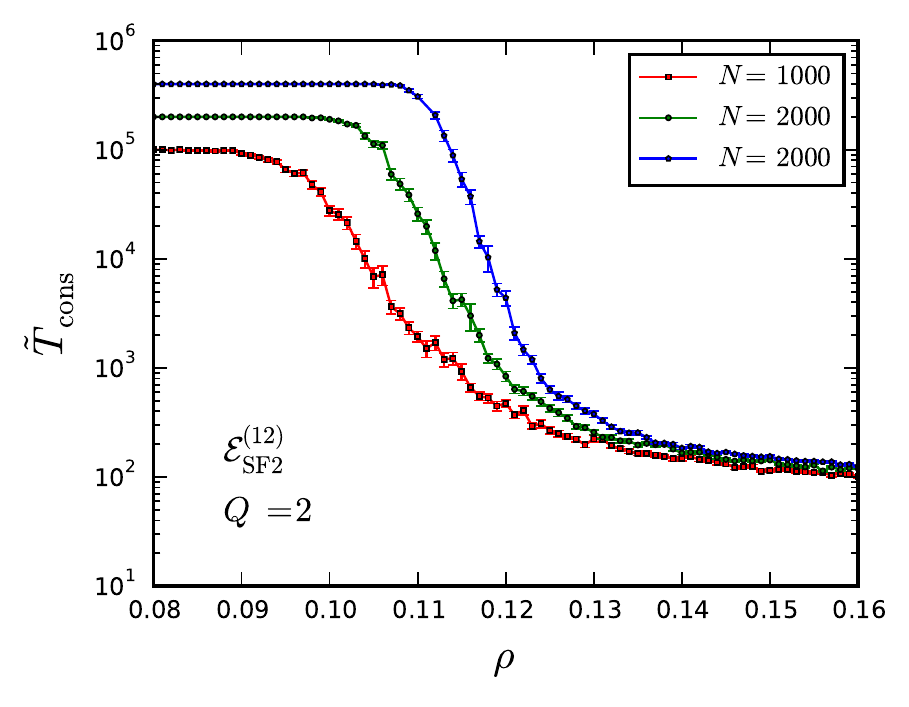}\,.
  \caption{\footnotesize Bounded time to consensus from simulations in M$_5$.}
  \label{fig:nuctopol}
  \vskip -0.2cm
\end{figure}

We leave for future research the precise computation of the critical connectedness $\gamma_\text{out/in,\,c}$ on heterogeneous networks via eqs.~(\ref{eq:hmfes}). Instead, we present here results obtained from numerical simulations. In particular, in Fig.~\ref{fig:nuctopol} we show the behaviour of the bounded time to consensus $\tilde T_\text{cons}$ as a function of $\rho$ for the network models introduced above. Although plots are qualitatively similar, the crossover region depends rather significantly on the topology of $\cE^{(12)}$. In this range of $\rho$ we can fit data to the curve described by eq.~(\ref{eq:thmodelT}), with $\nu_\text{c}$ replaced by $\rho_\text{c}$. We report our estimates of $\rho_\text{c}$ and $\beta$ in Table~\ref{tab:fitboundpars}. Pairwise comparisons suggest the following considerations.
\begin{itemize}[leftmargin=0.3in,rightmargin=0.0in]
\item{The network model with $\cE^{(12)}_\text{ER}$ differs from the PPM only in the internal structure of communities. A comparison of $\gamma_\text{out/in,c}$ in these models suggests that BA communities yield a more efficient opinion spread than ER ones. We know from ref.~\cite{Baronchelli:2} that $T_\text{cons}\propto N^{1.4}$ for both BA and ER networks (with no community structure). This is not in contradiction with our finding, which concerns indeed the effectiveness by which fluctuations break consensus within communities.}
\item{A comparison of $\rho_\text{c}$ in the network models with $\cE^{(12)}_\text{ER}$ and $\cE^{(12)}_\text{SF1}$ suggests that BA communities yield a more efficient opinion spread when interacting via random than via scale-free links, provided the latter have no correlation with the internal degree distribution. In other words, the effectiveness by which fluctuations break local consensus is largely reduced when intra- and inter-community links are heterogeneously distributed with no correlation to each other.}
\item{A comparison of $\rho_\text{c}$ in the network models with $\cE^{(12)}_\text{SF1}$ and $\cE^{(12)}_\text{SF2}$ shows that inter-community assortativity allows to restore the effectiveness by which fluctuations break local consensus. Indeed, $\gamma_\text{out/in,SF2,c}$ is very close to the critical connectedness observed in the PPM.}
\end{itemize}
In the end the variability of $\gamma_\text{out/in,\,c}$ is not dramatic. A glance to all network models we considered so far shows that, whenever  $\gamma_\text{out/in}$ is well defined, its critical value lies always in the range $0.1 \div 0.2$. Although it is not possible to parameterize the dependence of $\rho_\text{c}$ upon the topological structure of $\cE^{(ik)}$ in a simple way, such limited variation of $\rho_\text{c}$ is a clear signal of robustness of multi-language phases under variations of the network topology. 

%% file: sect9.tex
\section{Conclusions}
 
We studied the phase structure of the Naming Game (NG) on community-based networks. Prior to this paper it was known in the literature that communities of agents playing the NG tend to develop own long-lasting languages when sufficiently isolated. We showed that on infinitely extended networks the latter become everlasting. In other words, communities induce genuine multi-language phases in the thermodynamic limit. Our interest in studying these was both theoretical and practical. On the theoretical side, the NG is a non-trivial agent-based model ---designed to investigate the emergence of spoken languages in human societies--- whose phase structure is fully determined by the topology of the underlying network. On the practical side, the NG is an algorithm that, within limits, allows to detect communities in a given network. Either way, a first-principle analysis of which aspects of communities are more relevant or critical in order to guarantee the stability of local languages was lacking.

It should be clear that studying the phase structure of the NG on community-based networks is an ill-posed problem because the concept of community is not strictly defined by itself~\cite{Fortunato:2}. ECS conditions, introduced in sect.~2, define a huge \emph{ensemble} of networks for which no universal parameterization exists. Hence, the phase diagram of the system cannot be simply explored by varying a fistful of parameters. To bypass this difficulty we considered several distinct network models. Each of them was meant to highlight the dependence of multi-language phases upon specific features of communities. We studied the phase diagram in the stochastic block model to have a clear picture of how an increase in the fraction of links connecting communities triggers a sharp transition from local to global consensus. In connection with this, we derived exactly the cusp of the multi-language phase (region II of Fig.~\ref{fig:phases}) to show that in at least a simple case it is possible to get full insight into the phase transition. Then, we studied the NG on a network with two overlapping cliques to understand whether connectedness is more or less efficient than overlap in order to spread languages across communities. We finally examined the dependence of critical thresholds upon topological properties of the network, such as the number and relative size of communities or the structure of intra-/inter-community links, to investigate whether the model displays anomalous scaling somewhere in the phase space. 

The overall picture emerging from our study is that multi-language phases in the NG display a high degree of robustness against changes in the network topology. The characteristic scale of the connectedness parameter $\gamma_\text{out/in}$ (sect.~2), at which metastable equilibria become stable, lies at $\gamma_\text{out/in,c}\sim~0.1--0.2$, depending on specific features of the underlying network model. Such large values provide a full theoretical justification for using the NG as a community detection algorithm, in accordance with the original proposal of ref.~\cite{Lu:1} and the analysis performed in ref.~\cite{Gubanov}. Connectedness and overlap seem to contribute to a similar degree to break local equilibria and make the system converge to global consensus. The phase diagram appears to scale trivially under variations of the relative size of communities. Although the geometric structure of multi-language phases becomes increasingly complex as the number of communities $Q$ increases, at least the critical point corresponding to networks with fully symmetric communities displays a natural scaling with $Q$. Finally, the critical threshold $\gamma_\text{out/in,c}$ on two-community symmetric networks shows a mild dependence upon the topology of the intra-/inter-community links.

The phase structure induced by communities in the stochastic block model is remarkably similar to that induced by competing committed groups of agents on the fully connected graph~\cite{Xie:2}. The analytic structure of mean field equations is different in one case and the other, yet their solutions bear a strong resemblance. The rationale behind such similarity is not known. 

%% file: ackno.tex
\section*{Acknowledgments}

The computing resources used for our numerical study and the related technical support have been provided by the CRESCO/ENEA\-GRID High Performance Computing infrastructure and its staff \cite{Ponti}. CRESCO ({\color{red}C}omputational  {\color{red} RES}earch centre on {\color{red} CO}mplex systems) is funded by ENEA and by Italian and European research programmes.

%% file: apndx.tex
\begin{appendices}

\section{Derivation of MFEs in the SBM}

In order to work out the functions $f^{(i,\pm)}_D$ for $Q>2$, we need to introduce some additional notation.

{\mydef Given $k,n\ge 1$, and $D = \{A_{i_1},\ldots,A_{i_n}\}$ we let
\begin{equation}
  \theta_k\circ D = \left\{\begin{array}{ll}  \{A_{i_1},\ldots,A_{i_{k-1}},A_{i_{k+1}},\ldots,A_{i_n}\} & {\rm if }\ k\le n\,, \\[3.0ex]   D & {\rm otherwise}\,. \end{array} \right.
\end{equation}
In other words, the operator $\theta_k$ removes the $k$th name from a notebook. We also let
\begin{equation}
  \rho_k \circ D = A_{i_k}\,, \qquad {\rm for} \ k\le n\,,
\end{equation}
\ie the operator $\rho_k$ extracts the $k$th name out of a notebook. We finally let
\begin{equation}
  \Sigma_A(\cD) = \{D\in S(\cD):\ \ A\in D\}\,,
\end{equation}
that is to say $\Sigma_A$ is the set of all notebooks containing the name $A$.\hfill $\Box$
}
\vskip 0.1cm
We find it convenient to work out $f^{(i)}_D$ separately for notebooks with $|D|=1$ (single-name notebooks) and $|D|>1$ (multi-name notebooks). Indeed, if the initial conditions are chosen as in eq.~(\ref{eq:initcond}), densities $n^{(i)}_D$ with $|D|=1$ can increase throughout the game only owing to agent-agent interactions where a certain multi-name notebook collapses to $D$, whereas densities $n^{(i)}_D$ with $|D|>1$ can increase only when an agent adds a name to his/her notebook thus attaining $D$. We start from the latter.

\vskip 0.1cm

\noindent\underline{Case I : $|D|>1$}. In order to let $n^{(i)}_D$ increase, the only possibility is that a listener belonging to the $i$th community has initially a notebook differing from $D$ by the lack of one name and a speaker belonging to any community has a notebook containing that name. Therefore, we let $N_{+,D}=1$ for $|D|>1$. The contribution to MFEs is given by
\begin{equation}
  f^{(i,+,1)}_D(\bar n) = \sum_{\ell=1}^{|D|}  \  \  \sum_{\tilde D \in \Sigma_{\rho_\ell\circ D}({\cal D})} \ \ \frac{1}{|\tilde D|}\ \ \left [ \pi^{(ii)}n^{(i)}_{\tilde D} + \sum_{k\ne i}^{1\ldots Q} \pi^{(ik)}n^{(k)}_{\tilde D}\right] \ \ n^{(i)}_{\theta_\ell\circ D}\,.
\end{equation}
\vskip -0.2cm
\noindent The factor of $1/|\tilde D|$ represents the probability that the speaker chooses the name $\rho_\ell\circ D$ among those in his/her notebook.

In order to let $n^{(i)}_D$ decrease, a listener or a speaker in the $i$th community must have initially notebook $D$ and must modify it when interacting with another agent. The latter must have a notebook sharing at least one name with~$D$. Qualitatively, these are two different types of transitions, therefore  we let $N_{-,D}=2$ for $|D|>1$.
\begin{itemize}
\item{Type-I transitions lowering $n^{(i)}_{D}$ for $|D|>1$}
\end{itemize}
The listener belongs to the $i$th community and the speaker belongs to any community. The listener has initially notebook $D$, while the speaker has a notebook which has a non-vanishing overlap with $D$. The contribution to MFEs is given by
\begin{equation}
  f^{(i,-,1)}_D(\bar n) = \sum_{\ell=1}^{|D|} \ \ \sum_{\tilde D \in \Sigma_{\rho_\ell\circ D}({\cal D})} \frac{1}{|\tilde D|} \left[\pi^{(ii)}n^{(i)}_{\tilde D} + \sum_{k\ne i}^{1\ldots Q}\pi^{(ik)}n^{(k)}_{\tilde D}\right] n^{(i)}_D\,.
\end{equation}
\vskip -0.2cm
\noindent The factor of $1/|\tilde D|$ represents again the probability that the speaker chooses the name $\rho_\ell\circ D$ among those in his/her notebook.
\begin{itemize}
\item{Type-II transitions lowering $n^{(i)}_{D}$ for $|D|>1$}
\end{itemize}
The speaker belongs to the $i$th community and the listener belongs to any community. The speaker has initially notebook $D$, while the listener has a notebook which has a non-vanishing overlap with $D$. The contribution to MFEs is given by
\begin{equation}
  f^{(i,-,2)}_D(\bar n) = \sum_{\ell=1}^{|D|} \ \ \sum_{\tilde D \in \Sigma_{\rho_\ell\circ D}({\cal D})} \frac{1}{|D|} \left[\pi^{(ii)}n^{(i)}_{\tilde D} + \sum_{k\ne i}^{1\ldots Q}\pi^{(ik)}n^{(k)}_{\tilde D}\right] n^{(i)}_D\,,
\end{equation}
where the factor of $1/|D|$ represents once more the probability that the speaker chooses the name $\rho_\ell\circ D$ among those in his/her notebook.

\vskip 0.3cm

\noindent\underline{Case II : $|D|=1$}. In this case $D = \{A_\ell\}$ for some $\ell=1,\ldots,Q$. As mentioned above, $n^{(i)}_{A_\ell}$ can increase only because a listener or a speaker in the $i$th community has initially a multi-name notebook containing $A_\ell$, which collapses to $D=\{A_\ell\}$ when he/she interacts with another agent belonging to any community. The latter too must have initially a notebook containing $A_\ell$. There are three different types of transitions, thus we let $N_{+,A_\ell} = 3$.
\begin{itemize}
\item{Type-I transitions increasing $n^{(i)}_{A_\ell}$}
\end{itemize}
The listener belongs to the $i$th community and the speaker belongs to any community.  The speaker has notebook $D=\{A_\ell\}$, while the listener has any multi-name notebook $D_{\rm\scriptscriptstyle L}\in\Sigma_{A_\ell}({\cal D})$ with $|D_{\rm\scriptscriptstyle L}|>1$ (otherwise the interaction leaves the system unchanged!). The contribution to MFEs is given by
\begin{equation}
  f^{(i,+,1)}_{A_\ell}(\bar n) = \sum_{\substack{ D_{\rm\scriptscriptstyle L}\in \Sigma_{A_\ell}(\cD)\\[1.0ex] |D_{\rm\scriptscriptstyle L}|>1}} \left[\pi^{(ii)}n^{(i)}_{A_\ell} + \sum_{k\ne i}^{1\ldots Q}\pi^{(ik)}n^{(k)}_{A_\ell}\right]\,n^{(i)}_{D_{\rm\scriptscriptstyle L}}\,.
\end{equation}
\begin{itemize}
\item{Type-II transitions increasing $n^{(i)}_{A_\ell}$}
\end{itemize}
The speaker belongs to the $i$th community and the listener belongs to any community. The listener has notebook $D=\{A_\ell\}$, while the speaker has any multi-name notebook $D_{\rm\scriptscriptstyle S}\in\Sigma_{A_\ell}(\cD)$ with $|D_{\rm\scriptscriptstyle S}|>1$ (otherwise the interaction leaves again the system unchanged!). The contribution to MFEs is given by
\begin{equation}
  f^{(i,+,2)}_{A_\ell}(\bar n) = \sum_{\substack{D_{\rm\scriptscriptstyle S}\in \Sigma_{A_\ell}(\cD)\\[1.0ex] |D_{\rm\scriptscriptstyle S}|>1}} \frac{1}{|D_{\rm\scriptscriptstyle S}|} \left[\pi^{(ii)}n^{(i)}_{A_\ell} +\sum_{k\ne i}^{1\ldots Q}\pi^{(ik)}n^{(k)}_{A_\ell}\right]\,n^{(i)}_{D_{\rm\scriptscriptstyle S}}\,,
\end{equation}
  where the factor of $1/|D_S|$ represents the probability that the speaker chooses the name $A_\ell$ among those in his/her notebook. 
\vskip 0.2cm
\begin{itemize}
\item{Type-III transitions increasing $n^{(i)}_{A_\ell}$}
\end{itemize}
The speaker belongs to the $i$th community and the listener belongs to any community or the other way round. The speaker has a notebook $D_{\rm\scriptscriptstyle S}\in\Sigma_{A_k}(\cD)$ with $|D_{\rm \scriptscriptstyle S}|>1$ and the listener has a notebook $D_{\rm \scriptscriptstyle L}\in\Sigma_{A_k}(\cD)$ with $|D_{\rm\scriptscriptstyle L}|>1$. The contribution of this type of interaction to MFEs is given by
\begin{equation}
  f^{(i,+,3)}_{A_\ell}(\bar n) = \sum_{\substack{D_{\rm\scriptscriptstyle S},D_{\rm\scriptscriptstyle L}\in \Sigma_{A_\ell}(\cD) \\[1.0ex] |D_{\rm S}|,|D_{\rm L}|>1}} \frac{1}{|D_{\rm\scriptscriptstyle S}|} \,\left[2\pi^{(ii)}n^{(i)}_{D_{\rm\scriptscriptstyle S}}n^{(i)}_{D_{\rm\scriptscriptstyle L}} + \sum_{k\ne i}^{1\ldots Q} \pi^{(ik)}\left( n^{(i)}_{D_{\rm\scriptscriptstyle S}}\,n^{(k)}_{D_{\rm\scriptscriptstyle L}} + n^{(k)}_{D_{\rm\scriptscriptstyle S}}\,n^{(i)}_{D_{\rm\scriptscriptstyle L}}\right)\right]\,,
\end{equation}
where the factor of $1/|D_{\rm\scriptscriptstyle S}|$ represents the probability that the speaker chooses the name $A_\ell$ among those in his/her notebook and the factor of $2$ takes into account that $n^{(i)}_{A_\ell}$ increases by 2 fractional units following the transition if both speaker and listener belong to the $i$th community.
\vskip 0.3cm
We finally discuss transitions lowering $n^{(i)}_{A_\ell}$. For this to happen it is necessary that an agent belonging to the $i$th community, who has initially notebook $D=\{A_\ell\}$, switches to a multi-name notebook. This is possible only provided the agent adds a second name to his/her notebook and this can occur only if the agent is a listener. The speaker's notebook might either contain $A_\ell$ or not and we must take care of properly counting the probability of not choosing $A_\ell$ in the former case, otherwise we fall back into Case I/Type-II. Therefore,  we find it better to work out separately the two types of transitions and correspondingly we let $N_{-,A_\ell}=2$.
\vskip 0.2cm
\begin{itemize}
\item{Type-I transitions lowering $n^{(i)}_{A_\ell}$}
\end{itemize}
The listener belongs to the $i$th community and the speaker belongs to any community. The listener has notebook $D=\{A_\ell\}$ and the speaker has a notebook $D_{\rm\scriptscriptstyle S}\in \Sigma_{A_\ell}(\cD)$ with $|D_{\rm\scriptscriptstyle S}|>1$. The contribution to MFEs is given by
\begin{equation}
  f^{(i,-,1)}_{A_\ell}(\bar n) = \sum_{\substack{D_{\rm\scriptscriptstyle S}\in\Sigma_{A_\ell}(\cD) \\ |D_{\rm\scriptscriptstyle S}|>1}} \frac{|D_{\rm\scriptscriptstyle S}|-1}{|D_{\rm\scriptscriptstyle S}|}  \left[\pi^{(ii)}n^{(i)}_{D_{\rm\scriptscriptstyle S}} + \sum_{k\ne i}^{1\ldots Q}\pi^{(ik)} n^{(k)}_{D_{\rm\scriptscriptstyle S}}\right]n^{(i)}_{A_\ell}\,,
\end{equation}
where the factor of $\frac{|D_{\rm\scriptscriptstyle S}|-1}{|D_{\rm\scriptscriptstyle S}|}$ represents the probability that the speaker chooses a name different from $A_\ell$ among those in his/her notebook. 
\vskip 0.2cm
\begin{itemize}
\item{Type-II transitions lowering $n^{(i)}_{A_\ell}$}
\end{itemize}
The listener belongs to the $i$th community and the speaker belongs to any community. The listener has notebook $D=\{A_\ell\}$ and the speaker has a notebook $D_{\rm\scriptscriptstyle S}\in [\Sigma_{A_k}(\cD)]^{\rm c}$, where $[D]^{\rm c} = S(\cD)\setminus D$ denotes generically the complement of $D$ in $S(\cD)$. The contribution to MFEs is given by
\begin{equation}
  f^{(i,-,2)}_{A_\ell}(\bar n) = \sum_{D_{\rm\scriptscriptstyle S}\in[\Sigma_{A_\ell}(\cD)]^{\rm c}} n^{(i)}_{A_\ell} \left[\pi^{(ii)}n^{(i)}_{D_{\rm\scriptscriptstyle S}} + \sum_{k\ne i}^{1\ldots Q}\pi^{(ik)}n^{(k)}_{D_{\rm\scriptscriptstyle S}}\right]\,,
\end{equation}
where no probability coefficient is needed in front of the product of densities, such as in all previous cases, since here the speaker can choose equivalently any name among those in his/her notebook. 

When the above sums are expanded, lengthy and tedious algebraic expressions are generated, even for the case $Q=3$. To enable interested readers to write down complete MFEs, we provide in App.~B an essential and correctly working Maple$^\text{TM}$ code for the PPM (it generates MFEs for $Q=6$, as set at line 3). The code can be easily generalized to the SBM, while its output can be easily adapted for numerical analysis.

We apply in sect.~6 the formalism here presented. 

\section{Maple$^\textsuperscript{TM}$ code for generating MFEs in the PPM}

{\scriptsize

 \begin{minipage}{0.5\textwidth}
  \begin{Verbatim}[frame=single,rulecolor=\color{blue}]
with(CodeGeneration):

# Parameters
Q := 6:

###########################################################

sublist := proc(list, ini)
  [seq(list[i],i=ini..nops(list))]:
end proc:

addElt := proc(elt, list)
local templist, i:
  templist:=[]:
  for i from 1 to nops(list) do
    templist:=[op(templist),[elt, op(list[i])]]:
  end do:
  templist:
end proc:

listSubsets := proc(L, i)
local n, j, fl, temp:
  n := nops(L):
  if i=1 then
    fl:=[]:
    for j from 1 to n do
      fl:=[op(fl),[L[j]]]:
    end do:
  else
    fl:=[]:
    for j from 1 to n-i+1 do
      temp:=listSubsets(sublist(L,j+1),i-1):
      temp:=addElt(L[j],temp):
      fl:=[op(fl),op(temp)]:
    end do:
  end if:
  fl:
end proc:

listAllSubsets := proc(Q::posint)
description "generates $S(\\cal D)$ with Q names":
local L,i, temp, fl:
  L := [A||(1..Q)]:
  fl := []:
  for i from 1 to Q do
    fl:=[op(fl),op(listSubsets(L,i))]:
  end do:
  fl:
end proc:

SigmaAk := proc(L::list,item)
description "applies the operator $\\Sigma_{item}$ of "
            "eq. (3.12) to L":
local dd,outL:
  outL := NULL:
  for dd in L do
    if(member(item,dd)) then
      outL := outL,dd:
    end if:
  end do:
  outL:
end proc:

                           -- 1 --

\end{Verbatim}
\end{minipage}
\begin{minipage}{0.5\textwidth}
  \begin{Verbatim}[frame=single,rulecolor=\color{blue}]
TauAk := proc(L::list,item)
local dd,outL:
  outL := NULL:
  for dd in L do
    if(not member(item,dd)) then
      outL := outL,dd:
    end if:
  end do:
  outL:
end proc:

removeListItem := proc(L::list,item)
  remove(has,L,item):
end proc:

thetak := proc(L::list,k::posint)
description "applies the operator $\\theta_{k}$ of "
            "eq. (3.10) to L":
local seq1,seq2,i:
  if (k>nops(L)) then
    return L:
  end if:
  seq1 := seq(op(i,L),i=1..k-1):
  seq2 := seq(op(i,L),i=k+1..nops(L)):
  return [seq1,seq2]:
end proc:

rhok := proc(L::list,k::posint)
description "applies the operator $\\rho_{k}$ of "
            "eq. (3.11) to L":
local seq1,seq2,i:
  if (k>nops(L)) then
    return NULL:
  end if:
  return op(k,L):
end proc:

# ===========================================================
# Case I
# ===========================================================

CaseIrise := proc(Dset::list,i::posint)

global Q,NBList,pin,pout:
local out,j,k,Wk,Uk,NBSubList,Dtilde:
  if(nops(Dset)=1) then
    print("[CaseIrise]: WARNING -> |Dset|=1"):
    return:
  end if:
  if(i>Q) then
    print("[CaseIrise]: WARNING -> index i out of range"):
    return:
  end if:
  out := 0:
  for k from 1 to nops(Dset) do
    Wk := rhok(Dset,k):
    Uk := thetak(Dset,k):
    NBSubList := SigmaAk(NBList,Wk):
    for Dtilde in NBSubList do
      out := out +
      (1/nops(Dtilde))*n[i,Uk]*(pin*n[i,Dtilde] +
                            pout*sum(n[j,Dtilde],j=1..i-1) +
                            pout*sum(n[j,Dtilde],j=i+1..Q)):

                           -- 2 --
     
\end{Verbatim}
\end{minipage}

 \begin{minipage}{0.5\textwidth}
  \begin{Verbatim}[frame=single,rulecolor=\color{blue}]
    end do:
  end do:
  return out:
end proc:

CaseIlower1 := proc(Dset::list,i::posint)
global Q,NBList,pin,pout:
local out,j,k,Wk,NBSubList,Dtilde:
  if(nops(Dset)=1) then
    print("[CaseIlower1]: WARNING -> |Dset|=1"):
    return:
  end if:
  if(i>Q) then
    print("[CaseIlower1]: WARNING -> index i out of range"):
    return:
  end if:
  out := 0:
  for k from 1 to nops(Dset) do
    Wk := rhok(Dset,k):
    NBSubList := SigmaAk(NBList,Wk):
    for Dtilde in NBSubList do
      out := out +
      (1/nops(Dtilde))*n[i,Dset]*(pin*n[i,Dtilde] +
                           pout*sum(n[j,Dtilde],j=1..i-1) +
                           pout*sum(n[j,Dtilde],j=i+1..Q)):
    end do:
  end do:
  return out:
end proc:

CaseIlower2 := proc(Dset::list,i::posint)
global Q,NBList,pin,pout:
local out,j,k,Wk,NBSubList,Dtilde:
  if(nops(Dset)=1) then
    print("[CaseIlower2]: WARNING -> |Dset|=1"):
    return:
  end if:
  if(i>Q) then
    print("[CaseIlower2]: WARNING -> index i out of range"):
    return:
  end if:
  out := 0:
  for k from 1 to nops(Dset) do
    Wk := rhok(Dset,k):
    NBSubList := SigmaAk(NBList,Wk):
    for Dtilde in NBSubList do
      out := out +
      (1/nops(Dset))*n[i,Dset]*(pin*n[i,Dtilde] +
                            pout*sum(n[j,Dtilde],j=1..i-1) +
                            pout*sum(n[j,Dtilde],j=i+1..Q)):
    end do:
  end do:
  return out:
end proc:

# ===========================================================
# Case II
# ===========================================================

CaseIIrise1 := proc(Dset::list,i::posint)
global Q,NBList,pin,pout:
local out,j,k,Ak,NBSubList,DL:
  if(nops(Dset)>1) then
    print("[CaseIIrise1]: WARNING -> |Dset|>1"):
    return:
  end if:

                           -- 3 --

  \end{Verbatim}
 \end{minipage}
 \begin{minipage}{0.5\textwidth}
  \begin{Verbatim}[frame=single,rulecolor=\color{blue}]
  if(i>Q) then
    print("[CaseIIrise1]: WARNING -> index i out of range"):
    return:
  end if:
  out := 0:
  Ak := op(1,Dset):
  NBSubList := SigmaAk(NBList,Ak):
  for DL in NBSubList do
    if (nops(DL)>1) then
      out := out +
      n[i,DL]*(pin*n[i,Dset] +
               pout*sum(n[j,Dset],j=1..i-1) +
               pout*sum(n[j,Dset],j=i+1..Q)):
    end if:
  end do:
  return out:
end proc:

CaseIIrise2 := proc(Dset::list,i::posint)
global Q,NBList,pin,pout:
local out,j,k,Ak,NBSubList,DS:
  if(nops(Dset)>1) then
    print("[CaseIIrise2]: WARNING -> |Dset|>1"):
    return:
  end if:
  if(i>Q) then
    print("[CaseIIrise2]: WARNING -> index i out of range"):
    return:
  end if:
  out := 0:
  Ak := op(1,Dset):
  NBSubList := SigmaAk(NBList,Ak):
  for DS in NBSubList do
    if (nops(DS)>1) then
      out := out +
      (1/nops(DS))*n[i,DS]*(pin*n[i,Dset] +
                            pout*sum(n[j,Dset],j=1..i-1) +
                            pout*sum(n[j,Dset],j=i+1..Q)):
    end if:
  end do:
  return out:
end proc:

CaseIIrise3 := proc(Dset::list,i::posint)
global Q,NBList,pin,pout:
local out,j,k,Ak,NBSubList,DS,DL:
  if(nops(Dset)>1) then
    print("[CaseIIrise3]: WARNING -> |Dset|>1"):
    return:
  end if:
  if(i>Q) then
    print("[CaseIIrise3]: WARNING -> index i out of range"):
    return:
  end if:
  out := 0:
  Ak := op(1,Dset):
  NBSubList := SigmaAk(NBList,Ak):
  for DS in NBSubList do
    for DL in NBSubList do
      if (nops(DS)>1 and nops(DL)>1) then
        out := out +
       (1/nops(DS))*(2*pin*n[i,DS]*n[i,DL] +
       pout*sum(n[i,DS]*n[j,DL] + n[j,DS]*n[i,DL],j=1..i-1) +
       pout*sum(n[i,DS]*n[j,DL] + n[j,DS]*n[i,DL],j=i+1..Q)):
      end if:
    end do:
  end do:  
                           -- 4 --

  \end{Verbatim}
 \end{minipage}

 \begin{minipage}{0.5\textwidth}
  \begin{Verbatim}[frame=single,rulecolor=\color{blue}]
  return out:
end proc:

CaseIIlower1 := proc(Dset::list,i::posint)
global Q,NBList,pin,pout:
local out,j,k,Ak,NBSubList,DS:
  if(nops(Dset)>1) then
    print("[CaseIIlower1]: WARNING -> |Dset|>1"):
    return:
  end if:
  if(i>Q) then
    print("[CaseIIlower1]: WARNING -> index i out of range"):
    return:
  end if:
  out := 0:
  Ak := op(1,Dset):
  NBSubList := SigmaAk(NBList,Ak):
  for DS in NBSubList do
    if (nops(DS)>1) then
      out := out +
      ((nops(DS)-1)/nops(DS))*n[i,Dset]*(pin*n[i,DS] +
                               pout*sum(n[j,DS],j=1..i-1) +
                               pout*sum(n[j,DS],j=i+1..Q)):
    end if:
  end do:
  return out:
end proc:

CaseIIlower2 := proc(Dset::list,i::posint)
global Q,NBList,pin,pout:
local out,j,k,Ak,NBSubList,DS:
  if(nops(Dset)>1) then
    print("[CaseIIlower2]: WARNING -> |Dset|>1"):
    return:
  end if:
  if(i>Q) then
    print("[CaseIIlower2]: WARNING -> index i out of range"):
    return:
  end if:
  out := 0:
  Ak := op(1,Dset):
  NBSubList := [TauAk(NBList,Ak)]:
  for DS in NBSubList do
    out := out +
    n[i,Dset]*(pin*n[i,DS] +
               pout*sum(n[j,DS],j=1..i-1) +
               pout*sum(n[j,DS],j=i+1..Q)):
  end do:
  return out:
end proc:

#-------------------------------------------------------

NBList := listAllSubsets(Q):

f_intra := Array([seq(0,i=1..Q*(2^Q-1))]):
f_inter := Array([seq(0,i=1..Q*(2^Q-1))]):

SDctr := 1:
for qq from 1 to Q do
  Dctr := 1:
  for Dset in op(1..-2,NBList) do

    if(nops(Dset)>1) then
      f1       := expand(CaseIrise(Dset,qq)):
      f1_intra := coeff(f1,pin):

                           -- 5 --

  \end{Verbatim}
 \end{minipage}
 \begin{minipage}{0.5\textwidth}
  \begin{Verbatim}[frame=single,rulecolor=\color{blue}]
      f1_inter := coeff(f1,pout):
      f2       := expand(CaseIlower1(Dset,qq)):
      f2_intra := coeff(f2,pin):
      f2_inter := coeff(f2,pout):
      f3       := expand(CaseIlower2(Dset,qq)):
      f3_intra := coeff(f3,pin):
      f3_inter := coeff(f3,pout):

      T_intra  := f1_intra - f2_intra - f3_intra:
      T_inter  := f1_inter - f2_inter - f3_inter:
    else
      g1       := expand(CaseIIrise1(Dset,qq)):
      g1_intra := coeff(g1,pin):
      g1_inter := coeff(g1,pout):

      g2       := expand(CaseIIrise2(Dset,qq)):
      g2_intra := coeff(g2,pin):
      g2_inter := coeff(g2,pout):

      g3       := expand(CaseIIrise3(Dset,qq)):
      g3_intra := coeff(g3,pin):
      g3_inter := coeff(g3,pout):

      g4       := expand(CaseIIlower1(Dset,qq)):
      g4_intra := coeff(g4,pin):
      g4_inter := coeff(g4,pout):

      g5       := expand(CaseIIlower2(Dset,qq)):
      g5_intra := coeff(g5,pin):
      g5_inter := coeff(g5,pout):

      T_intra := g1_intra + g2_intra + g3_intra
               - g4_intra - g5_intra:
      T_inter := g1_inter + g2_inter + g3_inter
               - g4_inter - g5_inter:
    end if:

    # =======================================================
    #  remove Rosetta notebooks
    # =======================================================

    for ii from 1 to Q do
      for jj from 1 to (2**Q-2) do
        T_intra := eval(T_intra,
            n[ii,op(jj,NBList)]=xx[jj+(ii-1)*(2**Q-1)]):
        T_inter := eval(T_inter,
            n[ii,op(jj,NBList)]=xx[jj+(ii-1)*(2**Q-1)]):
      end do:
      T_intra := expand(eval(T_intra,n[ii,op(-1,NBList)]=
             1-add(xx[kk+(ii-1)*(2**Q-1)],kk=1..(2**Q-2)))):
      T_inter := expand(eval(T_inter,n[ii,op(-1,NBList)]=
             1-add(xx[kk+(ii-1)*(2**Q-1)],kk=1..(2**Q-2)))):
    end do:

    printf("#  clique = %2d \t state = %s \t index = %d\n\n",
          qq,convert(Dset,string),SDctr):
    printf("#  nops(intra) = %10d \t nops(inter) = %10d\n\n",
          nops(T_intra), nops(T_inter)):

    f_intra[SDctr] := T_intra:
    f_inter[SDctr] := T_inter:

    SDctr := SDctr + 1:
    Dctr := Dctr+1:

  end do:

                           -- 6 --

  \end{Verbatim}
 \end{minipage}

 \begin{minipage}{0.5\textwidth}
  \begin{Verbatim}[frame=single,rulecolor=\color{blue}]
  f_intra[SDctr] := 0:
  f_inter[SDctr] := 0:
  SDctr := SDctr + 1:

end do:

save f_intra,`f_intra.m`:
save f_inter,`f_inter.m`:

                           -- 7 --

  \end{Verbatim}
 \end{minipage}

}

\end{appendices}

%% file: main.bbl
\begin{thebibliography}{10}

\bibitem{fortcastlor}
C.~Castellano, S.~Fortunato, and V.~Loreto.
\newblock Statistical physics of social dynamics.
\newblock {\em Rev. Mod. Phys.}, 81:591--646, May 2009.

\bibitem{Niyogi}
P.~Niyogi and R.~C. Berwick.
\newblock {Evolutionary Consequences of Language Learning}.
\newblock {\em Linguistics and Philosophy}, 20(6):697--719, 1997.

\bibitem{Nowak:1}
M.~A. Nowak and D.~C. Krakauer.
\newblock {The evolution of language}.
\newblock {\em Proc. Natl. Acad. Sci. USA}, 96:8028--8033, 1999.

\bibitem{Nowak:2}
M.~A Nowak, J.~B. Plotkin, and D.~C. Krakauer.
\newblock {The Evolutionary Language Game}.
\newblock {\em J. of Theor. Biol.}, 200(2):147--162, 1999.

\bibitem{Nowak:3}
M.~A. Nowak, N.~L. Komarova, and P.~Niyogi.
\newblock {Evolution of Universal Grammar}.
\newblock {\em Science}, 291(5501):114--118, 2001.

\bibitem{Smith}
K.~Smith, S.~Kirby, and H.~Brighton.
\newblock {Iterated Learning: A Framework for the Emergence of Language}.
\newblock {\em Artif. Life}, 9(4):371--386, 2003.

\bibitem{Komarova}
N.~Komarova and P.~Niyogi.
\newblock {Optimizing the mutual intelligibility of linguistic agents in a
  shared world}.
\newblock {\em Artif. Intell.}, 154(1):1--42, 2004.

\bibitem{Baronchelli:1}
A.~Baronchelli, M.~Felici, E.~Caglioti, V.~Loreto, and L.~Steels.
\newblock {Sharp transition towards shared vocabularies in multi-agent
  systems}.
\newblock {\em J. Stat. Mech. Theory Exp.}, P06014, 2006.

\bibitem{Luc:1}
L.~Steels.
\newblock {A self--organizing spatial vocabulary}.
\newblock {\em Artif. Life}, 2(3):319--332, 1995.

\bibitem{Luc:2}
L.~Steels.
\newblock {Self--organizing vocabularies}.
\newblock In C.~G. Langton and K.~Shimohara, editors, {\em Artif. Life V},
  pages 179--184. Nara, Japan, 1996.

\bibitem{Wittgenstein}
L.~Wittgenstein.
\newblock {\em Philosophical Investigations}.
\newblock Wiley--Blackwell, 4th edition, 2009 [1953].

\bibitem{Baronchelli:5}
A.~Baronchelli, L.~Dall'Asta, A.~Barrat, and V.~Loreto.
\newblock {Strategies for fast convergence in semiotic dynamics}.
\newblock In L.~M. Rocha, editor, {\em Artif. Life X: Proceedings of the Tenth
  International Conference on the Simulation and Synthesis of Living Systems},
  pages 480--485, Cambridge, MA, US, 2006. The MIT Press.

\bibitem{Baronchelli:3}
A.~Baronchelli, L.~Dall'Asta, A.~Barrat, and V.~Loreto.
\newblock {Topology--induced coarsening in language games}.
\newblock {\em Phys. Rev. E}, 73:015102, 2006.

\bibitem{Baronchelli:2}
L.~Dall'Asta, A.~Baronchelli, A.~Barrat, and V.~Loreto.
\newblock {Nonequilibrium dynamics of language games on complex networks}.
\newblock {\em Phys. Rev. E}, 74:036105, 2006.

\bibitem{Baronchelli:6}
L.~Dall'Asta, A.~Baronchelli, A.~Barrat, and V.~Loreto.
\newblock {Agreement dynamics on small--world networks}.
\newblock {\em EPL}, 73(6):969, 2006.

\bibitem{Baronchelli:7}
A.~Baronchelli, L.~Dall'Asta, A.~Barrat, and V.~Loreto.
\newblock {Bootstrapping communication in language games}.
\newblock In A.~Cangelosi, A.~D.~M. Smith, and K.~Smith, editors, {\em The
  Evolution of Language, Proceedings of the 6th International Conference
  (EVOLANG6)}. World Scientific Publishing Company, 2006.

\bibitem{Centola}
  D.~Centola and A.~Baronchelli.
  \newblock {The spontaneous emergence of conventions: An experimental study of
    cultural evolution}.
  \newblock {\em PNAS}, 112(7):1989--1994, 2015.

\bibitem{Lu:1}
Q.~Lu, G.~Korniss, and B.~K. Szymanski.
\newblock {The Naming Game in social networks: community formation and
  consensus engineering}.
\newblock {\em J. Econ. Interact. Coord.}, 4(2):221--235, 2009.

\bibitem{Schaeffer}
S.~E. Schaeffer.
\newblock {Graph clustering}.
\newblock {\em Comp. Sci. Rev.}, 1(1):27 -- 64, 2007.

\bibitem{Porter}
M.~A. Porter, J.~P. Onnela, and P.~J. Mucha.
\newblock Communities in networks.
\newblock {\em Notices Amer. Math. Soc.}, 56(9):1082--1097, 2009.

\bibitem{Fortunato:1}
S.~Fortunato.
\newblock {Community detection in graphs}.
\newblock {\em Phys. Rep.}, 486(3–-5):75--174, 2010.

\bibitem{Coscia}
M.~Coscia, F.~Giannotti, and D.~Pedreschi.
\newblock {A Classification for Community Discovery Methods in Complex
  Networks}.
\newblock {\em Stat. Anal. Data Min.}, 4:512--546, 2011.

\bibitem{Newman:book}
M.~Newman.
\newblock {\em {Networks: An Introduction}}.
\newblock Oxford University Press, Inc., 2010.

\bibitem{Xie}
J.~Xie, S.~Kelley, and B.~K. Szymanski.
\newblock {Overlapping Community Detection in Networks: The State--of--the--art
  and Comparative Study}.
\newblock {\em ACM Comput. Surv.}, 45(4):43:1--43:35, 2013.

\bibitem{Fortunato:2}
S.~Fortunato and D.~Hric.
\newblock {Community detection in networks: A user guide}.
\newblock arXiv:1608.00163v1 [physics.soc-ph].

\bibitem{Gubanov}
D.~A. Gubanov, L.~I. Mikulich, and T.~S. Naumkina.
\newblock {Language games in investigation of social networks: Finding
  communities and influential agents}.
\newblock {\em Autom. Remote Control}, 77(1):144--158, 2016.

\bibitem{Newman}
M.~E.~J. Newman and M.~Girvan.
\newblock {Finding and evaluating community structure in networks}.
\newblock {\em Phys. Rev. E}, 69:026113, 2004.

\bibitem{Lancichinetti}
A.~Lancichinetti and S.~Fortunato.
\newblock {Benchmarks for testing community detection algorithms on directed
  and weighted graphs with overlapping communities}.
\newblock {\em Phys. Rev. E}, 80:016118, 2009.

\bibitem{Girvan:1}
M.~Girvan and M.~E.~J. Newman.
\newblock {Community structure in social and biological networks}.
\newblock {\em Proc. Natl. Acad. Sci. USA}, 99:7821--7826, 2002.

\bibitem{Flake}
G.~W. Flake, S.~Lawrence, C.~L. Giles, and F.~M. Coetzee.
\newblock {Self--organization and identification of Web communities}.
\newblock {\em Computer}, 35(3):66--70, 2002.

\bibitem{Lambiotte}
R.~Lambiotte and M.~Ausloos.
\newblock {Coexistence of opposite opinions in a network with communities}.
\newblock {\em J. Stat. Mech. Theory Exp.}, 2007(08):P08026, 2007.

\bibitem{Candia}
J.~Candia and K.~I. Mazzitello.
\newblock Mass media influence spreading in social networks with community
  structure.
\newblock {\em J. Stat. Mech. Theory Exp.}, 2008(07):P07007, 2008.

\bibitem{Baronchelli:4}
X.~Castell\'o, A.~Baronchelli, and V.~Loreto.
\newblock {Consensus and ordering in language dynamics}.
\newblock {\em Eur. Phys. J. B}, 71:557--564, 2009.

\bibitem{Xie:1}
J.~Xie, S.~Sreenivasan, G.~Korniss, W.~Zhang, C.~Lim, and B.~K. Szymanski.
\newblock {Social consensus through the influence of committed minorities}.
\newblock {\em Phys. Rev. E}, 84:011130, 2011.

\bibitem{Xie:2}
J.~Xie, J.~Emenheiser, M.~Kirby, S.~Sreenivasan, B.~K. Szymanski, and
  G.~Korniss.
\newblock Evolution of opinions on social networks in the presence of competing
  committed groups.
\newblock {\em PLoS ONE}, 7(3):e33215, 2012.

\bibitem{Palombi}
F.~Palombi and S.~Toti.
\newblock Stochastic dynamics of the multi--state voter model over a network
  based on interacting cliques and zealot candidates.
\newblock {\em J. Stat. Phys.}, 156(2):336--367, 2014.

\bibitem{Mobilia:1}
M.~Mobilia.
\newblock {Does a Single Zealot Affect an Infinite Group of Voters?}
\newblock {\em Phys. Rev. Lett.}, 91:028701, 2003.

\bibitem{Mobilia:2}
M.~Mobilia, A.~Petersen, and S.~Redner.
\newblock {On the role of zealotry in the voter model}.
\newblock {\em J. Stat. Mech. Theory Exp.}, 2007(8):P08029, 2007.

\bibitem{Holland}
P.~W. Holland, K.~B. Laskey, and S.~Leinhardt.
\newblock {Stochastic blockmodels: First steps}.
\newblock {\em Social Networks}, 5(2):109--137, 1983.

\bibitem{Condon}
A.~Condon and R.~M. Karp.
\newblock {Algorithms for graph partitioning on the planted partition model}.
\newblock {\em Random Structures \& Algorithms}, 18(2):116--140, 2001.

\bibitem{McSherry}
F.~McSherry.
\newblock {Spectral partitioning of random graphs}.
\newblock In {\em 42nd IEEE Symposium on Foundations of Computer Science
  (FOCS)}, pages 529--537, 2001.

\bibitem{BarabasiNS}
A.{--}L. Barab\'asi.
\newblock {\em Network Science}.
\newblock Cambridge University Press, 2013.

\bibitem{Baronchelli:9}
A.~Baronchelli, L.~Dall'Asta, A.~Barrat, and V.~Loreto.
\newblock Nonequilibrium phase transition in negotiation dynamics.
\newblock {\em Phys. Rev. E}, 76:051102, 2007.

\bibitem{Arnold}
V.~I. Arnold.
\newblock {\em Ordinary Differential Equations}.
\newblock The MIT Press, 1973.

\bibitem{Atkinson}
K.~E. Atkinson.
\newblock {\em An introduction to Numerical analysis, 2nd ed.}
\newblock Wiley India Pvt. Limited, 2008.

\bibitem{Stanley}
H.~E. Stanley.
\newblock {\em Introduction to Phase Transitions and Critical Phenomena}.
\newblock Oxford University Press, 1971.

\bibitem{Dickman:1}
R.~Dickman and R.~Vidigal.
\newblock {Quasi-stationary distributions for stochastic processes with an
  absorbing state}.
\newblock {\em J. Phys. A}, 35(5):1147, 2002.

\bibitem{Dickman:2}
R.~Dickman.
\newblock {Numerical analysis of the master equation}.
\newblock {\em Phys. Rev. E}, 65:047701, 2002.

\bibitem{Baronchelli:8}
L.~Dall'Asta and A.~Baronchelli.
\newblock {Microscopic activity patterns in the naming game}.
\newblock {\em J. Phys. A}, 39(48):14851, 2006.

\bibitem{forceatlas}
T.~Kamada and S.~Kawai.
\newblock {An algorithm for drawing general undirected graphs}.
\newblock {\em Inform. Process. Lett.}, 31(1):7--15, 1989.

\bibitem{BarabasiSF}
A.{--}L. Barab{\'a}si and A.~R{\'e}ka.
\newblock Emergence of scaling in random networks.
\newblock {\em Science}, 286(5439):509--512, 1999.

\bibitem{Vespignani}
R.~Pastor-Satorras and A.~Vespignani.
\newblock Epidemic spreading in scale-free networks.
\newblock {\em Phys. Rev. Lett.}, 86:3200--3203, 2001.

\bibitem{VespignaniTwo}
R.~Pastor-Satorras and A.~Vespignani.
\newblock Epidemic dynamics and endemic states in complex networks.
\newblock {\em Phys. Rev. E}, 63:066117, 2001.

\bibitem{Ponti}
G.~Ponti et~al.
\newblock {The role of medium size facilities in the HPC ecosystem: the case of
  the new CRESCO4 cluster integrated in the ENEAGRID infrastructure}.
\newblock In {\em Proceedings of the 2014 International Conference on High
  Performance Computing and Simulation - HPCS2014}, number 6903807, pages
  1030--1033, 2014.

\end{thebibliography}
